\documentclass[letterpaper]{article} 
\usepackage{pdflscape}
\usepackage{aaai25}  
\usepackage{times}  
\usepackage{helvet}  
\usepackage{courier}  
\usepackage[hyphens]{url}  
\usepackage{graphicx} 
\usepackage{natbib}  
\usepackage{caption} 
\usepackage{amsmath,amssymb,booktabs,microtype,xcolor}
\usepackage{enumitem}
\usepackage{subcaption}
\usepackage{graphicx}
\usepackage{tcolorbox}
\usepackage{cleveref}
\usepackage{xspace}
\usepackage{cleveref}
\usepackage{dialogue}
\usepackage{subcaption}
\usepackage{amssymb}
\usepackage{booktabs}
\usepackage{multirow}
\usepackage{glossaries}
\usepackage{listings}
\usepackage{duckuments}
\usepackage{xcolor}
\usepackage{colortbl}
\usepackage{longtable}
\usepackage[font=normalsize,labelfont=bf]{caption}
\usepackage{calc}
\usepackage{tabularray}

\usepackage{tikz}

\newcommand{\arxivUrl}
{\url{https://arxiv.org/abs/2508.08345}}

\newcommand\numindicators{30\xspace}
\newcommand\numcells{480\xspace}

\newcommand\numcompanies{16\xspace}

\newcommand\tldrDone[1]{}

\newcommand\eg{e.g.\xspace}

\newcommand{\tikzxmark}{%
\tikz[scale=0.23] {
    \draw[line width=0.7,line cap=round] (0,0) to [bend left=6] (1,1);
    \draw[line width=0.7,line cap=round] (0.2,0.95) to [bend right=3] (0.8,0.05);
}}
\newcommand{\tikzcmark}{%
\tikz[scale=0.23] {
    \draw[line width=0.7,line cap=round] (0.25,0) to [bend left=10] (1,1);
    \draw[line width=0.8,line cap=round] (0,0.35) to [bend right=1] (0.23,0);
}}

\begin{document}

\title{Do AI Companies Make Good on Voluntary Commitments to the White House?}


\author{
    Jennifer Wang\textsuperscript{\rm 1},
    Kayla Huang\textsuperscript{\rm 2},
    Kevin Klyman\textsuperscript{\rm 3},
    Rishi Bommasani\textsuperscript{\rm 3}
}
\affiliations{
    \textsuperscript{\rm 1}Brown University, Providence, RI, USA \\
    \textsuperscript{\rm 2}Harvard University, Cambridge, MA, USA \\
    \textsuperscript{\rm 3}Stanford University, Stanford, CA, USA \\
    jennifer\_wang2@brown.edu, kaylahuang@college.harvard.edu, kklyman@stanford.edu, 
    nlprishi@stanford.edu
}




\maketitle
\begin{abstract}
Voluntary commitments are central to international AI governance, as demonstrated by recent voluntary guidelines issued from the White House to the G7, from Bletchley Park to Seoul. 
But do AI companies actually make good on their commitments?
We score 16 companies based on their publicly disclosed behavior by developing a detailed rubric based on their eight voluntary commitments to the White House in 2023.
We find significant heterogeneity: while the highest-scoring company (OpenAI) scores 83.3\% overall on our rubric, the average score across all companies is just 53\%. 
The companies demonstrate systemically poor performance on their commitment to model weight security, with an average score of 17\%: 11 of the 16 companies receive 0\% for this commitment.
Our analysis highlights a clear structural shortcoming that future AI governance initiatives should correct: when companies make public commitments, they should proactively disclose how they meet their commitments to provide accountability, and these disclosures should be verifiable.
To advance policymaking on corporate AI governance, we provide three directed recommendations that address underspecified commitments, the role of complex AI supply chains, and public transparency that could be incorporated into AI governance initiatives worldwide.
\end{abstract}
\section{Introduction}

The growing importance of artificial intelligence (AI) has rapidly catalyzed global policymaking efforts.
Policymaking related to AI addresses many concerns including open innovation, market concentration, risk management, corporate governance, and geopolitics.
Since 2023, many AI policy efforts have centered on the interplay between corporate governance, given that prominent AI systems are developed by the world's most powerful companies, and risk reduction, due to the breadth of potential harms associated with AI systems.

The approach to global AI policy varies significantly across jurisdictions.
A key differentiator among jurisdictions that regulate AI companies is whether a policy imposes mandatory or voluntary obligations on companies.
Some jurisdictions have enacted mandatory requirements via legislative or executive action, such as the EU AI Act and the US Executive Order on the Safe, Secure, and Trustworthy Development and Use of Artificial Intelligence respectively.
However, much of global AI policy centers on voluntary actions taken by major companies in line with recommendations by government bodies.
Key examples include the NIST AI Risk Management Framework, the 2023 White House Voluntary Commitments on AI, the G7 International Code of Conduct, Canada's Voluntary Code of Conduct on the Responsible Development and Management of Advanced Generative AI Systems, the 2024 White House Voluntary Commitments to Combat Image-Based Sexual Abuse, and the Frontier AI Safety Commitments secured at 2024 AI Seoul Summit.
Voluntary measures offer flexibility in that they can allow companies to pilot different approaches to meeting commitments, optimize for objectives other than minimizing legal risk associated with regulatory compliance, and harmonize approaches across jurisdictions despite different legal and political systems.

But policy initiatives that rely on companies to voluntarily take action have a number of pitfalls.  
Voluntary measures do not come with penalties for noncompliance, meaning that companies may choose to not participate, claim they are participating but not implement the government's recommendations, opt for partial implementation, or implement recommendations in ways that are opaque or not verifiable.
Well-intentioned companies may have difficulty complying because voluntary measures are less likely to move markets and reorganize supply chains, meaning that measures requiring coordinated action may be less likely to succeed if voluntary.
Voluntary measures often lack any mechanism for monitoring implementation, presenting a potential loophole for noncompliance \cite{aragoncorrea2020effects}.
A company's public commitment that it will adhere to voluntary measures can give the illusion that the company is taking significant action to responsibly develop and deploy AI systems while it does not in fact make any changes.

To understand the impact and efficacy of voluntary commitments, we conduct the first comprehensive analysis of the first major commitments to governments made by top AI companies.
\footnote{This version of the paper has been abridged for the AIES format. Please find the full version on arXiv instead: \arxivUrl.}
In 2023, the White House secured voluntary commitments from 15 AI companies.\footnote{The commitments were secured in three phrases: (i) Amazon, Anthropic, Google, Inflection, Meta, Microsoft, and OpenAI committed in July 2023; (ii) Adobe, Cohere, IBM, Nvidia, Palantir, Salesforce, Scale AI, and Stability AI committed in September 2023; and (iii) Apple committed in July 2024, following the launch of its Apple Intelligence product.}
In announcing the commitments, the White House described their purpose as follows: ``These commitments, which the companies have chosen to undertake immediately, underscore three principles that must be fundamental to the future of AI – safety, security, and trust – and mark a critical step toward developing responsible AI. As the pace of innovation continues to accelerate, the Biden-Harris Administration will continue to remind these companies of their responsibilities and take decisive action to keep Americans safe.'' Although the Biden Administration’s AI Executive Order was later rescinded, the voluntary commitments secured from companies were not undone.

To reason about the companies and their behavior, we score companies based on how their public actions address their stated commitments.
We design a scoring rubric that transforms the eight commitments specified by the White House on product safety, system security, and public trust into \numindicators indicators.
Our rubric provides concrete and decidable criteria for determining if a company has satisfied its commitment.
To score the 16 companies that signed the 2023 White House Voluntary Commitments on AI, for each of the 480 (indicator, company) pairs, we gather relevant public information through December 31, 2024, assign a score, and provide evidence for our decision.

By compiling information about company practices and interpreting it via quantitative scores, we provide evidence for three key findings.
First, the scores demonstrate significant heterogeneity in companies' actions: the top-scoring company (OpenAI) scores 83\% on our rubric, whereas the bottom-scoring company (Apple) scores 13\%.
Of the eight commitments, there are six commitments where at least one company scores 100\%; at the same time, there are five commitments where at least one company scores 0\%.
Second, company-level scores demonstrate two clear, and interconnected, correlations: members of the Frontier Model Forum and earlier signatories tend to score higher.
The six highest scoring companies are the six members of the Frontier Model Forum (OpenAI, Anthropic, Google, Microsoft, Meta, Amazon) and each score at least 60\%.
Third, model weight security is a commitment with distinctively poor performance: companies score on average 17\%.
11 companies score 0\% on this commitment (Adobe, Apple, Cohere, IBM, Inflection, Meta, Nvidia, Palantir, Scale AI, Salesforce, Stability AI).

Beyond providing empirical insight into the relationship between company practices and stated commitments, our work reveals a key design flaw in the 2023 White House Voluntary Commitments on AI: companies made public commitments to the White House, but no mechanism was created to monitor implementation or provide the public with information about implementation.
To improve the design of future voluntary commitments related to corporate AI governance, we provide three recommendations to policymakers.\footnote{Unlike the 2023 White House Voluntary Commitments on AI, we find the 2024 White House Voluntary Commitments to Combat Image-Based Sexual Abuse adopt some of our recommendations.}
\begin{enumerate}
    \item \textbf{Commitments should be precise and specific.}
    The wording of the 2023 White House Voluntary Commitments is often vague, leading to significant ambiguity over the intent of a commitment and the steps required to satisfy a commitment.
    Commitments should be precise, specifying (i) what is the specific goal and (ii) what evidence is sufficient or satisfactory to indicate completion.
    \item \textbf{Commitments should be targeted.}
    Since the same commitments are directed towards companies with different business models and roles in the AI supply chain, some commitments appear inappropriate for some companies (\eg increased cybersecurity around model weights for companies that (largely) do not develop models). 
    In contrast, commitments should be tailored to either (i) specific companies (\eg if they operate across several levels of the supply chain) or (ii) a specific layer of the supply chain, clearly designating which companies belong to that layer.
    \item \textbf{Commitments should enable public verification.}
    Though the 2023 White House Voluntary Commitments on AI were issued more than two years ago, the actions that companies have taken in order to fulfill their stated commitments remains highly uncertain based on public information.
    Given that these commitments are made publicly, we recommend that commitments include accountability measures (\eg companies publish a transparency report six months after making commitments to indicate what actions they took for each commitment), especially to clarify whether companies changed their actions relative to what they may have done absent making such commitments.
\end{enumerate}

\section{The 2023 White House Voluntary Commitments on AI}
\textbf{Context.} In 2023, the White House secured eight voluntary commitments with 15 leading AI companies: they are ``commitments that companies are making to promote the safe, secure, and transparent development and use of generative AI (foundation) model technology'' \citep{whvc}. 
At a high level, these commitments indicate that companies who are signatories will uphold three duties: 
(i) ensure their products are safe before public release,
(ii) implement security practices for their AI models and systems, 
and 
(iii) earn public trust through responsible AI development.
The commitments stated that companies intend to follow these commitments, alongside existing laws, until regulations that cover the same issues come into force.
\newline \newline
\noindent \textbf{Scope.} In the initial July 2023 round of voluntary commitments, signed by seven companies at the time, the commitments were scoped to ``generative models that are overall more powerful than the current industry frontier (e.g. models that are overall more powerful than any currently released models, including GPT-4, Claude 2, PaLM 2, Titan and, in the case of image generation, DALL-E 2)''.
When the White House announced in September 2023 that eight additional companies had signed, it modified the scope of the commitments to ``generative
models that are overall more powerful than the current most advanced model produced by the company making the commitment''.
\newline \newline
\noindent \textbf{Commitments.}
The first commitment is to conduct internal and external red-teaming of models or systems, focusing on risks including chemical, biological, radiological, and nuclear threats, cyber capabilities, autonomous system control, societal risks, and broader national security concerns.
The second commitment addresses information sharing with different parties (\eg other companies and governments) around trust and safety concerns, dangerous or emergent capabilities, and attempts to circumvent safeguards. 
Together, these commitments address the topic of product safety.

The next two commitments address system security.
The third commitment covers the protection of proprietary and unreleased model weights through model-level cybersecurity, safeguards against insider threats, and personnel-level restricted access.
Building on these company-internal practices, the fourth commitment encourages external discovery of vulnerabilities via bounties for third-party reporting.

The final four commitments collectively address public trust.
These span commitments around content provenance methods and standards (commitment five), public reporting on capabilities and safety (commitment six), research on societal risks including empowering internal trust and safety teams (commitment seven), and prioritizing progress on society's greatest challenges as well as student, worker, and citizen engagement (commitment eight).
\section{Scoring Methodology}
To score companies, we define \numindicators indicators, gather public information on these indicators for each company, and use this information to support our score. 
Our methodology is inspired by the 2023 Foundation Model Transparency Index \citep{bommasani2023fmti}.

\subsection{Indicators}
\begin{figure*}
\centering
\includegraphics[keepaspectratio, width=0.95\textwidth]{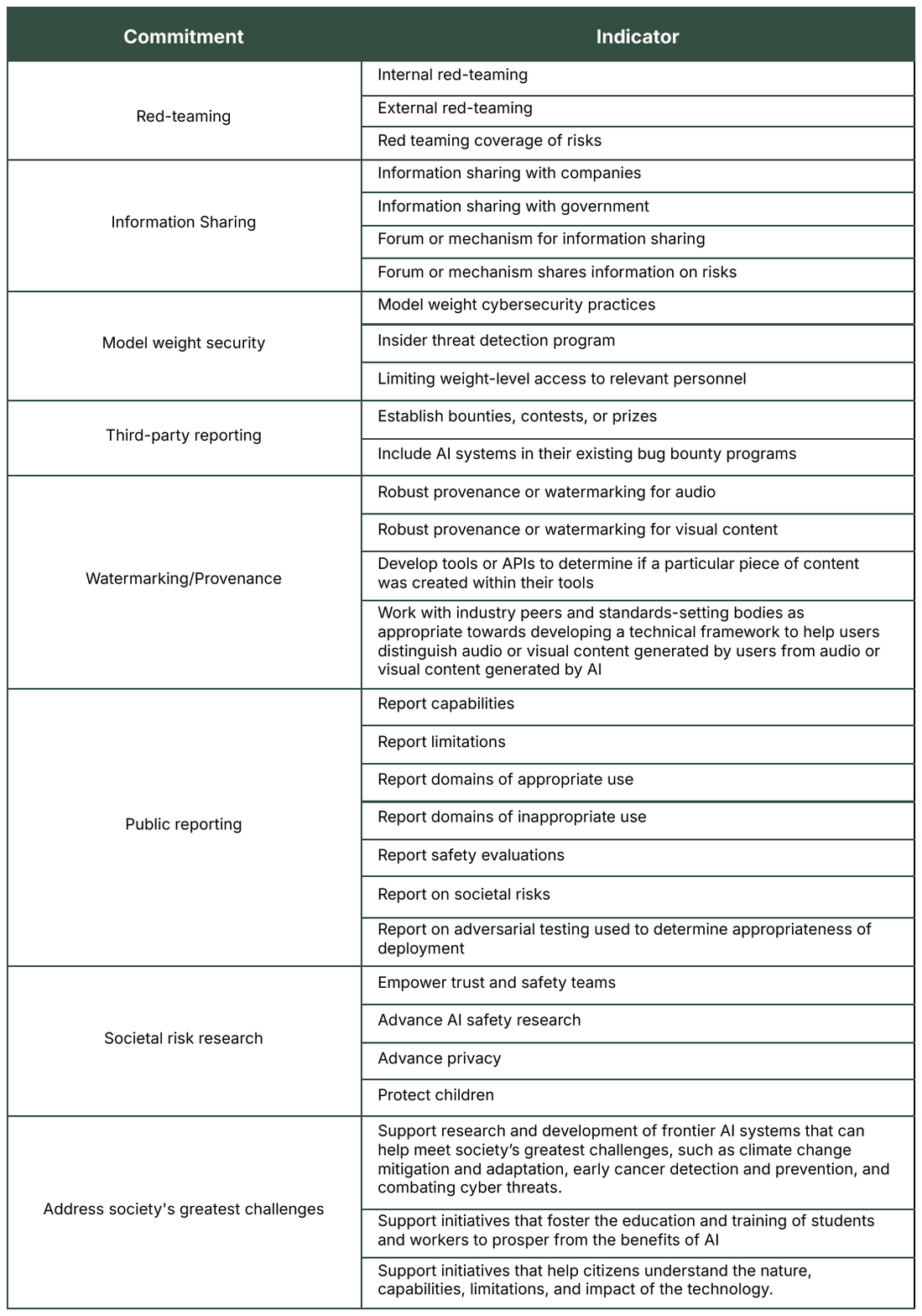}
\captionsetup{type=table}
\caption{Indicators. Table of the \numindicators indicators we use to score companies.
}
\label{fig:indicator-list}
\end{figure*}

The White House commitments \citep{whvc} are written as a combination of specific actions expected of companies and a more generic description of why these actions advance the public interest.
As written, the commitments do not provide decidable criteria for determining whether a company's actions are sufficient to state that they fulfilled the commitment.
Therefore, we define concrete \textit{indicators} that transform each high-level commitment into more specific, decidable criteria that we use to score companies. To maximize fidelity with the voluntary commitments, each indicator is a verbatim excerpt from the commitments. 
The reference text for each is in Appendix A.
Since the commitments vary in scope and content, we map each commitment to multiple indicators based on its wording.
The resulting mapping (see Figure \ref{fig:indicator-list}) yields 2--7 binary indicators per commitment and \numindicators indicators overall. 

As an example, consider the seventh voluntary commitment on public trust, which is entitled ``Prioritize research on societal risks posed by AI systems, including on avoiding harmful bias and discrimination, and protecting privacy''.
The commitment states: ``Companies commit generally to empowering trust and safety teams, advancing AI safety research, advancing privacy, protecting children, and working to proactively manage the risks of AI so that its benefits can be realized.''
We map this commitment to four indicators: (i) does the company empower its trust and safety teams? (ii) does the company advance AI safety research? (iii) does the company take steps to advance privacy? and (iv) does the company take steps to protect children?

We score each (company, indicator) pair on a binary basis. 
A score of 1 signifies that our search process surfaced publicly available documentation from the company that is sufficient to demonstrate that the company satisfied the portion of the 2023 White House Voluntary Commitments on AI captured by that indicator.
A score of 0 signifies that our search process did not surface such documentation, whether because the documents identified did not contain sufficient evidence to demonstrate the commitment was fulfilled or because no relevant documents were found through our search process.

We construct binary indicators for several reasons. 
First, our aim is to break the commitments down into distinct, decidable chunks that can be used to assess whether or not there is sufficient evidence that a specific sub-part of a commitment was or was not fulfilled.
Second, producing narrower criteria for scoring reduces subjectivity in assigning initial scores. 
Third, binary indicators simplify the scoring process by allowing scorers to focus on the sharp distinction between 0 and 1 point for each indicator \citep{bommasani2023fmti}.

We acknowledge that binary indicators are potentially reductive, leaving out valuable information that can be captured by more complex scoring schemes.
At the same time, a greater number of smaller, binary indicators can be aggregated to produce more complex scoring schemes, and the information we release associated with our scores could be used to produce alternate scores using different criteria.

\subsection{Information Gathering}
To score companies, we used public information released by the companies with no additional third-party sources.
In doing so, we highlight that companies, with the exception of commitment six on public reporting, did not commit to making such information publicly available.
It is therefore possible that companies do satisfy their voluntary commitments but do not provide any public evidence of implementation.
Given the high-profile and public nature of these commitments and companies' statements in support of public transparency \citep{bommasani2023fmti}, we believe it is appropriate to assess companies based on their public disclosures.

Nevertheless, companies may be motivated by values and interests other than public transparency.
For example, concerns regarding security may lead companies to not disclose information on their model-weight security practices and insider threat detection programs.
In some cases, companies may lack the authority to unilaterally disclose information related to their commitments, including information that has been shared with governments and/or other companies.\footnote{Potential motivations for a lack of transparency on matters like research into how frontier AI systems can help meet society's greatest challenges may be less well grounded, though absolute transparency could conflict with commercial interests.}
We emphasize the opportunity for Pareto improvement: companies likely can provide some additional information on their conduct to the public without any tradeoff with their financial, reputational, or security interests.

We score companies based on information we gathered by December 31, 2024---our scores do not reflect new information that was made available thereafter or models that have been released since.\footnote{Since some companies signed onto the commitments at different times, with Apple being a notable outlier in 2024 (compared to the other 15 companies in 2023), companies had varying amounts of time between their commitment and our scoring.}
We use information that is deliberately and directly disclosed by the company---other sources such as leaked information, media reporting, or external analysis is not used.
These decisions contribute to greater fairness when assessing companies and comparing their scores, as companies themselves control their scores by deciding what information to publish about their behavior.
\newline
\indent We gathered information in a three stage process.
First, we collected key reference documents for each company that describe their practices in relation to their generative AI models, systems, and products. 
These documents include (a) external-facing resources such as blog posts, press releases, and transparency reports, (b) resources useful for the research community such as research papers, technical reports, model cards, documentation for developers, and bug bounties, as well as (c) product policies and safety frameworks. 
These documents were identified through an initial review of publicly available materials for each company and then selected based on their relevance to the commitments. We prioritized materials that explicitly address how companies assess, mitigate, or communicate risks associated with their generative AI systems, as well as those that provide insight into internal governance structures or external accountability mechanisms.
\newline
\indent Second, we searched through these documents and produced additional resources by creating a search script and using a language model for standardized, automated search. 
For each (company, indicator) pair, we use the script to better narrow our search.
We query the Perplexity API with the following search string: ``What has \{COMPANY\_NAME\} done since the beginning of 2023 that might fit under: \{INDICATOR TEXT\}? Make sure to return links used to find this information. Keep it concise and make sure to return all links with no information from before 2023."\footnote{We considered various search APIs (including those from OpenAI, Anthropic, and Google) and prompts, eventually finding the Perplexity API performed best at surfacing new relevant documents.} For each link returned in the Perplexity response, we reviewed the source document for relevance. We note that Perplexity was used only to augment our information gathering process, not as a substitute for our manual search.
\newline
\indent Third, we compiled the sources resulting from the first two steps for every (company, indicator) pair as the basis for making scoring decisions.\footnote{The search scripts and compiled sources are released publicly under an MIT license at \url{https://github.com/rishibommasani/whvc}.}
While these compiled sources are not exhaustive---in significant part because companies often deprecate documents on their websites, bury important documentation several layers deep, or fail to adequately summarize their actions to fulfill public commitments---we reviewed hundreds of documents as part of this process.

\subsection{Scoring}
For each of the \numcompanies companies, we use the information gathered from the above process to produce initial scores for each of the \numindicators indicators. 

As we scored indicators and identified disagreements among scorers, we iteratively developed specific and measurable criteria to evaluate fulfillment of each indicator, requiring in every instance that evidence be publicly verifiable. These criteria reflect our interpretation of whether company actions align with the goals underlying the commitments, while remaining grounded in their language and scope. 

For instance, to assess if the company empowers its trust and safety team, we consider whether (1) the company explicitly identifies such a team and (2) the company's documentation indicates it adequately resources the team and/or provides it the authority to address potential risks. 
The criteria for every indicator can be found in Appendix D.

Two authors of this paper each independently assigned an initial score for every one of the \numcells (company, indicator) pairs. Both authors provided a source and a quote to justify each score. 
In the event of disagreement on a particular score, all of the authors of this work discussed, coming to agreement in assigning the final score.

The agreement rate was 75.6\% ($\frac{363}{480}$), reflecting substantial agreement. 
However, the ambiguity in the wording of the commitments and how they apply to each specific company was a core source of initial disagreement, as was the variation in the level of detail across companies' public documentation. 
We release the final score for all \numcells (company, indicator) pairs along with a justification for the score and associated reference(s) to public materials.

In the event that an indicator is related to a specific model or system (\eg whether the company implements model-weight cybersecurity practices), we score the company based on its flagship foundation model or system as of December 31, 2024.\footnote{The flagship model is defined as in \citet{bommasani2023transparency}: ``the foundation model that is most salient and/or capable
from the developer based on our judgment, which is directly informed by the company’s
public description of the model.''} 
We choose the flagship foundation model as an object of analysis because the September 2023 version of the commitments focus on the capabilities of the ``most advanced model'' for each company, while the July 2023 version explicitly named several companies' flagship models. 
In addition, many companies make their flagship foundation models (or derivatives) central to the bulk of their AI-based products and services due to their enhanced capabilities.
The mapping from companies to flagship models is provided in Appendix C. 
We acknowledge that other models and systems beyond the flagship models we consider may also fall in scope of the commitments.
\section{Results}
\begin{figure}
    \centering
    \includegraphics[keepaspectratio, width=0.4\textwidth]{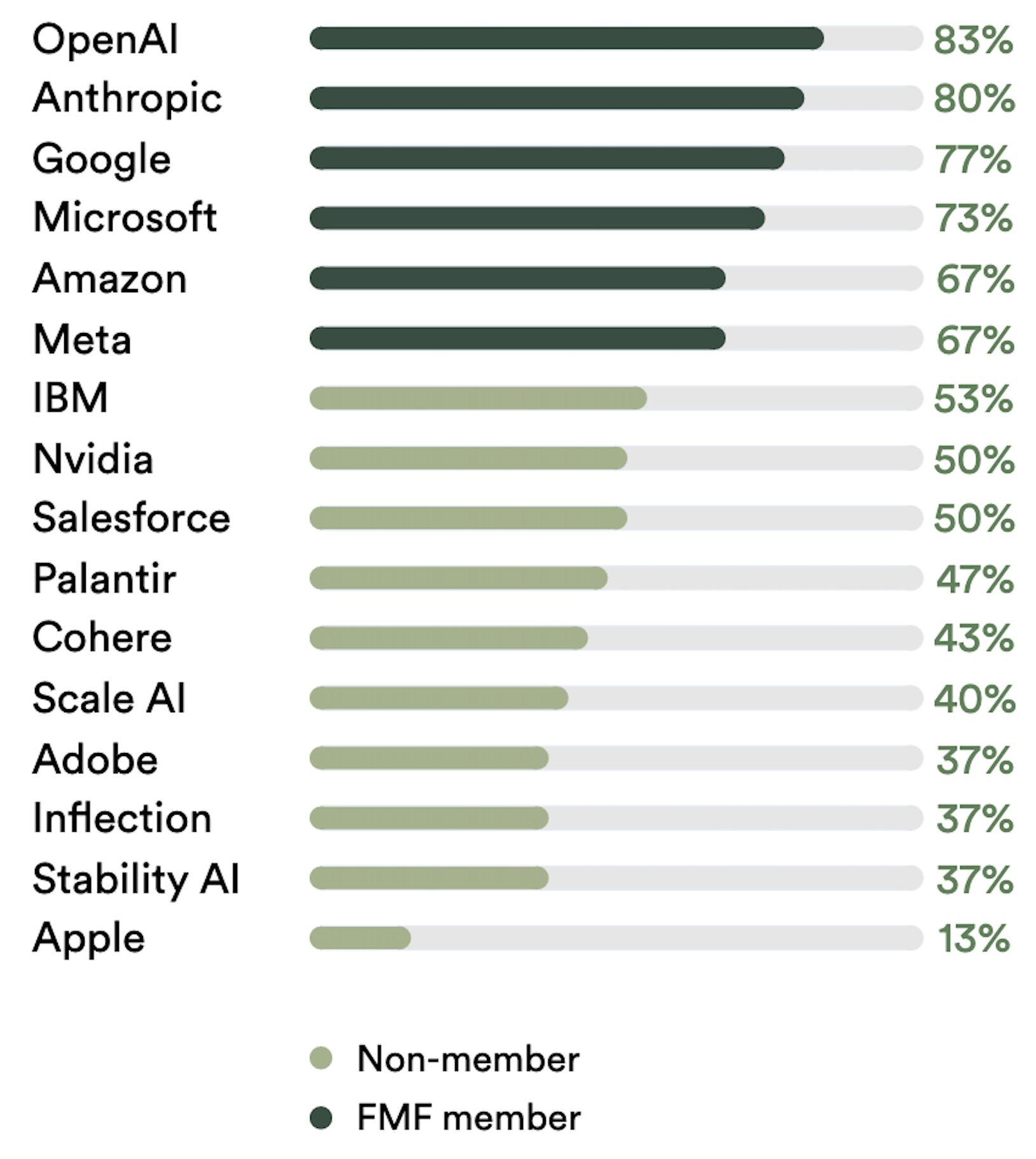}
    \caption{Aggregate scores by company.
    The score for each company, stratified by whether the company belongs to the Frontier Model Forum (FMF) as of December 31, 2024.}
    \label{fig:aggregate-company-scores}
\end{figure}

\noindent To organize our analysis, we apply three lenses:
(i) an overall company-level view,
(ii) a commitment-level view,
and (iii) a disaggregated indicator-level view. \newline
\indent In Figure \ref{fig:aggregate-company-scores}, we report the aggregate score as a percentage for each company across all the indicators.
The mean and median are 53.3\% and 50.0\% respectively, with a standard deviation of 19.5\%. 
The range is 70.0\% between the highest scoring company, OpenAI, at 83.3\% and the lowest scoring, Apple, at 13.3\%. 
While OpenAI satisfies 25 of the 30 indicators,\footnote{The indicators that OpenAI does not satisfy are: ``Insider threat detection program'', ``Report limitations'', ``Report domains of appropriate use'', ``Empower trust and safety teams'', and ``Support initiatives that help citizens understand the nature, capabilities, limitations, and impact of the technology''.} no company has a perfect score despite making these commitments to the White House over two years ago.\newline

\noindent \textbf{Significant variation in companies' scores.}
There is notable variation in how companies perform, with companies clustering into three distinct groups.
Four companies score at least one standard deviation above the mean: OpenAI (83.3\%), Anthropic (80.0\%), Google (76.7\%), and Microsoft (73.3\%).
The majority of companies fall within one standard deviation of the mean: Amazon (66.7\%), Meta (66.7\%), IBM (53.3\%), Nvidia (50.0\%), Salesforce (50.0\%), Adobe (36.7\%), Cohere (43.4\%), Palantir (36.7\%), Inflection (36.7\%), Stability AI (36.7\%), Scale AI (36.7\%).
The only company that scores at least one standard deviation below the mean is Apple (13.3\%).

\subsection{Company-Level Results}

\noindent \textbf{Frontier Model Forum members consistently score higher.}
Strikingly, the company-level scores clearly separate based on membership in the Frontier Model Forum (FMF).
The Frontier Model Forum is a non-profit industry association dedicated to advancing the safe development and deployment of frontier AI systems \citep{fmfaboutus}. 
Anthropic, Google, Microsoft, and OpenAI became the four founding members in July 2023, with Amazon and Meta joining in May 2024.
The six highest scoring companies, which all score at least 66\%, are the six FMF members. Their mean score is 74.4\% with a standard deviation of 6.9\%.
In contrast, the 10 other companies all score at or below 60\% with a mean of 40.7\% and a standard deviation of 11.4\%.
It is notable that FMF, in consultation with its members, has published technical reports intended in part to facilitate compliance with voluntary commitments \citep{fmfCassessments}. FMF states that its technical reports aim to ``examine how [Frontier AI] frameworks can be implemented effectively'' and acknowledges that such frameworks are the core component of the Frontier AI Safety Commitments at 2024 AI Seoul Summit \citep{fmfimplementingframeworks}. \\

\noindent \textbf{Earlier signatories generally score higher.}
The 16 companies signed onto the voluntary commitments in three phases of participation: seven companies in July 2023 (Amazon, Anthropic, Google, Inflection, Meta, Microsoft, and OpenAI), eight companies in September 2023 (Adobe, Cohere, IBM, Nvidia, Palantir, Salesforce, Scale AI, Stability AI) and one company in July 2024 (Apple).
We find company-level scores are clearly correlated with the timing of signature.
The first cohort has a mean of 69.0\% with a standard deviation of 15.6\%, while 
the second cohort has a mean of 44.6\% with a standard deviation of 6.4\%. 
It is possible this disparity reflects additional time the first cohort had before our scoring to publish documentation, but given that both cohorts had over 15 months prior to scoring, we hypothesize that the first cohort's business practices better align with the commitments.





\subsection{Commitment-Level Results}

\noindent \textbf{High scores for content provenance due to non-applicability.}
Based on the average per-commitment score for each company, the clear highest-scoring commitment is for (audiovisual) watermarking and provenance at 92.2\%.
14 companies receive 100\% for this commitment.\footnote{These companies are: Adobe, Amazon, Anthropic, Cohere, Google, IBM, Inflection, Meta, Microsoft, Nvidia, OpenAI, Palantir, Scale AI, Salesforce.}
In many cases, however, companies satisfy the associated indicators vacuously, because they do not develop audio or visual models, which are the subject of the provenance commitment.
Still, of the 8 companies that develop models with these output modalities, the average is 83.9\%, which exceeds that of all other commitments.
In particular, many of these companies follow industry standards associated with the Coalition for Content Provenance and Authenticity (C2PA) and their Content Credentials; 6 of the 8 companies are steering committee members of C2PA, while Apple and Stability AI are unaffiliated.
In contrast to the high scores for this commitment from most companies, Apple is the sole outlier as a company with audio and visual models that scores 0\% on these indicators. \\

\noindent \textbf{Low scores for model-weight security in spite of global emphasis.}
Based on the averages, the lowest scoring commitment is on model-weight security at 22.9\%.
Eleven companies score 0\% on this commitment, and none receives a full marks.
The high-scoring companies are OpenAI, Anthropic, and Microsoft at 66.7\%. Anthropic is the only company that indicates the existence of an insider risk program as part of its security standard. OpenAI and Microsoft, on the other hand, both state they create a secure research environment dedicated to model security and implement an access control protocol.
While transparency around model-weight security practices is valuable, we acknowledge that maximal transparency about security practices for model weights could undermine that very security.
However, the fact that every indicator is met by at least one company suggests that Pareto improvements are possible in how other companies navigate the transparency-security trade-off.
We emphasize the current results are particularly concerning given how model-weight security remains a clear challenge \citep{nevo2024securing} and features in many global AI policies (\eg the 
sixth commitment of the 
G7 International Code of Conduct \cite{g72023vc}, 
Section 3.1 of the 
US AI Safety Institute guidance on Managing Misuse Risk for Dual-Use Foundation Models \cite{nist2024managing}, 
Section 4 of 
US Executive Order 14141 on Advancing United States Leadership in Artificial Intelligence Infrastructure \cite{biden2025aiinfrastructure}). \\

\noindent \textbf{Low scores for third-party reporting align with concerns of chilling effects on third-party research.}
Alongside model-weight security, third-party reporting is another low-scoring commitment at 34.4\%.
Eight companies score 0\% on this commitment.
These low scores are especially surprising because 
the commitment is focused on providing bounties for reporting, and 
there are natural incentives for companies to make these bounties transparent to maximize external reporting.
Our finding aligns with those of \citet{longpre2024safeharbor}, who find that current company policies around AI-related bug bounties and protections for third-party research are unclear and uneven.
In particular, they argue that companies' policies suppress third-party reporting---given that researchers may be concerned with legal reprisal absent safe harbor (e.g. for responsible penetration testing)---instead of being supportive of such research, as required by this commitment.






\subsection{Indicator-Level Results}

\noindent \textbf{Extreme indicator-level scores align with commitment-level scores.} On average, each indicator is awarded to 8.5 of the 16 companies with a standard deviation of 4.9.
Seven indicators are satisfied by at least 14 companies (one standard deviation above the mean): four belong to the highest-scoring commitment on content provenance, two belong to the commitment on public reporting. These are ``Report capabilities'', which is satisfied by every company and is clearly incentivized by market forces, and ``Report domains of inappropriate use'', which is satisfied by every company except for Apple. The remaining indicator is to ``Establish or join a forum or mechanisms for information sharing'', which all companies receive on the basis of their membership in the US AI Safety Institute Consortium.

In contrast, five indicators are scored by at most three companies (one standard deviation below the mean): one is ``Insider threat detection program'' under the low-scoring model weight security commitment.
The other four are (i) ``Information sharing with government'', which only OpenAI and Anthropic satisfy by establishing memoranda of understanding with the US AI Safety Institute, (ii) ``Empower trust and safety teams``, which Google and Inflection satisfy by integrating trust and safety assessments into the model pre-launch processes and authorizing their teams to use a full range of tools to block malicious actors, (iii) ``Red teaming coverage of risks'', which OpenAI and Anthropic satisfied by conducting red-teaming exercises that address all the risk areas specified in the commitment, and (iv) ``Support initiatives that help citizens understand the nature, capabilities, limitations, and impact of the technology``, which none of the companies satisfied. \\

\noindent \textbf{Indicator-level analysis reveals substantial heterogeneity in information sharing.}
The information sharing commitment spans four indicators: information sharing with other companies (56.3\%), information sharing with governments (12.5\%), forum or mechanism for information sharing (100\%), and forum or mechanism that discloses information on risks (43.8\%).
While every company satisfies the indicator for a forum or mechanism for information sharing due to participation in the US AI Safety Institute Consortium, we do not automatically award the further point for sharing information on risks because it is not clear that this occurs in the Consortium.
Only seven companies are awarded this indicator, largely based on Frontier Model Forum membership.

Further, while the US AI Safety Institute Consortium was established by a governmental body in the National Institute of Standards and Technology, we do not automatically designate it as a means for information sharing with the government because our standard is that shared information should be non-public and do we not find evidence that companies share such information with the government through the Consortium.
As a result, only OpenAI and Anthropic score this indicator on the basis of their memoranda of understanding with the US AI Safety Institute, which  permit US AISI to directly access their models to perform risk assessments.
While companies do interface with the government in other ways---such as procurement of companies' AI systems, Congressional testimony from executives, and enforcement investigations into company practices---these are insufficient to satisfy this indicator. \\

\noindent \textbf{Certain indicators are overly vague, complicating consistent interpretation and meaningful implementation.}
While every indicator is only partially specified by the White House in its three-page document describing the voluntary commitments, some indicators are especially vague.
The clearest example is commitment seven, where ``Companies commit generally to empowering trust and safety teams, advancing AI safety research, advancing privacy, protecting children, and working to proactively manage the risks of AI so that its benefits can be realized''. 
All four of the resulting indicators are exceptionally broad and difficult to judge: what constitutes satisfactory privacy advancement or protection of children?
Even less clear is how these commitments are meant to relate with company practices on AI: for example, moderating the generation of child sexual abuse material and monitoring the use of language models by young children may both serve to protect children in very different senses.

Without concrete definitions to delineate what companies should do, companies and the public are highly unlikely to interpret the commitments in the same way.
In scoring these commitments, we chose to award points for constructive steps that met what we considered the minimum viable standard for public accountability and the maximally defensible standard absent greater clarification from the White House. 
Even if companies simultaneously demonstrated contradictory behavior, we credited them for taking steps aligned with the commitments (in order to establish a consistent baseline on company adherence).
For example, Meta received the point for ``Protecting children`` for its partnership with Thorn and the National Center for Missing \& Exploited Children, although the end-to-end encryption on its platforms prevents the detection of child sexual exploitation.
Taken together, the vagueness in how the commitments are articulated and the uncertainty regarding how to assess a company's practices in totality lead us to question whether such high-level commitments are meaningful.

\section{Related Work}
To contextualize our work, we discuss prior work that assesses major AI companies based on their public conduct and discuss other voluntary commitments.

\subsection{Assessments of AI Companies for 2023 WHVC}
Beyond our work, the most comprehensive analysis of company practices in relation to these commitments was conducted as part of a MIT Technology Review article published on the one-year anniversary of the commitments \citep{heikkila2024MIT}.
As part of this work, \citet{heikkila2024MIT} contacted the seven initial signatories and received responses from six of these companies, excluding Inflection, on how they addressed each of the commitments; external researchers also provided commentary.
Overall, the work found evidence that companies had taken steps to implement some technical model-level interventions(\eg red-teaming and watermarking) and made investments in safety research.
However, less evidence was found related to progress on information sharing, third-party reporting and public reporting.

\citet{heikkila2024MIT} indicates that no comprehensive evaluation had been performed of the commitments, company practices, or their relationship.
In light of this, our work not only provides a comprehensive assessment, but also introduces a concrete scoring system that yields quantitative findings.
In general, our findings largely agree with those of \citet{heikkila2024MIT} and \citet{roose2023NYT}, with the main difference being the depth and specificity of our results, though we highlight that our scores are based on public information from companies whereas the prior work only considered the brief responses companies provided to journalists. 
Further, our work expands the focus to the full set of 16 companies, rather than just the initial seven, which enables us to identify clear disparities between the initial signatories and the remaining signatories.

\subsection{Assessments of AI Companies}
As technology companies have grown in importance and become some of the world's most powerful entities, a multidisciplinary body of literature has emerged to assess these companies with a variety of methods.
In the space of quantitative assessments, several works have introduced scoring approaches either in the form of one-off analyses, akin to this work, or sustained indices, which score the same companies on a recurring cadence.
As an illustrative example, we highlight the Corporate Accountability Index that is maintained by Ranking Digital Rights (RDR), which has scored telecommunication and technology companies since 2015 for how they ``respect users’ fundamental rights, and on the mechanisms they have in place to ensure those promises are kept'' \citep{rdr2020index}.
\citet{kogen2024rdr} analyzed the 2018 Index and showed, by reviewing internal RDR documents and interviewing relevant stakeholders (\eg representatives from 11 companies and 14 civil society groups), that it usefully communicated legible, newsworthy, and flexible information that empowered social movements.

Drawing upon this tradition, several recent works have employed and developed similar scoring approaches for the assessment of AI companies \citep{bommasani2023eu-ai-act, bommasani2024foundationmodeltransparencyindex, klyman2024acceptableusepoliciesfoundation, longpre2024safeharbor, ailabwatchCommitments, ohEigeartaigh2023ai-safety, barrett2023AI-RMS, adamjonesCompaniesDoing}.
To our knowledge, \citet{bommasani2023eu-ai-act} provided the first assessment of major AI companies by scoring them on a rubric based on the European Parliament's proposal for the EU AI Act.
Based on the results, they made evidence-based recommendations aimed at (i) EU legislators on how the EU AI Act should be updated during the legislative negotiation and (ii) companies on how they could modify their practices to better align with the proposed requirements.
While some works similarly link scoring to specific governmental policies (\eg \citet{barrett2023AI-RMS} assess companies in relation to NIST's AI Risk Management Framework, \citet{ohEigeartaigh2023ai-safety} score in relation to the UK's recommendations), other works provide independent specification of the indicators or criteria of interest.
The Foundation Model Transparency Index is an annual index that scores foundation model developers for their transparency across the supply chain with 100 indicators that span the resources used to build a model (\eg data, compute), the properties of the model itself (\eg capabilities, risks), and the use of the model in society (\eg distribution, impact) \citep{bommasani2023fmti, bommasani2024foundationmodeltransparencyindex}. 

Cumulatively, these works all demonstrate a shared methodology of scoring companies with different approaches for sourcing the indicators, determining the scores, and theories of change for how the results and takeaways improve corporate governance and/or public policy.
Many of these works also share two key findings with our work.
While all of these works aim to increase public accountability, they all encounter limits due to the lack of transparency into company-internal practices. 
And, while the exact magnitudes and details often differ, these works almost always find considerable heterogeneity in company practices.
Together, they highlight the absence of clear norms, let alone more formal mechanisms, for ensuring public-facing transparency and standardizing industry-wide conduct.

\subsection{Voluntary Commitments From Governments}
Global AI policy reflects a broad constellation of efforts that spans long-standing policy in specific domains (\eg applying hiring discrimination laws to algorithmic hiring), more recent policy for digital technologies (\eg applying data protection laws to training data), and new policy for AI specifically (\eg new laws to govern AI).
While many jurisdictions face shared challenges, the overall global AI policy landscape reflects significant heterogeneity that indicates both region-specific considerations and idiosyncratic differences. 
In particular, when considering AI-specific policy, several jurisdictions currently employ voluntary approaches to corporate governance with the European Union's approach via the EU AI Act standing as a clear counter example.
At this juncture, given many of these voluntary and/or mandatory policies are very recent, little evidence exists to empirically validate the strengths and/or weaknesses of these two top-level approaches.

As a result, we briefly survey some of the voluntary commitments and approaches taken elsewhere in the world to contextualize the approach taken in the 2023 White House Voluntary Commitments on AI.
The U.S. NIST AI Risk Management Framework, as well as the associated profile on generative AI in particular, provides voluntary guidance to help organizations identify, assess, manage, and mitigate risks by emphasizing trustworthy AI principles such as fairness, transparency, accountability, security, and privacy \citep{tabassi2023airmf}.
The Canada Voluntary Code of Conduct on the Responsible Development and Management of Advanced Generative AI Systems introduces voluntary commitments applicable to the responsible development and deployment of foundation models, such as accountability, safety, fairness, human oversight, and robustness, as well as for developers and managers of generative AI systems \citep{canada2023vc}.
The G7 International Code of Conduct for Organizations Developing Advanced AI Systems articulates 11 commitments that span data protection, risk management, technical standard, and transparency reporting: while companies not signed on in the same way they have done for certain national-level commitments, these commitments may serve as the basis for global agreement \citep{g72023vc}.
Most recently, the Biden-Harris Administration secured voluntary commitments with AI model developers and data providers to prevent and mitigate the misuse of AI in creating and disseminating image-based sexual abuse content \citep{wh24vc}.

Beyond these standalone voluntary commitments, the ongoing series of international AI Summits have emerged as a key generative process for voluntary commitments as global policymakers work together to advance AI governance.
Beginning with the U.K. AI Safety Summit in November 2023, the Bletchley Declaration was signed by 29 world governments to foster international cooperation on AI policy through an agenda centered on (i) ``identifying AI safety risks of shared concern, building a shared scientific and evidence-based understanding of these risks, and sustaining that understanding \dots'' as well as  (ii) ``building respective risk-based policies across our countries to ensure safety \dots~alongside increased transparency by private actors developing frontier AI capabilities, appropriate evaluation metrics, tools for safety testing, and developing relevant public sector capability and scientific research''.
To advance this agenda, at the subsequent AI Seoul Summit in May 2024, 16 global companies (Amazon, Anthropic, Cohere, Google, G42, IBM, Inflection AI, Meta, Microsoft, Mistral AI, Naver, OpenAI, Samsung Electronics, Technology Innovation Institute, xAI, Zhipu.ai) signed onto the Frontier AI Safety Commitments.
The associated eight commitments address three outcomes: (i) improved risk management practices, (ii) increased accountability for safe development and deployment and (iii) sufficient transparency to external stakeholders.
Building on these efforts, the United States convened the growing global network of AI Safety Institutes in November 2024 for a working meeting on three high-priority topics (managing risks from synthetic content, testing foundation models, and conducting risk assessments for advanced AI systems) that articulated six principles for risk assessment (actionability, transparency, comprehensiveness, multi-stakeholder consideration, iterativity, and reproducibility).

\section{Discussion}
Our research into the voluntary commitments leads us to consider:
(i) future-looking commitment and policy design,
and
(ii) current corporate practices and governance.\\

\noindent \textbf{Commitments should be clearly worded.}
The White House Voluntary Commitments were first announced with a fact sheet and an accompanying three-page document.
While these documents are likely intended for public consumption and, therefore, provide generic high-level description, they are ambiguous.
In particular, some commitments are vague in terms of their intent (\eg language such as ``protect children''), especially when targeted at companies with large footprints and many roles in the AI supply chain.
Further, all commitments lack conditions for what constitutes satisfactory conduct.
While voluntary approaches permit flexibility to avoid being overly prescriptive or burdensome, these goals are achievable while still communicating about what is desired, especially for actions that can vary greatly in magnitude (\eg how much internal or external red-teaming is desired?). 
We recommend that commitments be precisely worded so that they articulate specific goals along with what constitutes sufficient evidence of completion.
Practically, these lower-level details may need to be split out into appendices or supporting documents, but the goal of broad intelligibility for the public need not be at odds with meaningful precision for deeply engaged stakeholders.  \\

\noindent \textbf{Commitments should be clearly targeted.}
The voluntary commitments, across their three phases of signing, specify essentially the same commitments for all 16 signatories.
However, these companies occupy significantly different positions in the AI ecosystem: they differ in their business models, their set of roles in the supply chain, and how their AI-related practices mediate public outcomes.
Given their uniform treatment under the commitments, some commitments generally made little sense for certain companies (\eg increased cybersecurity around model weights for companies that (largely) do not develop models).
While these commitments could have future-facing utility, we ultimately are skeptical of this one-size-fits-all approach, especially given our empirical findings that massive technology companies may take positive action in one part of their business practice while regressing in another.
We recommend that commitments either be tailored for each company or, when trying to standardize across companies, be tailored to a specific supply chain role.
The 2024 White House Voluntary Commitments to Combat Image-Based Sexual Abuse adopts this approach: for example, Meta and Microsoft have differentiated obligations that reflect how they operate different platforms downstream that contribute to the distribution of this imagery. \\

\noindent \textbf{Commitments should enable public verification.}
The voluntary commitments, except for commitment six on public reporting, specify no means for the public to understand or verify how companies took action to realize their commitments.
Empirically, our entire analysis and that of \citet{heikkila2024MIT} make clear that public insight is limited, even given more than a year has elapsed since the commitments were first made.
This directly contradicts one of the three stated goals of the voluntary commitments, which is to increase public trust.
Moving forward, commitments could be accompanied by accountability mechanisms (\eg a standardized transparency report that articulates how specific company actions address specific commitments) to address the clear gap we observe.
We recommend that public commitments, especially those made between very high profile institutions like the U.S. federal government and major AI companies, require periodic public transparency.\\

\noindent \textbf{Concerning practices.}
Beyond the specific practices we score, we highlight that some companies have released materials or otherwise discussed their conduct in relation to the commitments. 
These companies include Amazon \citep{amazonprogress}, Anthropic \citep{anthropicwh}, Google \citep{googlewhitehouse}, Meta \citep{metasafety}, Microsoft \citep{mssafety}, OpenAI \citep{openaisafety}, Inflection \citep{inflectionwh}, and Salesforce \citep{salesforceai}.
In reviewing these references, we at times disagreed with the company's claims that their conduct satisfactorily addresses the voluntary commitments. 
For example, Meta claims to have fulfilled the commitment on information sharing by publicly releasing artifacts about their models' capabilities and limitations. 
While these artifacts earned them points on public reporting, we only awarded points to companies for information sharing beyond public disclosure. Separately, Salesforce credits themselves for incentivizing third-party discovery through their bug bounty program to prevent AI-powered cyber threats. 
However, Salesforce does not specify that their AI systems are covered under the scope of this program, and therefore we did not award them the point on third-party reporting. 
In part, this reflects that these statements are often simultaneously high-level (\eg ``we’re prioritizing cybersecurity safeguards to protect proprietary and unreleased models and we’re participating in industry-
wide events to support broader protections...'') and are made without accompanying proof.\looseness=-1

These statements compound the issues we raise on commitment design.
If companies not only do not demonstrate how they addresses public commitments, but also broadly claim they satisfied the commitments based on their unilateral judgment, then the overall integrity of the commitments is further compromised.
In turn, this further substantiates our recommendation for why standardized and timely reporting in response to public commitments is especially vital for these commitments to meaningfully advance corporate governance. \\

\noindent \textbf{Promising practices.}
As a positive demonstration of how companies can communicate about their commitments, we point to the webpage Anthropic published on tracking their progress.\footnote{See \url{https://www.anthropic.com/voluntary-commitments}.} 
On the page, Anthropic enumerates every commitment they have made and how they map to actions they have taken.
In particular, such a page also clarifies how overlapping commitments (\eg commitments to conduct internal and external risk assessment that overlap across the White House Voluntary Commitments, the G7 International Code of Conduct, and the Frontier AI Safety Commitments) are streamlined by global companies operating in many jurisdictions.
While this does not imply whether or not Anthropic meets our per-indicator standard, nor any standard the White House envisioned, it claries how Anthropic sees the correspondence between their actions and their commitments.
All major AI companies could implement a similar approach to track how companies' internal practices and external commitments evolve.\looseness=-1

\section{Conclusion}
We present the first comprehensive analysis of how leading AI companies have implemented major voluntary commitments made to governments. Our findings reveal substantial variation in implementation, with scores ranging from 83.3\% for highest-scoring company to just 13.3\% for the lowest. High-scoring companies tended to be early signatories and members of the Frontier Model Forum. 
However, overall performance was particularly weak in model weight security and third-party reporting.
These findings underscore the need for future voluntary commitments to be clearly defined, appropriately targeted to specific roles of companies, and supported by mechanisms for public verification in order to meaningfully influence corporate behavior. 
\section{Acknowledgments}
We thank Melissa Heikkilä, Miles Brundage, Miranda Bogen, Sayash Kapoor, Shayne Longpre, Percy Liang, Daniel E. Ho, and Daniel Zhang for discussions on this topic.
RB is funded by the Stanford Lieberman Fellowship. 
This work is entirely unrelated to the involvement of RB in the EU AI Act Codes of Practice.
KK completed this work at Stanford prior to his work in government.
\clearpage
\appendix
\onecolumn

\clearpage
\edef\originalwidth{\the\pdfpagewidth}
\edef\originalheight{\the\pdfpageheight}

\eject
\newlength{\myheight}
\setlength{\myheight}{\originalwidth-4cm}
\pdfpageheight=\myheight
\newlength{\mywidth}
\setlength{\mywidth}{50cm}
\pdfpagewidth=\mywidth

\vspace*{-2cm}
\begin{figure*}[t]
\centering
\begin{minipage}{\pdfpagewidth}
\centering
\includegraphics[keepaspectratio, height=30cm, width=\linewidth]{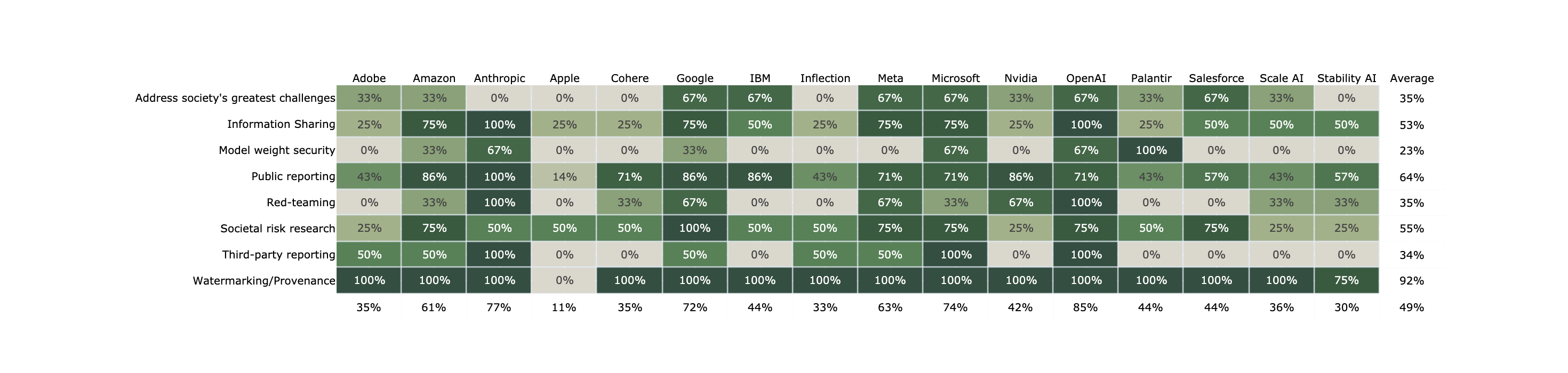}
    \caption{\textbf{Scores for each of the eight commitments in the 2023 White House Voluntary Commitments on AI.} Each cell represents a company's average score across all of the indicators for a given commitment.}
    \end{minipage}
    \label{fig:disaggregate-commitment-scores}
\end{figure*}

\clearpage
\pdfpageheight=\originalheight
\pdfpagewidth=\originalwidth




\clearpage
\edef\originalwidth{\the\pdfpagewidth}
\edef\originalheight{\the\pdfpageheight}

\eject
\setlength{\myheight}{\originalwidth+5cm}
\pdfpageheight=\myheight
\setlength{\mywidth}{50cm}
\pdfpagewidth=\mywidth

\begin{figure*}[t]
\begin{minipage}{\pdfpagewidth}
\includegraphics[keepaspectratio, width=0.9\pdfpagewidth]{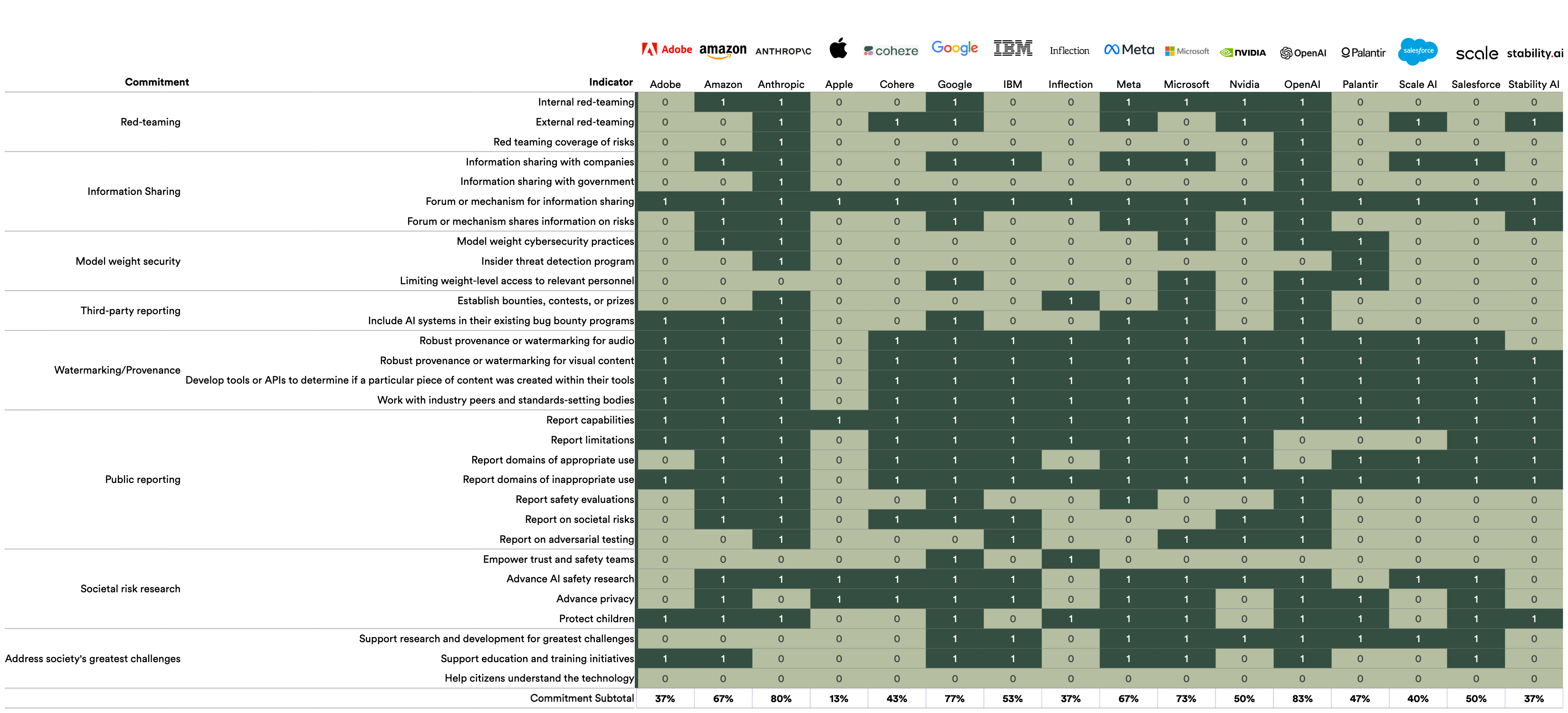}
    \caption{\textbf{Per-indicator scores.} The score for each (indicator, company) pair.}
    \end{minipage}
    \label{fig:disaggregate-indicator-scores}
\end{figure*}

\clearpage
\pdfpageheight=\originalheight
\pdfpagewidth=\originalwidth
\newpage
\section{Appendix A: Indicators}
\begin{enumerate}
\item Internal red-teaming
\begin{itemize}
    \item Indicator under Commitment 1 on Red Teaming
    \item Reference text from 2023 WHVC: ``Commit to internal ... red-teaming of models or systems in areas including misuse, societal risks, and national security concerns, such as bio, cyber, and other safety areas.''
\end{itemize}

\item External red-teaming
\begin{itemize}
    \item Indicator under Commitment 1 on Red Teaming
    \item Reference text from 2023 WHVC: ``Companies commit to ... developing a multi-faceted, specialized, and detailed red-teaming regime, including drawing on independent domain experts, for all major public releases of new models within scope.''
\end{itemize}

\item Red teaming coverage of risks
\begin{itemize}
    \item Indicator under Commitment 1 on Red Teaming
    \item Reference text from 2023 WHVC: ``In designing the regime, they will ensure that they give significant attention to the following:
    \begin{itemize}
        \item Bio, chemical, and radiological risks, such as the ways in which systems can lower barriers to entry for weapons development, design, acquisition, or use
        \item Cyber capabilities, such as the ways in which systems can aid vulnerability discovery, exploitation, or operational use, bearing in mind that such capabilities could also have useful defensive applications and might be appropriate to include in a system
        \item The effects of system interaction and tool use, including the capacity to control physical systems
        \item The capacity for models to make copies of themselves or `self-replicate'
        \item Societal risks, such as bias and discrimination''
    \end{itemize}
\end{itemize}

\item Information sharing with companies
\begin{itemize}
    \item Indicator under Commitment 2 on Information Sharing
    \item Reference text from 2023 WHVC: ``Work toward information sharing among companies and governments regarding trust and safety risks, dangerous or emergent capabilities, and attempts to circumvent safeguards''
    \item Notes: information shared with companies should be information beyond public disclosure.
\end{itemize}

\item Information sharing with government
\begin{itemize}
    \item Indicator under Commitment 2 on Information Sharing
    \item Reference text from 2023 WHVC: ``Work toward information sharing among companies and governments regarding trust and safety risks, dangerous or emergent capabilities, and attempts to circumvent safeguards''
\end{itemize}

\item Forum or mechanism for information sharing
\begin{itemize}
    \item Indicator under Commitment 2 on Information Sharing
    \item Reference text from 2023 WHVC: ``They commit to establish or join a forum or mechanism through which they can develop, advance, and adopt shared standards and best practices for frontier AI safety, such as the NIST AI Risk Management Framework or future standards related to red-teaming, safety, and societal risks''
\end{itemize}

\item Forum or mechanism shares information on risks
\begin{itemize}
    \item Indicator under Commitment 2 on Information Sharing
    \item Reference text from 2023 WHVC: ``The forum or mechanism can facilitate the sharing of information on advances in frontier capabilities and emerging risks and threats, such as attempts to circumvent safeguards, and can facilitate the development of technical working groups on priority areas of concern.''
    \item Notes: A forum that facilitates this kind of information must be a forum that restricts who can join and what they do that with the shared information.
\end{itemize}

\item Model weight cybersecurity practices
\begin{itemize}
    \item Indicator under Commitment 3 on Model Weight Security
    \item Reference text from 2023 WHVC: ``In addition, it requires storing and working with the weights in an appropriately secure environment to reduce the risk of unsanctioned release.''
\end{itemize}

\item Insider threat detection program
\begin{itemize}
    \item Indicator under Commitment 3 on Model Weight Security
    \item Reference text from 2023 WHVC: ``This includes ... establishing a robust insider threat detection program consistent with protections provided for their most valuable intellectual property and trade secrets.''
\end{itemize}

\item Limiting weight-level access to relevant personnel
\begin{itemize}
    \item Indicator under Commitment 3 on Model Weight Security
    \item Reference text from 2023 WHVC: ``This includes limiting access to model weights to those whose job function requires it...''
\end{itemize}

\item Establish bounties, contests, or prizes
\begin{itemize}
    \item Indicator under Commitment 4 on Third-Party Reporting
    \item Reference text from 2023 WHVC: ``They commit to establishing for systems within scope bounties systems, contests, or prizes to incent the responsible disclosure of weaknesses, such as unsafe behaviors...''
\end{itemize}

\item Include AI systems in their existing bug bounty programs
\begin{itemize}
    \item Indicator under Commitment 4 on Third-Party Reporting
    \item Reference text from 2023 WHVC: ``They commit ... to include AI systems in their existing bug bounty programs''
\end{itemize}

\item Robust provenance or watermarking for audio
\begin{itemize}
    \item Indicator under Commitment 5 on Content Provenance
    \item Reference text from 2023 WHVC: ``To further this goal, they agree to develop robust mechanisms, including provenance and/or watermarking systems for audio or visual content created by any of their publicly available systems within scope introduced after the watermarking system is developed.''
\end{itemize}

\item Robust provenance or watermarking for visual content
\begin{itemize}
    \item Indicator under Commitment 5 on Content Provenance
    \item Reference text from 2023 WHVC: ``To further this goal, they agree to develop robust mechanisms, including provenance and/or watermarking systems for audio or visual content created by any of their publicly available systems within scope introduced after the watermarking system is developed.''
\end{itemize}

\item Develop tools or APIs to determine if a particular piece of content was created within their tools
\begin{itemize}
    \item Indicator under Commitment 5 on Content Provenance
    \item Reference text from 2023 WHVC: ``They will also develop tools or APIs to determine if a particular piece of content was created with their system.''
\end{itemize}

\item Work with industry peers and standards-setting bodies as appropriate towards developing a technical framework to help users distinguish audio or visual content generated by users from audio or visual content generated by AI
\begin{itemize}
    \item Indicator under Commitment 5 on Content Provenance
    \item Reference text from 2023 WHVC: ``More generally, companies making this commitment pledge to work with industry peers and standards-setting bodies as appropriate towards developing a technical framework to help users distinguish audio or visual content generated by users from audio or visual content generated by AI.''
\end{itemize}

\item Report capabilities
\begin{itemize}
    \item Indicator under Commitment 6 on Public Reporting
    \item Reference text from 2023 WHVC: ``Publicly report model or system capabilities, limitations, and domains of appropriate and inappropriate use, including discussion of societal risks, such as effects on fairness and bias.''
\end{itemize}

\item Report limitations
\begin{itemize}
    \item Indicator under Commitment 6 on Public Reporting
    \item Reference text from 2023 WHVC: ``These reports should include ... significant limitations in performance that have implications for the domains of appropriate use...''
    \item Notes: Limitations must be specific to the model and not to AI generally.
\end{itemize}

\item Report domains of appropriate use
\begin{itemize}
    \item Indicator under Commitment 6 on Public Reporting
    \item Reference text from 2023 WHVC: ``Publicly report ... domains of appropriate and inappropriate use, including discussion of societal risks, such as effects on fairness and bias''
\end{itemize}

\item Report domains of inappropriate use
\begin{itemize}
    \item Indicator under Commitment 6 on Public Reporting
    \item Reference text from 2023 WHVC: ``Publicly report ... domains of appropriate and inappropriate use, including discussion of societal risks, such as effects on fairness and bias''
\end{itemize}

\item Report safety evaluations
\begin{itemize}
    \item Indicator under Commitment 6 on Public Reporting
    \item Reference text from 2023 WHVC: ``These reports should include the safety evaluations conducted (including in areas such as dangerous capabilities, to the extent that these are responsible to publicly disclose) ...''
\end{itemize}

\item Report on societal risks
\begin{itemize}
    \item Indicator under Commitment 6 on Public Reporting
    \item Reference text from 2023 WHVC: ``These reports should include ... discussion of the model's effects on societal risks such as fairness and bias ...''
\end{itemize}

\item Report on adversarial testing used to determine appropriateness of deployment
\begin{itemize}
    \item Indicator under Commitment 6 on Public Reporting
    \item Reference text from 2023 WHVC: ``These reports should include ... the results of adversarial testing conducted to evaluate the model's fitness for deployment.''
\end{itemize}

\item Empower trust and safety teams
\begin{itemize}
    \item Indicator under Commitment 7 on Societal Risk Research
    \item Reference text from 2023 WHVC: ``Companies commit generally to empowering trust and safety teams, advancing AI safety research, advancing privacy, protecting children, and working to proactively manage the risks of AI so that its benefits can be realized.''
\end{itemize}

\item Advance AI safety research
\begin{itemize}
    \item Indicator under Commitment 7 on Societal Risk Research
    \item Reference text from 2023 WHVC: ``Companies commit generally to empowering trust and safety teams, advancing AI safety research, advancing privacy, protecting children, and working to proactively manage the risks of AI so that its benefits can be realized.''
\end{itemize}

\item Advance privacy
\begin{itemize}
    \item Indicator under Commitment 7 on Societal Risk Research
    \item Reference text from 2023 WHVC: ``Companies commit generally to empowering trust and safety teams, advancing AI safety research, advancing privacy, protecting children, and working to proactively manage the risks of AI so that its benefits can be realized.''
\end{itemize}

\item Protect children
\begin{itemize}
    \item Indicator under Commitment 7 on Societal Risk Research
    \item Reference text from 2023 WHVC: ``Companies commit generally to empowering trust and safety teams, advancing AI safety research, advancing privacy, protecting children, and working to proactively manage the risks of AI so that its benefits can be realized.''
\end{itemize}

\item Support R\&D of frontier AI to address society's greatest challenges
\begin{itemize}
    \item Indicator under Commitment 8 on Address Society's Greatest Challenges
    \item Reference text from 2023 WHVC: ``Companies making this commitment agree to support research and development of frontier AI systems that can help meet society's greatest challenges, such as climate change mitigation and adaptation, early cancer detection and prevention, and combating cyber threats.''
    \item Notes: There is a distinction between supporting research and development of AI in service of societal goals, compared to providing commercial services to public interest companies and advancing AI research in specific domains. The level of engagement and initiative varies.
\end{itemize}

\item Foster the education and training of students and workers to prosper from the benefits of AI
\begin{itemize}
    \item Indicator under Commitment 8 on Address Society's Greatest Challenges
    \item Reference text from 2023 WHVC: ``Companies also commit to supporting initiatives that foster the education and training of students and workers to prosper from the benefits of AI ...''
    \item Notes: The initiatives covered should be accessible to all students and works regardless of prior technical training.
\end{itemize}

\item Help citizens understand the nature, capabilities, limitations, and impact of the technology
\begin{itemize}
    \item Indicator under Commitment 8 on Address Society's Greatest Challenges
    \item Reference text from 2023 WHVC: ``Companies also commit to ... helping citizens understand the nature, capabilities, limitations, and impact of the technology.''
\end{itemize}

\end{enumerate}

\newpage
\section{Appendix B: Indicator Scores for Companies}
\small
\setlongtables
\begin{longtable}{p{0.32\textwidth} c p{0.45\textwidth} p{0.08\textwidth}}
\caption{\normalsize \textbf{Indicator Scores for Adobe}} \\
\textbf{Indicator} & \textbf{Score} & \textbf{Justification} & \textbf{Source} \\
\toprule
\endfirsthead

\multicolumn{4}{c}{Indicator Scores for Adobe -- \textit{Continued from previous page}} \\
\textbf{Indicator} & \textbf{Score} & \textbf{Justification} & \textbf{Source} \\
\toprule
\endhead

\multicolumn{4}{r}{\textit{Continued on next page}} \\
\endfoot

\bottomrule
\endlastfoot

Internal red-teaming & \tikzxmark & Adobe conducts internal red teaming and penetration testing for Firefly, but it is unclear what risk areas are covered. & \cite{adoberedt} \\
\midrule
External red-teaming & \tikzxmark & While Adobe engages with third party security researchers, including through a bug bounty, we do not consider this as an organized initiative dedicated to external red-teaming. & \cite{adoberedt} \\
\midrule
Red teaming coverage of risks & \tikzxmark & No relevant evidence found. & - \\
\midrule
Information sharing with companies & \tikzxmark & While Adobe is involved in C2PA, which involves other companies, this (to our knowledge) does not involve information beyond public information at present. & \cite{adobecontentauth} \\
\midrule
Information sharing with government & \tikzxmark & While Adobe conducts an annual Forum event with government, this does not entail sharing specific information about their models beyond what is public to our knowledge. & - \\
\midrule
Forum or mechanism for information sharing & \tikzcmark & Adobe is a member of the US AI Safety Institute Consortium. & \cite{nistaisic} \\
\midrule
Forum or mechanism shares information on risks & \tikzxmark & Based on public information, the NIST AISIC does not share sensitive information on risks that must be carefully controlled and restricted to forum members to prevent misuse or security vulnerabilities. & \cite{nistaisic} \\
\midrule
Model weight cybersecurity practices & \tikzxmark & No relevant evidence found. & - \\
\midrule
Insider threat detection program & \tikzxmark & No relevant evidence found. & - \\
\midrule
Limiting weight-level access to relevant personnel & \tikzxmark & No relevant evidence found. & - \\
\midrule
Establish bounties, contests, or prizes & \tikzxmark & Adobe did not establish bounties but has an existing bug bounty that includes AI systems. & \cite{adoberedt} \\
\midrule
Include AI systems in their existing bug bounty programs & \tikzcmark & Adobe expanded their bug bounty program to include their implementation of Content Credentials and Adobe Firefly. & \cite{adoberedt} \\
\midrule
Robust provenance or watermarking for audio & \tikzcmark & We award this point due to the Adobe Content Authenticity web app and the associated Content Credentials. Creators can easily apply Content Credentials in batch to sign their digital work — including images, audio and video files. & \cite{adobeaca} \\
\midrule
Robust provenance or watermarking for visual content & \tikzcmark & We award this point due to the Adobe Content Authenticity web app and the associated Content Credentials. Creators can easily apply Content Credentials in batch to sign their digital work — including images, audio and video files. Content Credentials are supported by Firefly. & \cite{adobeaca} \\
\midrule
Develop tools or APIs to determine if a particular piece of content was created within their tools & \tikzcmark & Adobe developed the C2PA Verify tool that determines if a Content Credential was issued by a known source. & \cite{adobeaca} \\
\midrule
Work with industry peers and standards-setting bodies towards developing a technical framework & \tikzcmark & Adobe formed the Content Authenticity Initiative, a global coalition of over 1,500 members across industries, united to promote trust and transparency in digital content. & \cite{adobecommit} \\
\midrule
Report capabilities & \tikzcmark & Adobe describes model capabilities in the press release associated with Firefly Image 3. & \cite{adobefirefly3} \\
\midrule
Report limitations & \tikzcmark & Adobe maintains a list of Firefly's known limitations. & \cite{adobelimit} \\
\midrule
Report domains of appropriate use & \tikzxmark & No relevant evidence found. & - \\
\midrule
Report domains of inappropriate use & \tikzcmark & Adobe's user guidelines enumerates many prohibited uses. & \cite{adobeguide} \\
\midrule
Report safety evaluations & \tikzxmark & No relevant evidence found. & - \\
\midrule
Report on societal risks & \tikzxmark & No specific information provided on societal risks, beyond the concern of deep fakes in relation to content provenance initiatives. & \cite{adoberoad} \\
\midrule
Report on adversarial testing & \tikzxmark & No relevant evidence found. & - \\
\midrule
Empower trust and safety teams & \tikzxmark & While Adobe discusses how they created an AI Ethics engineering team four years ago, and some of its actions, there is no clear evidence that they empower their trust \& safety teams. & \cite{adoberoad} \\
\midrule
Advance AI safety research & \tikzxmark & While Adobe conducts AI fairness research, we do not consider this as AI safety research. & \cite{adobefair} \\
\midrule
Advance privacy & \tikzxmark & No relevant evidence found. & - \\
\midrule
Protect children & \tikzcmark & Adobe enforces a zero-tolerance policy on CSAM, using tools like PhotoDNA and CSAI Match to detect and block known content via hash matching. & \cite{adobechild} \\
\midrule
Support research and development of frontier AI systems that can help meet society's greatest challenges & \tikzxmark & No relevant evidence found. & - \\
\midrule
Support initiatives that foster the education and training of students and workers & \tikzcmark & Adobe launched a global initiative to empower 30 million next-generation learners with AI literacy, content creation, and digital marketing skills by 2030. & \cite{adobeskill} \\
\midrule
Support initiatives that help citizens understand the technology & \tikzxmark & No relevant evidence found. & - \\
\end{longtable}

\newpage
\newpage
\small
\setlongtables
\begin{longtable}{p{0.32\textwidth} c p{0.45\textwidth} p{0.08\textwidth}}
\caption{\normalsize \textbf{Indicator Scores for Amazon}} \\
\textbf{Indicator} & \textbf{Score} & \textbf{Justification} & \textbf{Source} \\
\toprule
\endfirsthead

\multicolumn{4}{c}{Indicator Scores for Amazon -- \textit{Continued from previous page}} \\
\textbf{Indicator} & \textbf{Score} & \textbf{Justification} & \textbf{Source} \\
\toprule
\endhead

\multicolumn{4}{r}{\textit{Continued on next page}} \\
\endfoot

\bottomrule
\endlastfoot

Internal red-teaming & \tikzcmark & Amazon describes that its staff conduct manual red-teaming on AI systems, including Amazon Titan. & \cite{amazonprogress} \\
\midrule
External red-teaming & \tikzxmark & No relevant evidence found. & - \\
\midrule
Red teaming coverage of risks & \tikzxmark & While Amazon describes that it conducts multiple iterations of red-teaming on issues including safety, security, privacy, veracity, and fairness, there is no evidence that Amazon conducts testing on self-replication or the effects of system interaction and tool use. & \cite{amazontitan} \\
\midrule
Information sharing with companies & \tikzcmark & Amazon is a member of the Frontier Model Forum, which facilitates information-sharing among companies. & \cite{amazonprogress} \\
\midrule
Information sharing with government & \tikzxmark & No relevant evidence found. & \cite{amazonprogress} \\
\midrule
Forum or mechanism for information sharing & \tikzcmark & Amazon joined the U.S. AI Safety Institute Consortium. & \cite{amazonprogress} \\
\midrule
Forum or mechanism shares information on risks & \tikzcmark & The Frontier Model Forum shares information as described. & \cite{amazonprogress} \\
\midrule
Model weight cybersecurity practices & \tikzcmark & Amazon describes its security practices for model weights on AWS. & \cite{amazonsecure} \\
\midrule
Insider threat detection program & \tikzxmark & No evidence specific to Nova found. & - \\
\midrule
Limiting weight-level access to relevant personnel & \tikzxmark & No relevant evidence found indicating access is restricted to necessary personnel for training Amazon models. & - \\
\midrule
Establish bounties, contests, or prizes & \tikzxmark & No relevant evidence found. & - \\
\midrule
Include AI systems in their existing bug bounty programs & \tikzcmark & Amazon's Vulnerability Research program includes generative AI assessments and submissions. & \cite{amazonbugbounty} \\
\midrule
Robust provenance or watermarking for audio & \tikzcmark & Amazon Nova does not provide audio capabilities. & \cite{amazonnova} \\
\midrule
Robust provenance or watermarking for visual content & \tikzcmark & Amazon has stated that every image or video generated by Nova has a digital watermark. & \cite{amazonprogress, amazonnova} \\
\midrule
Develop tools or APIs to determine if a particular piece of content was created within their tools & \tikzcmark & Amazon cites a detection solution for identifying images created by Nova with corresponding watermarks. Amazon has previously introduced an API that detects its Titan Image Generator watermark and stated that a new API update is rolling out for Nova. & \cite{amazonprogress, amazonscience} \\
\midrule
Work with industry peers and standards-setting bodies towards developing a technical framework & \tikzcmark & Amazon is part of the Coalition for Content Provenance and Authenticity steering committee. & \cite{amazonc2pa} \\
\midrule
Report capabilities & \tikzcmark & Amazon reports core capabilities of Nova and provides benchmark results in its technical report for the Nova family of models. & \cite{amazontech} \\
\midrule
Report limitations & \tikzcmark & Amazon describes the limitations of Nova models in its service card. & \cite{amazoncard} \\
\midrule
Report domains of appropriate use & \tikzcmark & Amazon reports intended use cases for Nova models in its service cards. & \cite{amazoncard} \\
\midrule
Report domains of inappropriate use & \tikzcmark & Amazon specifies that its Nova models are not designed to provide legal/medical/financial opinions or advice, among other domains of inappropriate use. & \cite{amazoncard} \\
\midrule
Report safety evaluations & \tikzcmark & Amazon reports that they performed evaluations on CBRN threats. & \cite{amazontech} \\
\midrule
Report on societal risks & \tikzcmark & Amazon describes their testing on societal risks, including hate speech, political misinformation, and extremism. & \cite{amazontech} \\
\midrule
Report on adversarial testing & \tikzxmark & Amazon reports the use of human, automated, internal, and external red teaming mechanisms but not the results. & \cite{amazontech} \\
\midrule
Empower trust and safety teams & \tikzxmark & While Amazon describes training its employees on considerations around fairness, privacy, and model explainability, there is no relevant evidence of empowering its trust and safety team. & \cite{amazonprogress} \\
\midrule
Advance AI safety research & \tikzcmark & Amazon developed and shared tools to implement safeguards, prevent harm content, and conduct safety evaluations. & \cite{amazonprogress} \\
\midrule
Advance privacy & \tikzcmark & Amazon conducts public research on privacy and security related to generative AI, such as private text generation. & \cite{amazonprivacy} \\
\midrule
Protect children & \tikzcmark & Amazon works with Thorn to design generative AI services that reduce risk of misuse for child exploitation. & \cite{amazonprogress} \\
\midrule
Support research and development of frontier AI systems that can help meet society's greatest challenges & \tikzxmark & While Amazon provides services to businesses that build generative AI systems for drug development and energy usage optimization, these services are commercial by nature and do not represent a proactive effort to support the research and development of frontier AI systems that can help meet society's greatest challenges. & \cite{amazonprogress} \\
\midrule
Support initiatives that foster the education and training of students and workers & \tikzcmark & Amazon launched free courses about safe and responsible AI to provide AI skills training to 2 million people by 2025. & \cite{amazonprogress} \\
\midrule
Support initiatives that help citizens understand the technology & \tikzxmark & No relevant evidence found. & - \\

\end{longtable}

\newpage
\small
\setlongtables
\begin{longtable}{p{0.32\textwidth} c p{0.45\textwidth} p{0.08\textwidth}}
\caption{\normalsize \textbf{Indicator Scores for Anthropic}} \\
\textbf{Indicator} & \textbf{Score} & \textbf{Justification} & \textbf{Source} \\
\toprule
\endfirsthead

\multicolumn{4}{c}{Indicator Scores for Anthropic -- \textit{Continued from previous page}} \\
\textbf{Indicator} & \textbf{Score} & \textbf{Justification} & \textbf{Source} \\
\toprule
\endhead

\multicolumn{4}{r}{\textit{Continued on next page}} \\
\endfoot

\bottomrule
\endlastfoot

Internal red-teaming & \tikzcmark & Anthropic describes internally red-teaming their systems prior to deployment that cover areas, including misuse, societal risks, and national security concerns. & \cite{anthropicredteam} \\
\midrule
External red-teaming & \tikzcmark & Anthropic describes working with external biosecurity experts to red-team their systems. They also participated in the Generative AI Red Teaming Challenge. & \cite{anthropicfrontier, humaneintelligence2023grt} \\
\midrule
Red teaming coverage of risks & \tikzcmark & Anthropic described conducting red-teaming in the domains of national security, CBRN, trust and safety. & \cite{anthropicredteam} \\
\midrule
Information sharing with companies & \tikzcmark & Anthropic is a member of the Frontier Model Forum, which facilitates information-sharing among companies. & \cite{frontiermodelforum} \\
\midrule
Information sharing with government & \tikzcmark & Anthropic provides the U.S. AI Safety Institute with new models before and following their releases. & \cite{nistsharing} \\
\midrule
Forum or mechanism for information sharing & \tikzcmark & Anthropic is a member of the U.S. AI Safety Institute Consortium and the Frontier Model Forum. & \cite{nistaisic, frontiermodelforum} \\
\midrule
Forum or mechanism shares information on risks & \tikzcmark & The Frontier Model Forum shares information as desired. & \cite{frontiermodelforum} \\
\midrule
Model weight cybersecurity practices & \tikzcmark & Anthropic is implementing two-party controls, secure software development framework, supply chain levels for software artifacts, and other cybersecurity best practices with the specific aim of protecting model weights. & \cite{anthropicsecurity} \\
\midrule
Insider threat detection program & \tikzcmark & Anthropic's Responsible Scaling Policy indicates the existence of an insider risk program for ASL-2 Security Standard. & \cite{anthropicsecurity} \\
\midrule
Limiting weight-level access to relevant personnel & \tikzxmark & Anthropic's Responsible Scaling Policy discusses access management tools under its ASL-2 Security Standard but does not specifically indicate that access is restricted to model weights. & \cite{anthropicsecurity} \\
\midrule
Establish bounties, contests, or prizes & \tikzcmark & Anthropic operates an invite-only bug bounty program for identifying model safety issues. & \cite{anthropicbounty} \\
\midrule
Include AI systems in their existing bug bounty programs & \tikzcmark & Anthropic expanded its bug bounty program to include a new initiative focused on identifying and mitigating universal jailbreak attacks. & \cite{anthropicbounty} \\
\midrule
Robust provenance or watermarking for audio & \tikzcmark & Anthropic does not produce audiovisual models, and so we default their score to 1 for this commitment. & - \\
\midrule
Robust provenance or watermarking for visual content & \tikzcmark & Anthropic does not produce audiovisual models, and so we default their score to 1 for this commitment. & - \\
\midrule
Develop tools or APIs to determine if a particular piece of content was created within their tools & \tikzcmark & Anthropic does not produce audiovisual models, and so we default their score to 1 for this commitment. & - \\
\midrule
Work with industry peers and standards-setting bodies towards developing a technical framework & \tikzcmark & Anthropic does not produce audiovisual models, and so we default their score to 1 for this commitment. & - \\
\midrule
Report capabilities & \tikzcmark & Anthropic describes the capabilities of Claude 3.5 in their Claude 3.5 addendum through benchmark results. & \cite{anthropicaddendum} \\
\midrule
Report limitations & \tikzcmark & Anthropic describes the limitations of Claude 3.5 in their Claude 3 model card under areas for improvement. & \cite{anthropiccard} \\
\midrule
Report domains of appropriate use & \tikzcmark & Anthropic reports the intended uses of Claude 3.5 in their Claude 3 model card. & \cite{anthropiccard} \\
\midrule
Report domains of inappropriate use & \tikzcmark & Anthropic reports the unintended uses of Claude 3.5 in their Claude 3 model card. & \cite{anthropiccard} \\
\midrule
Report safety evaluations & \tikzcmark & Anthropic reports evaluation results around catastrophic harms, including autonomous replication and adaption, and cybersecurity risks in their Claude 3.5 addendum. & \cite{anthropicaddendum} \\
\midrule
Report on societal risks & \tikzcmark & Anthropic describes their testing on societal risks, including discrimination, stereotype bias, and CBRN threats in their Claude 3.5 addendum. & \cite{anthropicaddendum} \\
\midrule
Report on adversarial testing & \tikzcmark & Anthropic reports on results of multimodal policy red teaming and other evaluation results in their Claude 3.5 addendum. & \cite{anthropicaddendum} \\
\midrule
Empower trust and safety teams & \tikzxmark & While Anthropic explicitly indicates that a key goal at the company level is to accelerate safety work, there is no relevant evidence of empowering its trust and safety team. & \cite{anthropicviews} \\
\midrule
Advance AI safety research & \tikzcmark & Anthropic describes their research directions of scaling supervision, mechanistic interpretability, and process-oriented learning, with the goal of building safety systems. & \cite{anthropicviews} \\
\midrule
Advance privacy & \tikzxmark & No relevant evidence found. & - \\
\midrule
Protect children & \tikzcmark & Anthropic reports child sexual abuse material to the National Center for Missing \& Exploited Children. & \cite{anthropicchild} \\
\midrule
Support research and development of frontier AI systems that can help meet society's greatest challenges & \tikzxmark & No relevant evidence found. & - \\
\midrule
Support initiatives that foster the education and training of students and workers & \tikzxmark & Anthropic provides educational courses on using Claude, but they are not targeted specifically to students, workers, or broader accessibility. & \cite{anthropiccourses} \\
\midrule
Support initiatives that help citizens understand the technology & \tikzxmark & Although Anthropic and the Collective Intelligence Project ran a public input process involving ~1,000 Americans to draft a constitution for an AI system, this initiative does not improve citizens' understanding of the nature, capabilities, limitations, and impact of AI. & \cite{anthropiccollective} \\

\end{longtable}

\newpage
\newpage
\small
\setlongtables
\begin{longtable}{p{0.32\textwidth} c p{0.45\textwidth} p{0.08\textwidth}}
\caption{\normalsize \textbf{Indicator Scores for Apple}} \\
\textbf{Indicator} & \textbf{Score} & \textbf{Justification} & \textbf{Source} \\
\toprule
\endfirsthead

\multicolumn{4}{c}{Indicator Scores for Apple -- \textit{Continued from previous page}} \\
\textbf{Indicator} & \textbf{Score} & \textbf{Justification} & \textbf{Source} \\
\toprule
\endhead

\multicolumn{4}{r}{\textit{Continued on next page}} \\
\endfoot

\bottomrule
\endlastfoot

Internal red-teaming & \tikzxmark & While Apple describes employing both manual and automatic red-teaming to evaluate their models, there is no mention of the risk areas that these red-teaming efforts cover in its technical report. & \cite{applefoundation} \\
\midrule
External red-teaming & \tikzxmark & While Apple describes running red-teaming projects with both internal and external participants, its approach does not constitute a detailed red-teaming regime based on public information. & \cite{applefoundation} \\
\midrule
Red teaming coverage of risks & \tikzxmark & No relevant evidence found. & - \\
\midrule
Information sharing with companies & \tikzxmark & No relevant evidence found. & - \\
\midrule
Information sharing with government & \tikzxmark & No relevant evidence found. & - \\
\midrule
Forum or mechanism for information sharing & \tikzcmark & Apple is a member of the U.S. AI Safety Institute Consortium. & \cite{nistaisic} \\
\midrule
Forum or mechanism shares information on risks & \tikzxmark & Based on public information, the NIST AISIC does not share sensitive information on risks that must be carefully controlled and restricted to forum members to prevent misuse or security vulnerabilities. & - \\
\midrule
Model weight cybersecurity practices & \tikzxmark & Apple's Private Cloud Compute system does not address model weight security. & \cite{applepcc} \\
\midrule
Insider threat detection program & \tikzxmark & No relevant evidence found. & - \\
\midrule
Limiting weight-level access to relevant personnel & \tikzxmark & No relevant evidence found. & - \\
\midrule
Establish bounties, contests, or prizes & \tikzxmark & While Apple has an existing bug bounty and expanded it to include their compute system, this bounty does not include AI systems. & \cite{applebounty} \\
\midrule
Include AI systems in their existing bug bounty programs & \tikzxmark & While Apple has an existing bug bounty and expanded it to include their compute system, this bounty does not include AI systems. & \cite{applebounty} \\
\midrule
Robust provenance or watermarking for audio & \tikzxmark & No relevant evidence found. & - \\
\midrule
Robust provenance or watermarking for visual content & \tikzxmark & Apple intends to label AI-generated images in their metadata but has not provided evidence of implementation. & - \\
\midrule
Develop tools or APIs to determine if a particular piece of content was created within their tools & \tikzxmark & No relevant evidence found. & - \\
\midrule
Work with industry peers and standards-setting bodies towards developing a technical framework & \tikzxmark & No relevant evidence found. & - \\
\midrule
Report capabilities & \tikzcmark & Apple reports the language and reasoning capabilities of their foundation models in their technical report. & \cite{applefoundation} \\
\midrule
Report limitations & \tikzxmark & The limitation discussed is generically about the use of LMs for grading, not their specific models. & \cite{applefoundation} \\
\midrule
Report domains of appropriate use & \tikzxmark & No relevant evidence found. & - \\
\midrule
Report domains of inappropriate use & \tikzxmark & No relevant evidence found. & - \\
\midrule
Report safety evaluations & \tikzxmark & Apple reports safety evaluations, including response rate to adversarial prompts and human evaluation of output harmfulness. & \cite{applefoundation} \\
\midrule
Report on societal risks & \tikzxmark & No relevant evidence found. & - \\
\midrule
Report on adversarial testing & \tikzxmark & Apple described their methods for manual and automatic red-teaming but does not report the results. & \cite{applefoundation} \\
\midrule
Empower trust and safety teams & \tikzxmark & While Apple has a Responsible AI team, there's no more information about how it empowers its trust and safety teams. & \cite{applefoundation} \\
\midrule
Advance AI safety research & \tikzcmark & Apple publishes a number of papers on AI safety. & \cite{applewhisper} \\
\midrule
Advance privacy & \tikzcmark & Apple publishes privacy research and created Private Cloud Compute designed for private AI processing. & \cite{applepcc, appleprivacy} \\
\midrule
Protect children & \tikzxmark & No relevant evidence found. & - \\
\midrule
Support research and development of frontier AI systems that can help meet society's greatest challenges & \tikzxmark & No relevant evidence found. & - \\
\midrule
Support initiatives that foster the education and training of students and workers & \tikzxmark & While Apple Developer Academy includes coursework in AI, it is not targeted specifically to workers or the broader public. & \cite{appleacademy} \\
\midrule
Support initiatives that help citizens understand the technology & \tikzxmark & No relevant evidence found. & - \\

\end{longtable}

\newpage
\newpage
\small
\setlongtables
\begin{longtable}{p{0.32\textwidth} c p{0.45\textwidth} p{0.08\textwidth}}
\caption{\normalsize \textbf{Indicator Scores for Cohere}} \\
\textbf{Indicator} & \textbf{Score} & \textbf{Justification} & \textbf{Source} \\
\toprule
\endfirsthead

\multicolumn{4}{c}{Indicator Scores for Cohere -- \textit{Continued from previous page}} \\
\textbf{Indicator} & \textbf{Score} & \textbf{Justification} & \textbf{Source} \\
\toprule
\endhead

\multicolumn{4}{r}{\textit{Continued on next page}} \\
\endfoot

\bottomrule
\endlastfoot

Internal red-teaming & \tikzxmark & No relevant evidence found. & - \\
\midrule
External red-teaming & \tikzcmark & Cohere participated in the Generative AI Red Teaming Challenge. & \cite{humaneintelligence2023grt} \\
\midrule
Red teaming coverage of risks & \tikzxmark & No relevant evidence found. & - \\
\midrule
Information sharing with companies & \tikzxmark & While Lakera and Cohere partnered to define new LLM security standards, this partnership is commercial in nature. & \cite{coherelakera} \\
\midrule
Information sharing with government & \tikzxmark & Although Cohere participated in the U.S. Senate AI Insight Forum, we do not award them this point because they did not initiate engagement and this engagement is not ongoing. & \cite{coheresenate} \\
\midrule
Forum or mechanism for information sharing & \tikzcmark & Cohere is a member of the U.S. AI Safety Institute Consortium. & \cite{nistaisic} \\
\midrule
Forum or mechanism shares information on risks & \tikzxmark & Based on public information, the NIST AISIC does not share sensitive information on risks that must be carefully controlled and restricted to forum members to prevent misuse or security vulnerabilities. & - \\
\midrule
Model weight cybersecurity practices & \tikzxmark & No details about cybersecurity specific to model weights, just use of SOC 2. & \cite{coheresecurity} \\
\midrule
Insider threat detection program & \tikzxmark & No relevant evidence found. & - \\
\midrule
Limiting weight-level access to relevant personnel & \tikzxmark & While Cohere states that "access to cloud infrastructure and other sensitive tools are limited to authorized employees who require it for their role," securing access to cloud infrastructure is not equivalent to securing access to model weights. & \cite{coheresecurity} \\
\midrule
Establish bounties, contests, or prizes & \tikzxmark & Cohere Responsible Disclosure Policy describes an invite-only bug bounty program with BugCrowd, but there is insufficient information to evaluate the activities of the bug bounty. & \cite{coheredisclosure} \\
\midrule
Include AI systems in their existing bug bounty programs & \tikzxmark & No relevant evidence found. & - \\
\midrule
Robust provenance or watermarking for audio & \tikzcmark & Cohere's flagship models do not generate audio or images, and so we default their score to \tikzcmark{} for this commitment. & - \\
\midrule
Robust provenance or watermarking for visual content & \tikzcmark & Cohere's flagship models do not generate audio or images, and so we default their score to \tikzcmark{} for this commitment. & - \\
\midrule
Develop tools or APIs to determine if a particular piece of content was created within their tools & \tikzcmark & Cohere's flagship models do not generate audio or images, and so we default their score to \tikzcmark{} for this commitment. & - \\
\midrule
Work with industry peers and standards-setting bodies towards developing a technical framework & \tikzcmark & Cohere's flagship models do not generate audio or images, and so we default their score to \tikzcmark{} for this commitment. & - \\
\midrule
Report capabilities & \tikzcmark & Cohere describes the capabilities of Command R in their documentation. & \cite{coheredocs} \\
\midrule
Report limitations & \tikzcmark & Cohere discloses the language limitations of its model by specifying which languages it supports in the model card. & \cite{cohereuse} \\
\midrule
Report domains of appropriate use & \tikzcmark & Cohere describes the intended use cases of Command R as text generation and RAG and tool-use tasks in its model card. & \cite{cohereuse} \\
\midrule
Report domains of inappropriate use & \tikzcmark & Cohere discloses the unintended and prohibited uses in decision-making in Command R's model card. & \cite{cohereuse} \\
\midrule
Report safety evaluations & \tikzxmark & While Cohere reports testing for model toxicity and bias, their safety evaluations do not target severe risks such as CBRN and child safety. & \cite{cohereuse} \\
\midrule
Report on societal risks & \tikzcmark & Cohere publishes the evaluation results on the BOLD dataset in its model card, testing for model toxicity and bias. & \cite{cohereuse} \\
\midrule
Report on adversarial testing & \tikzxmark & While Cohere reports evaluation results for BOLD, which can be construed as part of safety, these results do not constitute adversarial testing to determine appropriateness of deployment. & \cite{cohereuse} \\
\midrule
Empower trust and safety teams & \tikzxmark & While Cohere created a Responsibility Council to inform their product and business decisions, there are no relevant mentions of a trust and safety team. & \cite{cohereuse} \\
\midrule
Advance AI safety research & \tikzcmark & Cohere supports the Cohere for AI Scholars program, which provides research grants to researchers advancing safe, responsible LLM capabilities and applications. They also allow the "intentional stress testing of the API and adversarial attacks." & \cite{coheregrants} \\
\midrule
Advance privacy & \tikzcmark & Cohere conducts research on privacy and is a member of the Coalition for Secure AI. & \cite{utpala2023dprompt} \\
\midrule
Protect children & \tikzxmark & No relevant evidence found. & - \\
\midrule
Support research and development of frontier AI systems that can help meet society's greatest challenges & \tikzxmark & While Cohere's support of the Aya initiative is highly commendable, we do not award a point as providing multilingual AI capabilities at present does not seem akin to the named societal challenges that are more established and fundamental (e.g. cancer treatment). & \cite{cohereaya} \\
\midrule
Support initiatives that foster the education and training of students and workers & \tikzxmark & While Cohere operates the Cohere for AI Scholars program, this program is structured as a research apprenticeship geared towards those with existing technical backgrounds. It is not designed to foster the education and training of students and workers in general, and so we do not award this point. This is also the case for its LLM University. & \cite{coherescholars, coherellmu} \\
\midrule
Support initiatives that help citizens understand the technology & \tikzxmark & No evidence found of citizen-specific programs. & - \\

\end{longtable}

\newpage
\newpage
\small
\setlongtables
\begin{longtable}{p{0.32\textwidth} c p{0.45\textwidth} p{0.08\textwidth}}
\caption{\normalsize \textbf{Indicator Scores for Google}} \\
\textbf{Indicator} & \textbf{Score} & \textbf{Justification} & \textbf{Source} \\
\toprule
\endfirsthead

\multicolumn{4}{c}{Indicator Scores for Google -- \textit{Continued from previous page}} \\
\textbf{Indicator} & \textbf{Score} & \textbf{Justification} & \textbf{Source} \\
\toprule
\endhead

\multicolumn{4}{r}{\textit{Continued on next page}} \\
\endfoot

\bottomrule
\endlastfoot

Internal red-teaming & \tikzcmark & Google mentions their internal, company-wide red team that attempts different types of attacks. Its technical report for Gemini also describes their internal red-teaming efforts focused on security, safety, and privacy failures. & \cite{fabian2023redteam, geminiteam2024gemini} \\
\midrule
External red-teaming & \tikzcmark & Google participated in the Generative AI Red Teaming Challenge. They also mention working with a small set of independent external groups to conduct unstructured red teaming. & \cite{humaneintelligence2023grt, geminiteam2024gemini} \\
\midrule
Red teaming coverage of risks & \tikzxmark & While the Gemini technical report discusses red-teaming across a broad range of risks, including CBRN and societal risks, these risk areas do not cover the effects of system interaction and tool use. & \cite{geminiteam2024gemini} \\
\midrule
Information sharing with companies & \tikzcmark & Google is a member of the Frontier Model Forum, which facilitates information-sharing among companies. & \cite{frontiermodelforum} \\
\midrule
Information sharing with government & \tikzxmark & No specific sustained information sharing mechanism is described, though conferences and other convenings are mentioned without clarification on whether information beyond what is made public is shared. & \cite{googlewhitehouse} \\
\midrule
Forum or mechanism for information sharing & \tikzcmark & Google is a member of the U.S. AI Safety Institute Consortium and the Frontier Model Forum. & \cite{nistaisic, frontiermodelforum} \\
\midrule
Forum or mechanism shares information on risks & \tikzcmark & The Frontier Model Forum shares information as desired. & \cite{frontiermodelforum} \\
\midrule
Model weight cybersecurity practices & \tikzxmark & No information is given about model weight security in the training of Gemini models. & - \\
\midrule
Insider threat detection program & \tikzxmark & No clear indication is provided that Google implements insider threat detection in relation to Gemini. & - \\
\midrule
Limiting weight-level access to relevant personnel & \tikzcmark & Google applies access controls to ensure only necessary personnel have access for business purposes. & \cite{googlesecurity} \\
\midrule
Establish bounties, contests, or prizes & \tikzxmark & Google did not establish bounties but has an existing bug bounty that includes AI systems. & \cite{googlebounty} \\
\midrule
Include AI systems in their existing bug bounty programs & \tikzcmark & Google expanded their VRP to reward for attack scenarios specific to generative AI. & \cite{googlebounty} \\
\midrule
Robust provenance or watermarking for audio & \tikzcmark & Google developed SynthID, which watermarks and identifies AI-generated content. & \cite{synthid} \\
\midrule
Robust provenance or watermarking for visual content & \tikzcmark & Google developed SynthID, which watermarks and identifies AI-generated content. & \cite{synthid} \\
\midrule
Develop tools or APIs to determine if a particular piece of content was created within their tools & \tikzcmark & SynthID is a tool to determine if content is generated using Google models as described in the associated scientific paper. & \cite{synthid} \\
\midrule
Work with industry peers and standards-setting bodies towards developing a technical framework & \tikzcmark & Google contributed to the Partnership on AI efforts on a Synthetic Media Framework to establish best practices. & \cite{googleaiprogress} \\
\midrule
Report capabilities & \tikzcmark & Google reports model capabilities through benchmark results in the Gemini technical report. & \cite{geminiteam2024gemini} \\
\midrule
Report limitations & \tikzcmark & Google notes that there are limitations in a short section at the end of their Gemini technical report. & \cite{geminiteam2024gemini} \\
\midrule
Report domains of appropriate use & \tikzcmark & Google enumerates domains of intended use in their model card for Gemini. & \cite{gemmacard} \\
\midrule
Report domains of inappropriate use & \tikzcmark & The prohibited use policy specifies such domains. & \cite{gemmapolicy} \\
\midrule
Report safety evaluations & \tikzcmark & The Gemini technical report includes safety evaluations around potential misuse. & \cite{geminiteam2024gemini} \\
\midrule
Report on societal risks & \tikzcmark & The Gemini technical report includes evaluations on societal risks. & \cite{geminiteam2024gemini} \\
\midrule
Report on adversarial testing & \tikzxmark & Google cites three types of external testing in the Gemini technical report but does not report the results from these exercises. & \cite{geminiteam2024gemini} \\
\midrule
Empower trust and safety teams & \tikzcmark & Google instituted AI principles ethics reviews and impact assessments conducted by their Trust and Safety team as part of the pre-launch process. & \cite{googleaiprogress, googlewhitehouse} \\
\midrule
Advance AI safety research & \tikzcmark & Google regularly publishes and funds AI safety research. & \cite{googleprivacy, frontiersafety} \\
\midrule
Advance privacy & \tikzcmark & Google conducts privacy research around AI. & \cite{googleaiprogress, googleprivacy} \\
\midrule
Protect children & \tikzcmark & We award this point for their partnerships with Thorn and All Tech is Human on CSAM. & \cite{googlechild} \\
\midrule
Support research and development of frontier AI systems that can help meet society's greatest challenges & \tikzcmark & Google has worked on AI projects to support researchers in healthcare and energy management. & \cite{googlewhitehouse} \\
\midrule
Support initiatives that foster the education and training of students and workers & \tikzcmark & Google announced funding to equip educators and students with foundational AI skills through the development of AI curriculums. & \cite{googleorg} \\
\midrule
Support initiatives that help citizens understand the technology & \tikzxmark & No evidence found of citizen-specific programs. & - \\

\end{longtable}

\newpage
\newpage
\small
\setlongtables
\begin{longtable}{p{0.32\textwidth} c p{0.45\textwidth} p{0.08\textwidth}}
\caption{\normalsize \textbf{Indicator Scores for IBM}} \\
\textbf{Indicator} & \textbf{Score} & \textbf{Justification} & \textbf{Source} \\
\toprule
\endfirsthead

\multicolumn{4}{c}{Indicator Scores for IBM -- \textit{Continued from previous page}} \\
\textbf{Indicator} & \textbf{Score} & \textbf{Justification} & \textbf{Source} \\
\toprule
\endhead

\multicolumn{4}{r}{\textit{Continued on next page}} \\
\endfoot

\bottomrule
\endlastfoot

Internal red-teaming & \tikzxmark & While IBM recruits company volunteers to conduct red-teaming, there is no evidence that these efforts cover the areas of misuse, societal risks, and national security concerns. & \cite{ibmredteam,ibmguide} \\
\midrule
External red-teaming & \tikzxmark & While IBM partnered with a third-party company to conduct external red-teaming of Granite, they do not provide information on how these red-teaming operations are conducted. As such, we do not consider this partnership to be sufficient for a detailed regime. & \cite{ibmguide} \\
\midrule
Red teaming coverage of risks & \tikzxmark & While the Granite 3.0 technical report details evaluations for many risk areas, these do not constitute a detailed red-teaming regime with full coverage of the risk areas outlined in the voluntary commitments. & \cite{ibmtech} \\
\midrule
Information sharing with companies & \tikzcmark & IBM is a member of the AI Alliance. & \cite{ibmalliance} \\
\midrule
Information sharing with government & \tikzxmark & No evidence found of sustained mechanism for providing information about Granite models to government. & - \\
\midrule
Forum or mechanism for information sharing & \tikzcmark & IBM is a member of the U.S. AI Safety Institute Consortium. & \cite{nistaisic} \\
\midrule
Forum or mechanism shares information on risks & \tikzxmark & Based on public information, the NIST AISIC does not share sensitive information on risks that must be carefully controlled and restricted to forum members. & - \\
\midrule
Model weight cybersecurity practices & \tikzxmark & While IBM provides related services to clients, they do not clearly disclose that they implement these practices in training their Granite models. & \cite{ibmsecurity} \\
\midrule
Insider threat detection program & \tikzxmark & While IBM provides related services to clients, they do not clearly disclose that they implement these practices in training their Granite models. & \cite{ibmsecurity} \\
\midrule
Limiting weight-level access to relevant personnel & \tikzxmark & While IBM provides related services to clients, they do not clearly disclose that they implement these practices in training their Granite models. & \cite{ibmverify} \\
\midrule
Establish bounties, contests, or prizes & \tikzxmark & No relevant evidence found. & - \\
\midrule
Include AI systems in their existing bug bounty programs & \tikzxmark & No relevant evidence found. & - \\
\midrule
Robust provenance or watermarking for audio & \tikzcmark & IBM's flagship models do not generate audio or images, and so we default their score to \tikzcmark{} for this commitment. & - \\
\midrule
Robust provenance or watermarking for visual content & \tikzcmark & IBM's flagship models do not generate audio or images, and so we default their score to \tikzcmark{} for this commitment. & - \\
\midrule
Develop tools or APIs to determine if a particular piece of content was created within their tools & \tikzcmark & IBM's flagship models do not generate audio or images, and so we default their score to \tikzcmark{} for this commitment. & - \\
\midrule
Work with industry peers and standards-setting bodies towards developing a technical framework & \tikzcmark & IBM's flagship models do not generate audio or images, and so we default their score to \tikzcmark{} for this commitment. & - \\
\midrule
Report capabilities & \tikzcmark & Technical report for Granite 3.0 details model capabilities. & \cite{ibmtech} \\
\midrule
Report limitations & \tikzcmark & Model card for Granite 3.0 details model limitations. & \cite{ibmgranitecard} \\
\midrule
Report domains of appropriate use & \tikzcmark & Model card for Granite 3.0 details appropriate uses such as summarization and text classification. & \cite{ibmgranitecard} \\
\midrule
Report domains of inappropriate use & \tikzcmark & IBM outlines AI use restrictions in its service descriptions for foundation models in connection with the use of its Cloud Service. & \cite{ibmterms} \\
\midrule
Report safety evaluations & \tikzxmark & Technical report for Granite 3.0 details safety evaluation, but does not include CBRN results. & \cite{ibmtech} \\
\midrule
Report on societal risks & \tikzcmark & Technical report for Granite 3.0 describes socio-technical harms and risks. & \cite{ibmtech} \\
\midrule
Report on adversarial testing & \tikzcmark & Technical report for Granite 3.0 details red-teaming results. & \cite{ibmtech} \\
\midrule
Empower trust and safety teams & \tikzxmark & While IBM provides the WatsonX Governance service, there is no relevant evidence of empowering its trust and safety team. & \cite{ibmwatsonx} \\
\midrule
Advance AI safety research & \tikzcmark & IBM regularly publishes AI safety research. & \cite{ibmguide} \\
\midrule
Advance privacy & \tikzcmark & IBM conducts AI privacy research and created the AI Privacy Toolkit, a set of open-source tools for developers to ensure the privacy and compliance of their models. & \cite{ibmprivacy} \\
\midrule
Protect children & \tikzxmark & No relevant evidence found. & - \\
\midrule
Support research and development of frontier AI systems that can help meet society's greatest challenges & \tikzcmark & IBM committed \$30 million worth of technology and services to improve water management. & \cite{ibmclimate} \\
\midrule
Support initiatives that foster the education and training of students and workers & \tikzcmark & IBM is training two million learners in AI by 2026, with a focus on underrepresented communities. & \cite{ibmtraining} \\
\midrule
Support initiatives that help citizens understand the technology & \tikzxmark & No citizen-specific program found. & - \\

\end{longtable}

\newpage
\newpage
\small
\setlongtables
\begin{longtable}{p{0.32\textwidth} c p{0.45\textwidth} p{0.08\textwidth}}
\caption{\normalsize \textbf{Indicator Scores for Inflection}} \\
\textbf{Indicator} & \textbf{Score} & \textbf{Justification} & \textbf{Source} \\
\toprule
\endfirsthead

\multicolumn{4}{c}{Indicator Scores for Inflection -- \textit{Continued from previous page}} \\
\textbf{Indicator} & \textbf{Score} & \textbf{Justification} & \textbf{Source} \\
\toprule
\endhead

\multicolumn{4}{r}{\textit{Continued on next page}} \\
\endfoot

\bottomrule
\endlastfoot

Internal red-teaming & \tikzxmark & Inflection mentions hosting an internal Safety team which pressure tests their models, but there is no evidence that its internal red-teaming efforts cover areas such as misuse, societal risks, and national security concerns. & \cite{inflectionwh} \\
\midrule
External red-teaming & \tikzxmark & While Inflection mentions commissioning outside experts to conduct red-teaming, they do not provide further information on their red-teaming operations. & \cite{inflectionwh} \\
\midrule
Red teaming coverage of risks & \tikzxmark & Inflection notes that its CBRN red-teaming efforts are expanding, but its current efforts do not cover the full range of risks outlined in the voluntary commitments. & \cite{inflectionsafety} \\
\midrule
Information sharing with companies & \tikzxmark & While Inflection is involved with the Partnership on AI, this does not appear to involve private information-sharing between companies. & \cite{inflectionpai, inflectionml} \\
\midrule
Information sharing with government & \tikzxmark & While Inflection responded to the UK government before the AI Safety Summit, there is no evidence of non-public information sharing. & \cite{inflectionsafety} \\
\midrule
Forum or mechanism for information sharing & \tikzcmark & Inflection is a member of the U.S. AI Safety Institute Consortium. & \cite{nistaisic} \\
\midrule
Forum or mechanism shares information on risks & \tikzxmark & Based on public information, the NIST AISIC does not share sensitive risk information. & - \\
\midrule
Model weight cybersecurity practices & \tikzxmark & No relevant evidence found. & - \\
\midrule
Insider threat detection program & \tikzxmark & No relevant evidence found. & - \\
\midrule
Limiting weight-level access to relevant personnel & \tikzxmark & Inflection restricts access to data and code, but not model weights specifically. & \cite{inflectionsafety} \\
\midrule
Establish bounties, contests, or prizes & \tikzcmark & Inflection implemented a closed pilot bug bounty program inviting security researchers to identify vulnerabilities. & \cite{inflectionsafety} \\
\midrule
Include AI systems in their existing bug bounty programs & \tikzxmark & Inflection did not have an existing bug bounty. & \cite{inflectionsafety} \\
\midrule
Robust provenance or watermarking for audio & \tikzcmark & Inflection's flagship models do not generate audio or images. & - \\
\midrule
Robust provenance or watermarking for visual content & \tikzcmark & Inflection's flagship models do not generate audio or images. & - \\
\midrule
Develop tools or APIs to determine if a particular piece of content was created within their tools & \tikzcmark & Inflection's flagship models do not generate audio or images. & - \\
\midrule
Work with industry peers and standards-setting bodies on content traceability & \tikzcmark & Inflection's flagship models do not generate audio or images. & - \\
\midrule
Report capabilities & \tikzcmark & Inflection outlines the capabilities of Pi 3.0 through prompt-response examples. & \cite{inflectiondocs} \\
\midrule
Report limitations & \tikzcmark & Inflection’s documentation includes a section on known limitations. & \cite{inflectionsafety} \\
\midrule
Report domains of appropriate use & \tikzxmark & No relevant evidence found. & - \\
\midrule
Report domains of inappropriate use & \tikzcmark & Inflection’s Terms of Service detail illegal and harmful uses. & \cite{inflectionpolicy} \\
\midrule
Report safety evaluations & \tikzxmark & No relevant evidence found. & - \\
\midrule
Report on societal risks & \tikzxmark & No relevant evidence found. & - \\
\midrule
Report on adversarial testing & \tikzxmark & No relevant evidence found. & - \\
\midrule
Empower trust and safety teams & \tikzcmark & Inflection's safety team is authorized to block malicious actors and implement mitigations. & \cite{inflectionsafety} \\
\midrule
Advance AI safety research & \tikzxmark & Inflection is a member of MLCommons' safety working group but does not provide evidence of its own contributions. & \cite{inflectionml} \\
\midrule
Advance privacy & \tikzxmark & No relevant evidence found. & - \\
\midrule
Protect children & \tikzcmark & Inflection partners with Thorn to mitigate CSAM risks. & \cite{inflectionsafety, inflectionthorn} \\
\midrule
Support frontier AI for societal challenges & \tikzxmark & No relevant evidence found. & - \\
\midrule
Support education/training of students and workers & \tikzxmark & No relevant evidence found. & - \\
\midrule
Support citizen understanding of AI & \tikzxmark & No citizen-specific program found. & - \\

\end{longtable}

\newpage
\newpage
\small
\setlongtables
\begin{longtable}{p{0.32\textwidth} c p{0.45\textwidth} p{0.08\textwidth}}
\caption{\normalsize \textbf{Indicator Scores for Meta}} \\
\textbf{Indicator} & \textbf{Score} & \textbf{Justification} & \textbf{Source} \\
\toprule
\endfirsthead

\multicolumn{4}{c}{Indicator Scores for Meta -- \textit{Continued from previous page}} \\
\textbf{Indicator} & \textbf{Score} & \textbf{Justification} & \textbf{Source} \\
\toprule
\endhead

\multicolumn{4}{r}{\textit{Continued on next page}} \\
\endfoot

\bottomrule
\endlastfoot

Internal red-teaming & \tikzcmark & Meta conducted red-teaming exercises with their internal teams covering types of misues, societal risks, and national security concerns. & \cite{metallama} \\
\midrule
External red-teaming & \tikzcmark & Meta participated in the Generative AI Red Teaming Challenge and conducted red-teaming with exteneral experts. & \cite{humaneintelligence2023grt, metadavos} \\
\midrule
Red teaming coverage of risks & \tikzxmark & While Meta describes its red-teaming efforts across a range of risk categories, including the production of weapons and privacy violations, there is no evidence that they they conduct red-teaming on self-replication or the effects of system interaction and tool use. & \cite{metasafety} \\
\midrule
Information sharing with companies & \tikzcmark & Meta is a member of the Frontier Model Forum, which facilitates information-sharing among companies. & \cite{frontiermodelforum} \\
\midrule
Information sharing with government & \tikzxmark & While Meta provides services procured by the federal government, no information is disclosed about what is provided as part of the procurement process, nor are separate mechanisms discussed for information sharing about the models & \cite{metasafety} \\
\midrule
Forum or mechanism for information sharing & \tikzcmark & Meta is a member of the U.S. AI Safety Institute Consortium and the Frontier Model Forum. & \cite{nistaisic, frontiermodelforum} \\
\midrule
Forum or mechanism shares information on risks & \tikzcmark & The Frontier Model Forum shares information as desired. & \cite{frontiermodelforum} \\
\midrule
Model weight cybersecurity practices & \tikzxmark & Meta does not specifically indicate they implement security practices for model weights in relation to their Llama models. & \cite{metasafety} \\
\midrule
Insider threat detection program & \tikzxmark & Meta indicates their intention to implement insider threat detection without providing evidence of implementation. & \cite{metasafety} \\
\midrule
Limiting weight-level access to relevant personnel & \tikzxmark & Meta indicates their intention to implement insider threat detection without providing evidence of implementation. & \cite{metasafety} \\
\midrule
Establish bounties, contests, or prizes & \tikzxmark & Meta has an existing bug bounty program for privacy and security issues that includes Meta's large language models; no additional bounties, contests, or prizes found. & \cite{metabounty} \\
\midrule
Include AI systems in their existing bug bounty programs & \tikzcmark & Meta has an existing bug bounty program for privacy and security issues that includes Meta's large language models. & \cite{metabounty} \\
\midrule
Robust provenance or watermarking for audio & \tikzcmark & Meta introduced AudioSeal, a watermarking technique for localized detection of AI-generated speech. & \cite{metaaudio} \\
\midrule
Robust provenance or watermarking for visual content & \tikzcmark & Meta added visible indicators on photorealistic images generated by AI. & \cite{metasafety, meta2023stablesignature} \\
\midrule
Develop tools or APIs to determine if a particular piece of content was created within their tools & \tikzcmark & Meta mentions building tools that can identify visible markers at scale, following the C2PA and IPTC technical standards. & \cite{meta2024labeling} \\
\midrule
Work with industry peers and standards-setting bodies as appropriate towards developing a technical framework to help users distinguish audio or visual content generated by users from audio or visual content generated by AI & \tikzcmark & Meta has worked with industry partners in PAI to develop provenance standards. & \cite{metasafety, meta2023stablesignature} \\
\midrule
Report capabilities & \tikzcmark & Model card for Llama 3.3 reports capabilities through benchmark results & \cite{meta2024llamacard} \\
\midrule
Report limitations & \tikzcmark & Model card for Llama 3.3 reports limitations & \cite{meta2024llamacard} \\
\midrule
Report domains of appropriate use & \tikzcmark & Model card for Llama 3.3 reports intended use cases & \cite{meta2024llamacard} \\
\midrule
Report domains of inappropriate use & \tikzcmark & Model card for Llama 3.3 enumerates prohibited uses. & \cite{meta2024llamacard} \\
\midrule
Report safety evaluations & \tikzcmark & Llama 3.3 model card reports evaluations around cybersecurity, child safety, and CBRN risks. & \cite{meta2024llamacard} \\
\midrule
Report on societal risks & \tikzxmark & While the Llama 3.3 model card includes testing for cybersecurity, child safety, and CBRN risks, it does not report societal risk such as fairness and bias. & \cite{meta2024llamacard} \\
\midrule
Report on adversarial testing used to determine appropriateness of deployment & \tikzxmark & The Llama 3.3 Model card describes their methods but does not disclose results of adversarial testing. & \cite{meta2024llamacard} \\
\midrule
Empower trust and safety teams & \tikzxmark & While Meta supports teams developing Purple Llama, Prompt Guard, and Llama Guard 3, all safety tools, there is no relevant evidnece of empowering its trust and safety team, which is responsible for monitoring potential risks like bias, misinformation, and security threats. & \cite{metallama} \\
\midrule
Advance AI safety research & \tikzcmark & Meta developed trust and safety tools and evaluations for AI developers and regularly publishes AI safety research. & \cite{metapurple} \\
\midrule
Advance privacy & \tikzcmark & Meta launched a generative AI privacy guide and has grown their privacy teams substantially in the last years. & \cite{meta2024genaiguide, meta_privacy_progress} \\
\midrule
Protect children & \tikzcmark & We award this point for their partnerships with Thorn and the National Center for Missing \& Exploited Children on combatting CSAM. & \cite{metallama} \\
\midrule
Support research and development of frontier AI systems that can help meet society’s greatest challenges & \tikzcmark & The partnership between Meta and CMU advances the use of AI for addressing climate change. & \cite{metaclimate} \\
\midrule
Support initiatives that foster the education and training of students and workers & \tikzcmark & As part of its AI Learning Alliance, Meta built a deep learning course curriculum with Georgia Tech that is free to all. They are also working with professors at HBCUs and HSIs to further develop and teach the curriculum. & \cite{metaalliance} \\
\midrule
Support initiatives that help citizens understand the nature, capabilities, limitations, and impact of the technology & \tikzxmark & Although Meta hosted a "Community Forum on Generative AI," which gave randomly selected member of the public the chance to share their views on the principles governing the use of AI chatbots, this initiative does not improve citizens' understanding of the nature, capabilities, limitations, and impact of the technology. & \cite{meta2024communityforums} \\

\end{longtable}

\newpage
\newpage
\small
\setlongtables
\begin{longtable}{p{0.32\textwidth} c p{0.45\textwidth} p{0.08\textwidth}}
\caption{\normalsize \textbf{Indicator Scores for Microsoft}} \\
\textbf{Indicator} & \textbf{Score} & \textbf{Justification} & \textbf{Source} \\
\toprule
\endfirsthead

\multicolumn{4}{c}{Indicator Scores for Microsoft -- \textit{Continued from previous page}} \\
\textbf{Indicator} & \textbf{Score} & \textbf{Justification} & \textbf{Source} \\
\toprule
\endhead

\multicolumn{4}{r}{\textit{Continued on next page}} \\
\endfoot

\bottomrule
\endlastfoot

Internal red-teaming & \tikzcmark & Microsoft has an AI Red Team that identifies safety and security vulnerabilities in their latest generative AI models. & \cite{msredteam} \\
\midrule
External red-teaming & \tikzxmark & Microsoft reports building external red-teaming capacity but does not make clear whether external experts have red-teamed their models. & \cite{msreport} \\
\midrule
Red teaming coverage of risks & \tikzxmark & While Microsoft's PyRit framework covers a broad range of risk areas, it does not include the risk of model autonomy and self-replication. & \cite{mspyrit} \\
\midrule
Information sharing with companies & \tikzcmark & Microsoft is a member of the Frontier Model Forum, which facilitates information-sharing among companies. & \cite{frontiermodelforum} \\
\midrule
Information sharing with government & \tikzxmark & While Microsoft services are procured by the federal government, no information is disclosed about what is provided as part of the procurement process, nor are separate mechanisms discussed for information sharing about Microsoft's models (as opposed to, say, their offerings of OpenAI models as an Azure service). & \cite{msgovai, msfederal} \\
\midrule
Forum or mechanism for information sharing & \tikzcmark & Microsoft is a member of the U.S. AI Safety Institute Consortium and the Frontier Model Forum. & \cite{nistaisic, frontiermodelforum} \\
\midrule
Forum or mechanism shares information on risks & \tikzcmark & The Frontier Model Forum shares information as desired. & \cite{frontiermodelforum} \\
\midrule
Model weight cybersecurity practices & \tikzcmark & Microsoft implements model-specific security controls. & \cite{mssafety} \\
\midrule
Insider threat detection program & \tikzxmark & Microsoft does not describe an insider threat detection model in relation to Phi or their AI services. & - \\
\midrule
Limiting weight-level access to relevant personnel & \tikzcmark & Microsoft implements strong identify and access control to their AI technology and logs access requests to identify anomalies and/or unauthorized access. & \cite{mssafety} \\
\midrule
Establish bounties, contests, or prizes & \tikzcmark & Microsoft launched a new AI bug bounty program. & \cite{mssafety} \\
\midrule
Include AI systems in their existing bug bounty programs & \tikzcmark & Microsoft created a Copilot bounty program that adheres to the terms of existing programs, namely Microsoft Bounty Terms and Conditions and their bounty Safe Harbor policy. & \cite{microsoft2025copilotbounty} \\
\midrule
Robust provenance or watermarking for audio & \tikzcmark & Microsoft watermarks to the speech outputs created with the personal voice feature. & \cite{msaudio} \\
\midrule
Robust provenance or watermarking for visual content & \tikzcmark & Microsoft attaches provenance metadata generated by their AI services & \cite{mssafety} \\
\midrule
Develop tools or APIs to determine if a particular piece of content was created within their tools & \tikzcmark & Content Credentials Verify is a tool for users to inspect the Content Credentials of AI-generated content via DALL-E through the Azure OpenAI endpoint. & \cite{mssafety} \\
\midrule
Work with industry peers and standards-setting bodies as appropriate towards developing a technical framework to help users distinguish audio or visual content generated by users from audio or visual content generated by AI & \tikzcmark & Microsoft co-founded C2PA and co-developed the C2PA technical specification. & \cite{mssafety} \\
\midrule
Report capabilities & \tikzcmark & Technical report for Phi-4 reports capabilities through benchmark results. & \cite{msphi4} \\
\midrule
Report limitations & \tikzcmark & Technical report for Phi-4 discloses limitations such as limited skill scope, bias in generation-based benchmarks, and limitations of multiple-choice tasks. & \cite{msphi4} \\
\midrule
Report domains of appropriate use & \tikzcmark & The Phi 4 model card enumerates intended and out-of-scope use cases. & \cite{msphicard} \\
\midrule
Report domains of inappropriate use & \tikzcmark & The Phi 4 model card enumerates intended and out-of-scope use cases. & \cite{msphicard} \\
\midrule
Report safety evaluations & \tikzxmark & Technical report for Phi-4 discloses safety evaluation results, but not for CBRN. & \cite{msphi4} \\
\midrule
Report on societal risks & \tikzxmark & No relevant evidence found. & - \\
\midrule
Report on adversarial testing used to determine appropriateness of deployment & \tikzcmark & Technical report for Phi-4 describes the results from their red-teaming exercise. & \cite{msphi4} \\
\midrule
Empower trust and safety teams & \tikzxmark & While Microsoft has created a responsible AI council, there is no relevant evidence around empowering its trust and safety team. & \cite{crampton2023reflecting} \\
\midrule
Advance AI safety research & \tikzcmark & Microsoft Azure AI releases tooling to advance AI safety. & \cite{msazure} \\
\midrule
Advance privacy & \tikzcmark & Microsoft Azure releases tooling to advance AI privacy. & \cite{msprivacy} \\
\midrule
Protect children & \tikzcmark & We award this point for their partnerships with Thorn on combatting CSAM. & \cite{mschild} \\
\midrule
Support research and development of frontier AI systems that can help meet society’s greatest challenges, such as climate change mitigation and adaptation, early cancer detection and prevention, and combating cyber threats & \tikzcmark & Microsoft launched AI for Health, a philantropic program to support researchers tackling global health challenges. & \cite{mshealth} \\
\midrule
Support initiatives that foster the education and training of students and workers to prosper from the benefits of AI & \tikzcmark & Microsoft launched an AI Skills Initiative to help students and workers develop AI skills. They also developed an Education AI Toolkit is intended for educators. & \cite{msskills, mstoolkit} \\
\midrule
Support initiatives that help citizens understand the nature, capabilities, limitations, and impact of the technology & \tikzxmark & While Microsoft hosted a series featuring Trevor Noah to discuss the potential of AI in addressing global issues, this initiative is a one-time occurence and not sustained. & \cite{msnoah} \\

\end{longtable}

\newpage
\newpage
\small
\setlongtables
\begin{longtable}{p{0.32\textwidth} c p{0.45\textwidth} p{0.08\textwidth}}
\caption{\normalsize \textbf{Indicator Scores for Nvidia}} \\
\textbf{Indicator} & \textbf{Score} & \textbf{Justification} & \textbf{Source} \\
\toprule
\endfirsthead

\multicolumn{4}{c}{Indicator Scores for Nvidia -- \textit{Continued from previous page}} \\
\textbf{Indicator} & \textbf{Score} & \textbf{Justification} & \textbf{Source} \\
\toprule
\endhead

\multicolumn{4}{r}{\textit{Continued on next page}} \\
\endfoot

\bottomrule
\endlastfoot

Internal red-teaming & \tikzcmark & Nvidia has an internal AI red team whose testing covers types of misues, societal risks, and security concerns. & \cite{nvidiared} \\
\midrule
External red-teaming & \tikzcmark & Nvidia participated in the Generative AI Red Teaming Challenge & \cite{humaneintelligence2023grt} \\
\midrule
Red teaming coverage of risks & \tikzxmark & While Nvidia describes the high-level risks their red teaming targets, there is nsufficient coverage of risk areas, such as CBRN and cyber capabilities. & \cite{nvidiared} \\
\midrule
Information sharing with companies & \tikzxmark & No relevant evidence found. & - \\
\midrule
Information sharing with government & \tikzxmark & While Nvidia has made generous contributions to NAIRR and they are collaborating with DARPA to identify AI-generated images, we do not count these towards fulfilling the commitment. There is no evidence that Nvidia has shared non-public information with governments about trust and safety risks, AI system capabilities, or attempts to circumvent safeguards. & \cite{nvidiaaisic, nvidiawashington} \\
\midrule
Forum or mechanism for information sharing & \tikzcmark & Nvidia is a member of the AI Safety Institute Consortium. & \cite{nistaisic} \\
\midrule
Forum or mechanism shares information on risks & \tikzxmark & Based on public information, the NIST AISIC does not share sensitive information on risks that must be carefully controlled and restricted to forum members to prevent misuse or security vulnerabilities. & - \\
\midrule
Model weight cybersecurity practices & \tikzxmark & While Nvidia provides related services to clients, they do not clearly disclose that they implement these practices in training their models. & \cite{nvidiasecurity} \\
\midrule
Insider threat detection program & \tikzxmark & While Nvidia provides related services to clients, they do not clearly disclose that they implement this threat detection internally. & \cite{nvidiamorpheus} \\
\midrule
Limiting weight-level access to relevant personnel & \tikzxmark & No relevant evidence found. & - \\
\midrule
Establish bounties, contests, or prizes & \tikzxmark & No relevant evidence found. & - \\
\midrule
Include AI systems in their existing bug bounty programs & \tikzxmark & No relevant evidence found. & - \\
\midrule
Robust provenance or watermarking for audio & \tikzcmark & Nvidia's flagship models do not generate audio or images, and so we default their score to 1 for this commitment. & - \\
\midrule
Robust provenance or watermarking for visual content & \tikzcmark & Nvidia's flagship models do not generate audio or images, and so we default their score to 1 for this commitment. & - \\
\midrule
Develop tools or APIs to determine if a particular piece of content was created within their tools & \tikzcmark & Nvidia's flagship models do not generate audio or images, and so we default their score to 1 for this commitment. & - \\
\midrule
Work with industry peers and standards-setting bodies as appropriate towards developing a technical framework to help users distinguish audio or visual content generated by users from audio or visual content generated by AI & \tikzcmark & Nvidia's flagship models do not generate audio or images, and so we default their score to 1 for this commitment. & - \\
\midrule
Report capabilities & \tikzcmark & Technical report for he Nemotron-4-340B Instruct reports capabilities through benchmark results. & \cite{nvidiatech} \\
\midrule
Report limitations & \tikzcmark & The Nemotron Hugging Face model card report limitations in toxic language, unsafe content, and societal biases. & \cite{nvidiabase} \\
\midrule
Report domains of appropriate use & \tikzcmark & The Nemotron Hugging Face model card reports intended uses. & \cite{nvidiainstruct} \\
\midrule
Report domains of inappropriate use & \tikzcmark & The NemoTron license outlines inappropriate uses. & \cite{nvidialicense} \\
\midrule
Report safety evaluations & \tikzxmark & Technical report for he Nemotron-4-340B Instruct details safety evaluation results, but none for CBRN. & \cite{nvidiatech} \\
\midrule
Report on societal risks & \tikzcmark & Technical report covers evaluations for safety and societal risks. & \cite{nvidiatech} \\
\midrule
Report on adversarial testing used to determine appropriateness of deployment & \tikzcmark & Technical report for he Nemotron-4-340B Instruct details safety evaluation results from red-teaming. & \cite{nvidiatech} \\
\midrule
Empower trust and safety teams & \tikzxmark & No relevant evidence found. & - \\
\midrule
Advance AI safety research & \tikzcmark & Nvidia developed an open source toolkit for developing safe and trustworhy LLM conversational systems. & \cite{nvidiasafe} \\
\midrule
Advance privacy & \tikzxmark & They do not publicly indicate that they concretely take actions to advance privacy. & - \\
\midrule
Protect children & \tikzxmark & No relevant evidence found. & - \\
\midrule
Support research and development of frontier AI systems that can help meet society’s greatest challenges, such as climate change mitigation and adaptation, early cancer detection and prevention, and combating cyber threats & \tikzcmark & Nvidia detailed partnerships to improve diagnostics and healthcare delivery and advance climate modeling efforts. & \cite{nvidiasummit} \\
\midrule
Support initiatives that foster the education and training of students and workers to prosper from the benefits of AI & \tikzxmark & While NVIDIA maintains a certification for educators through its Deep Learning Institute University Ambassador Program, it does not target these efforts to workers or the broader public. & \cite{nvidiaedu, nvidiaglobal} \\
\midrule
Support initiatives that help citizens understand the nature, capabilities, limitations, and impact of the technology & \tikzxmark & No relevant evidence found. & - \\

\end{longtable}

\newpage
\newpage
\small
\setlongtables
\begin{longtable}{p{0.32\textwidth} c p{0.45\textwidth} p{0.08\textwidth}}
\caption{\normalsize \textbf{Indicator Scores for OpenAI}} \\
\textbf{Indicator} & \textbf{Score} & \textbf{Justification} & \textbf{Source} \\
\toprule
\endfirsthead

\multicolumn{4}{c}{Indicator Scores for OpenAI -- \textit{Continued from previous page}} \\
\textbf{Indicator} & \textbf{Score} & \textbf{Justification} & \textbf{Source} \\
\toprule
\endhead

\multicolumn{4}{r}{\textit{Continued on next page}} \\
\endfoot

\bottomrule
\endlastfoot

Internal red-teaming & \tikzcmark & OpenAI documents internal red-teaming in o1 system card in areas including misuse, societal risks, and national security concerns. & \cite{openaio1} \\
\midrule
External red-teaming & \tikzcmark & OpenAI launched the OpenAI Red Teaming Network, a community of experts to inform their risk assessment. & \cite{openaired} \\
\midrule
Red teaming coverage of risks & \tikzcmark & The o1 system card details full coverage of the risk areas in the voluntary commitments under their external red-teaming efforts & \cite{openaio1} \\
\midrule
Information sharing with companies & \tikzcmark & OpenAI is a member of the Frontier Model Forum, which facilitates information-sharing among companies. & \cite{frontiermodelforum} \\
\midrule
Information sharing with government & \tikzcmark & OpenAI provides the U.S. AI Safety Institute with new models before and and following their releases. & \cite{nistsharing} \\
\midrule
Forum or mechanism for information sharing & \tikzcmark & OpenAI is a member of the U.S. AI Safety Institute Consortium and the Frontier Model Forum. & \cite{nistaisic, frontiermodelforum} \\
\midrule
Forum or mechanism shares information on risks & \tikzcmark & The Frontier Model Forum shares information as desired. & \cite{frontiermodelforum} \\
\midrule
Model weight cybersecurity practices & \tikzcmark & OpenAI created secure research environments dedicated to model security, including protecting model weights. & \cite{openaisecure, openaifrontier} \\
\midrule
Insider threat detection program & \tikzxmark & No evidence found. & \cite{openaiinsider} \\
\midrule
Limiting weight-level access to relevant personnel & \tikzcmark & OpenAI built a service called AccessManager to manage internal authorization and access to sensitive resources, including model weights. & \cite{openaisecure} \\
\midrule
Establish bounties, contests, or prizes & \tikzcmark & OpenAI launched a bug bounty program for their systems. & \cite{openaibounty} \\
\midrule
Include AI systems in their existing bug bounty programs & \tikzcmark & OpenAI expanded their bug bounty to include model issues. & \cite{openaibounty} \\
\midrule
Robust provenance or watermarking for audio & \tikzcmark & OpenAI has implemented tamper-resistant watermarking on their generated audiovisual content. & \cite{openaiwater} \\
\midrule
Robust provenance or watermarking for visual content & \tikzcmark & OpenAI has implemented tamper-resistant watermarking on their generated audiovisual content. & \cite{openaiwater} \\
\midrule
Develop tools or APIs to determine if a particular piece of content was created within their tools & \tikzcmark & OpenAI built a tool that predicts the likelihood an image was generated by DALL-E 3. & \cite{openaiwater} \\
\midrule
Work with industry peers and standards-setting bodies as appropriate towards developing a technical framework to help users distinguish audio or visual content generated by users from audio or visual content generated by AI & \tikzcmark & OpenAI joined C2PA and is working towards developing the C2PA standard. & \cite{openaiwater} \\
\midrule
Report capabilities & \tikzcmark & OpenAI describes o1's capabilities in the model release announcement. & \cite{openaio1} \\
\midrule
Report limitations & \tikzxmark & While the o1 system card details a number of safety challenges, it does not provide a comprehensive description of the model limitations. & \cite{openaio1} \\
\midrule
Report domains of appropriate use & \tikzxmark & While OpenAI's usage policies describes prohibited uses, it does not report domains of appropriate use. & \cite{openaiusage} \\
\midrule
Report domains of inappropriate use & \tikzcmark & OpenAI's usage policies describes prohibited uses. & \cite{openaiusage} \\
\midrule
Report safety evaluations & \tikzcmark & o1 system card includes safety evaluation. & \cite{openaio1} \\
\midrule
Report on societal risks & \tikzcmark & o1 system card includes assessment on measuring CBRN risks and impact on different industries / occupations according to prepardness framework. & \cite{openaio1} \\
\midrule
Report on adversarial testing used to determine appropriateness of deployment & \tikzcmark & o1 system card details adversarial testing results for external red-teaming. & \cite{openaio1} \\
\midrule
Empower trust and safety teams & \tikzxmark & While OpenAI formed a Safety and Security Committee, there is no relevant mention of a trust and safety team. & \cite{openaicommittee} \\
\midrule
Advance AI safety research & \tikzcmark & OpenAI developed a Prepardness Framework as a proactive, risk-based approach to AI development. & \cite{openaisafety} \\
\midrule
Advance privacy & \tikzcmark & The OpenAI consumer privacy documentation details how they aim to minimize personal information in model training and in model generations. & \cite{openaiprivacy} \\
\midrule
Protect children & \tikzcmark & We award this point for their partnerships with Thorn and the National Center for Missing \& Exploited Children on combatting CSAM. & \cite{openaiapproach, openaisafety} \\
\midrule
Support research and development of frontier AI systems that can help meet society’s greatest challenges, such as climate change mitigation and adaptation, early cancer detection and prevention, and combating cyber threats & \tikzcmark & OpenAI is working with Color Health to accelerate cancer patients' access to treatement. & \cite{openaihealth} \\
\midrule
Support initiatives that foster the education and training of students and workers to prosper from the benefits of AI & \tikzcmark & OpenAI partners with Common Sense media to provide free AI training courses for K-12 educators, helping them understand and responsibly implement the basics of AI into their work. & \cite{openaicommonsense} \\
\midrule
Support initiatives that help citizens understand the nature, capabilities, limitations, and impact of the technology & \tikzxmark & Microsoft and OpenAI launched a \$2 million Societal Resilience Fund to further AI education and literacy among voters and vulnerable communities. However, we do not award this point because the fund is described to support initiatives that create better understanding of AI capabilities, and not the nature, limitations, and impact of the technology. & \cite{openairesilience} \\

\end{longtable}

\newpage
\newpage
\small
\setlongtables
\begin{longtable}{p{0.32\textwidth} c p{0.45\textwidth} p{0.08\textwidth}}
\caption{\normalsize \textbf{Indicator Scores for Palantir}} \\
\textbf{Indicator} & \textbf{Score} & \textbf{Justification} & \textbf{Source} \\
\toprule
\endfirsthead

\multicolumn{4}{c}{Indicator Scores for Palantir -- \textit{Continued from previous page}} \\
\textbf{Indicator} & \textbf{Score} & \textbf{Justification} & \textbf{Source} \\
\toprule
\endhead

\multicolumn{4}{r}{\textit{Continued on next page}} \\
\endfoot

\bottomrule
\endlastfoot

Internal red-teaming & \tikzxmark & No relevant evidence found. & - \\
\midrule
External red-teaming & \tikzxmark & No relevant evidence found. & - \\
\midrule
Red teaming coverage of risks & \tikzxmark & No relevant evidence found. & - \\
\midrule
Information sharing with companies & \tikzxmark & No relevant evidence found. & - \\
\midrule
Information sharing with government & \tikzxmark & While Palantir extensively engages the US government in procurement relationships, we find no evidence of information sharing with the government outside of these procurement relationships and the procurement relations/contracts are also not described publicly. & \cite{palantiromb} \\
\midrule
Forum or mechanism for information sharing & \tikzcmark & Palantir is a member of the U.S. AI Safety Institute Consortium. & \cite{nistaisic} \\
\midrule
Forum or mechanism shares information on risks & \tikzxmark & Based on public information, the NIST AISIC does not share information on risks. & - \\
\midrule
Model weight cybersecurity practices & \tikzcmark & Palantir is not a frontier model developer to our knowledge, so we default their score to \tikzcmark{} for this commitment. & - \\
\midrule
Insider threat detection program & \tikzcmark & Palantir is not a frontier model developer to our knowledge, so we default their score to \tikzcmark{} for this commitment. & - \\
\midrule
Limiting weight-level access to relevant personnel & \tikzcmark & Palantir is not a frontier model developer to our knowledge, so we default their score to \tikzcmark{} for this commitment. & - \\
\midrule
Establish bounties, contests, or prizes & \tikzxmark & No relevant evidence found. & - \\
\midrule
Include AI systems in their existing bug bounty programs & \tikzxmark & No mention of AI in existing bug bounties. & - \\
\midrule
Robust provenance or watermarking for audio & \tikzcmark & Palantir's flagship models do not generate audio or images, and so we default their score to 1 for this commitment. & - \\
\midrule
Robust provenance or watermarking for visual content & \tikzcmark & Palantir's flagship models do not generate audio or images, and so we default their score to 1 for this commitment. & - \\
\midrule
Develop tools or APIs to determine if a particular piece of content was created within their tools & \tikzcmark & Palantir's flagship models do not generate audio or images, and so we default their score to 1 for this commitment. & - \\
\midrule
Work with industry peers and standards-setting bodies towards developing a technical framework & \tikzcmark & Palantir's flagship models do not generate audio or images, and so we default their score to 1 for this commitment. & - \\
\midrule
Report capabilities & \tikzcmark & Documentation details platform capabilities. & \cite{palantirai} \\
\midrule
Report limitations & \tikzxmark & No relevant evidence found. & - \\
\midrule
Report domains of appropriate use & \tikzcmark & Palantir platform describes many domains of use. & \cite{palantiraip} \\
\midrule
Report domains of inappropriate use & \tikzcmark & Palantir's terms and conditions enumerate illegal and improper uses for their products & \cite{palantirterms} \\
\midrule
Report safety evaluations & \tikzxmark & No relevant evidence found. & - \\
\midrule
Report on societal risks & \tikzxmark & While high-level ethical principles are discussed, no specifics on risks associated with the platform in particular. & \cite{palantirethics} \\
\midrule
Report on adversarial testing & \tikzxmark & No relevant evidence found. & - \\
\midrule
Empower trust and safety teams & \tikzxmark & No relevant evidence found. & - \\
\midrule
Advance AI safety research & \tikzxmark & No relevant evidence found. & - \\
\midrule
Advance privacy & \tikzcmark & Palantir has a Privacy and Civil Liberties Engineering team and released a series of writing around their privacy-by-design engineering approach. & \cite{palantirprivacy} \\
\midrule
Protect children & \tikzcmark & We award this point for the Palantir-NCMEC partnership. & \cite{palantirncmec} \\
\midrule
Support research and development of frontier AI systems that can help meet society's greatest challenges & \tikzcmark & Palantir and Tree Energy Solutions partnered to leverage Palantir AI software to drive the green energy transition. & \cite{palantirtes} \\
\midrule
Support initiatives that foster the education and training of students and workers & \tikzxmark & No evidence found; we do not award this for a generic scholarship program on STEM subjects. & \cite{palantirscholar} \\
\midrule
Support initiatives that help citizens understand the technology & \tikzxmark & No relevant evidence found. & - \\

\end{longtable}

\newpage
\newpage
\small
\setlongtables
\begin{longtable}{p{0.32\textwidth} c p{0.45\textwidth} p{0.08\textwidth}}
\caption{\normalsize \textbf{Indicator Scores for Salesforce}} \\
\textbf{Indicator} & \textbf{Score} & \textbf{Justification} & \textbf{Source} \\
\toprule
\endfirsthead

\multicolumn{4}{c}{Indicator Scores for Salesforce -- \textit{Continued from previous page}} \\
\textbf{Indicator} & \textbf{Score} & \textbf{Justification} & \textbf{Source} \\
\toprule
\endhead

\multicolumn{4}{r}{\textit{Continued on next page}} \\
\endfoot

\bottomrule
\endlastfoot

Internal red-teaming & \tikzxmark & Salesforce has conducted 19 internal red teaming exercises, but there is no evidence that these exercise covers types of misuse, societal risks, and national security concerns. & \cite{salesforceai} \\
\midrule
External red-teaming & \tikzxmark & While Salesforce has conducted two external red teaming exercises, we cannot gauge whether these are structured, organized operations that constitute a detailed red teaming regime based on available information. & \cite{salesforceai} \\
\midrule
Red teaming coverage of risks & \tikzxmark & No relevant evidence found. & - \\
\midrule
Information sharing with companies & \tikzcmark & Salesforce is a member of the AI Alliance. & \cite{salesforce_ibm_2024} \\
\midrule
Information sharing with government & \tikzxmark & While Salesforce publishes many (20+) articles publicly that provide value to the government and to other companies, given this information is public, we do not consider it as eligible to award this point and find no other evidence. & \cite{salesforceai} \\
\midrule
Forum or mechanism for information sharing & \tikzcmark & Salesforce is a member of the U.S. AI Safety Institute Consortium. & \cite{nistaisic} \\
\midrule
Forum or mechanism shares information on risks & \tikzxmark & Based on public information, the NIST AISIC does not share information on risks. & - \\
\midrule
Model weight cybersecurity practices & \tikzxmark & No information found on model weight security specifically. & - \\
\midrule
Insider threat detection program & \tikzxmark & While Salesforce provides insider threat detection services, there is no clear indication that these programs are implemented in relation to Salesforce's AI systems. & \cite{salesforce_insider_threats_2024} \\
\midrule
Limiting weight-level access to relevant personnel & \tikzxmark & No relevant evidence found. & - \\
\midrule
Establish bounties, contests, or prizes & \tikzxmark & While Salesforce has a bug bounty program to prevent AI-powered cyber threats, Salesforce does not specify that their AI systems are covered under the scope of this program. & \cite{salesforcehack} \\
\midrule
Include AI systems in their existing bug bounty programs & \tikzxmark & While Salesforce has a bug bounty program to prevent AI-powered cyber threats, Salesforce does not specify that their AI systems are covered under the scope of this program. & \cite{salesforcehack} \\
\midrule
Robust provenance or watermarking for audio & \tikzcmark & Salesforce's flagship models do not generate audio or images, and so we default their score to 1 for this commitment. & - \\
\midrule
Robust provenance or watermarking for visual content & \tikzcmark & Salesforce's flagship models do not generate audio or images, and so we default their score to 1 for this commitment. & - \\
\midrule
Develop tools or APIs to determine if a particular piece of content was created within their tools & \tikzcmark & Salesforce's flagship models do not generate audio or images, and so we default their score to 1 for this commitment. & - \\
\midrule
Work with industry peers and standards-setting bodies towards developing a technical framework & \tikzcmark & Salesforce's flagship models do not generate audio or images. & - \\
\midrule
Report capabilities & \tikzcmark & The XGen-7B technical report describes model capabilities. & \cite{salesforcexgen} \\
\midrule
Report limitations & \tikzcmark & The XGen-7B technical report describes limitations. & \cite{salesforcexgen} \\
\midrule
Report domains of appropriate use & \tikzcmark & The Salesforce Einstein model cards include intended uses. & \cite{salesforceeinstein} \\
\midrule
Report domains of inappropriate use & \tikzcmark & The Salesforce Einstein model cards include out-of-scope uses. & \cite{salesforceeinstein} \\
\midrule
Report safety evaluations & \tikzxmark & No relevant evidence found. & - \\
\midrule
Report on societal risks & \tikzxmark & No relevant evidence found. & - \\
\midrule
Report on adversarial testing & \tikzxmark & No relevant evidence found. & - \\
\midrule
Empower trust and safety teams & \tikzxmark & While Salesforce has funded user research headcount to focus on trust and responsible AI, there is no relevant evidence of empowering its trust and safety team. & \cite{salesforceai} \\
\midrule
Advance AI safety research & \tikzcmark & Salesforce AI Research has published research on trust and safety evaluations for LLMs and have open-sourced a library for auditing generative AI for trustworthiness. & \cite{salesforceai} \\
\midrule
Advance privacy & \tikzcmark & We award this point for the Einstein Trust Layer, which reduces the presence of PII in prompts to LMs, which may in turn reduce privacy risks associated with prompt-conditioned LM generation and/or the storage of prompts. & \cite{salesforceai} \\
\midrule
Protect children & \tikzcmark & We award this point for their role in the Technology Task Force for CSAM. & \cite{salesforcethorn} \\
\midrule
Support research and development of frontier AI systems that can help meet society's greatest challenges & \tikzcmark & Salesforce launched the Salesforce Accelerator to support purpose-driven nonprofits in developing AI-powered climate solutions. & \cite{salesforceai} \\
\midrule
Support initiatives that foster the education and training of students and workers & \tikzcmark & Salesforce provides grants to U.S. school districts and global education nonprofits to introduce AI literacy. & \cite{salesforcegrants} \\
\midrule
Support initiatives that help citizens understand the technology & \tikzxmark & While Salesforce provides information that could be intelligible to laypersons, we find no evidence of initiatives specifically aimed at increasing citizen's AI literacy. & - \\

\end{longtable}

\newpage
\newpage
\small
\setlongtables
\begin{longtable}{p{0.32\textwidth} c p{0.45\textwidth} p{0.08\textwidth}}
\caption{\normalsize \textbf{Indicator Scores for Scale AI}} \\
\textbf{Indicator} & \textbf{Score} & \textbf{Justification} & \textbf{Source} \\
\toprule
\endfirsthead

\multicolumn{4}{c}{Indicator Scores for Scale AI -- \textit{Continued from previous page}} \\
\textbf{Indicator} & \textbf{Score} & \textbf{Justification} & \textbf{Source} \\
\toprule
\endhead

\multicolumn{4}{r}{\textit{Continued on next page}} \\
\endfoot

\bottomrule
\endlastfoot

Internal red-teaming & \tikzxmark & No internal red-teaming described for Donovan or other models built by Scale AI, though we recognize that Scale does conduct such pre-deployment red-teaming for other developers' models and creates related services. & \cite{scalevision, scalewhite} \\
\midrule
External red-teaming & \tikzcmark & Scale AI participated in the Generative AI Red Teaming Challenge & \cite{humaneintelligence2023grt} \\
\midrule
Red teaming coverage of risks & \tikzxmark & No relevant evidence found of red-teaming for specific risks; we do not award this point for creating safety benchmarks like WMDP nor for red-teaming OpenAI's GPT-4 model. & \cite{scalewmdp, scaleevaluation} \\
\midrule
Information sharing with companies & \tikzxmark & While Scale AI engages in commercial partnerships with other companies,  we do not award this point given that the primary objective of these partnerships is not to advance the trust and safety of AI. & \cite{scalecsis, scalellm} \\
\midrule
Information sharing with government & \tikzxmark & While ScaleAI provides services procured by the federal government, no information is disclosed about what is provided as part of the procurement process, nor are separate mechanisms discussed for information sharing about the models & \cite{scaledefense} \\
\midrule
Forum or mechanism for information sharing & \tikzcmark & Scale AI is a member of the U.S. AI Safety Institute Consortium. & \cite{nistaisic} \\
\midrule
Forum or mechanism shares information on risks & \tikzxmark & Based on public information, the NIST AISIC does not share sensitive information on risks that must be carefully controlled and restricted to forum members to prevent misuse or security vulnerabilities. & - \\
\midrule
Model weight cybersecurity practices & \tikzxmark & No relevant evidence found. & - \\
\midrule
Insider threat detection program & \tikzxmark & No relevant evidence found. & - \\
\midrule
Limiting weight-level access to relevant personnel & \tikzxmark & No relevant evidence found. & - \\
\midrule
Establish bounties, contests, or prizes & \tikzxmark & While Scale AI launched the Humanity's Last Exam with a prize pool, this project is not aimed at incentivizing the responsible disclosure of weaknesses, such as unsafe behaviors. & \cite{scaleexam} \\
\midrule
Include AI systems in their existing bug bounty programs & \tikzxmark & No relevant evidence found. & - \\
\midrule
Robust provenance or watermarking for audio & \tikzcmark & Scale AI's flagship models do not generate audio or images, and so we default their score to 1 for this commitment. & - \\
\midrule
Robust provenance or watermarking for visual content & \tikzcmark & Scale AI's flagship models do not generate audio or images, and so we default their score to 1 for this commitment. & - \\
\midrule
Develop tools or APIs to determine if a particular piece of content was created within their tools & \tikzcmark & Scale AI's flagship models do not generate audio or images, and so we default their score to 1 for this commitment. & - \\
\midrule
Work with industry peers and standards-setting bodies towards developing a technical framework & \tikzcmark & Scale AI's flagship models do not generate audio or images, and so we default their score to 1 for this commitment. & - \\
\midrule
Report capabilities & \tikzcmark & Scale reports that Donovan has capabilities for military and intelligence operations planning and for target analysis. & \cite{scaledonovan} \\
\midrule
Report limitations & \tikzxmark & While Scale describes limitations of LMs in the context of their work evaluation other developers' LMs, we do not find any description of the limitations of their own tools. & \cite{scalemodel} \\
\midrule
Report domains of appropriate use & \tikzcmark & We award this point for the description of uses of Donovan. & \cite{scaledonovan} \\
\midrule
Report domains of inappropriate use & \tikzcmark & The Scale Acceptable Use Policy outlines inappropriate uses of its services. & \cite{scaleaup} \\
\midrule
Report safety evaluations & \tikzxmark & While Scale conducts safety evaluations of other developers' models, we do not find evidence of safety evaluations for Donovan. & \cite{scalemodel} \\
\midrule
Report on societal risks & \tikzxmark & While Scale conducts safety evaluations, and provides a holistic framework for testing and evaluation that addresses societal risks, they do not report on such matters for Donovan. & \cite{scaleevaluation, scalemodel} \\
\midrule
Report on adversarial testing & \tikzxmark & No relevant evidence found. & - \\
\midrule
Empower trust and safety teams & \tikzxmark & Scale AI launched the Safety, Evaluations, and Alignment Lab aimed at enhancing transparency and standardization in AI safety. & \cite{scaleseal} \\
\midrule
Advance AI safety research & \tikzcmark & We award this point for their safety evaluation benchmark: the Weapons of Mass Destruction Proxy. & \cite{scaleresponsible} \\
\midrule
Advance privacy & \tikzxmark & No relevant evidence found. & - \\
\midrule
Protect children & \tikzxmark & No relevant evidence found. & - \\
\midrule
Support research and development of frontier AI systems that can help meet society’s greatest challenges & \tikzcmark & We award this point for the CSIS partnership on international security. & \cite{scalecsis, scaleukraine} \\
\midrule
Support initiatives that foster the education and training of students and workers & \tikzxmark & While Scale AI partnered with Chegg to develop proprietary LLMs for personalized learning, this initiative is not accessible to all students and workers and does not benefit them in developing AI skills. & \cite{scalechegg} \\
\midrule
Support initiatives that help citizens understand the nature, capabilities, limitations, and impact of the technology & \tikzxmark & While Scale provides public leaderboards, we find no evidence of direct methods for increasing lay citizen AI literacy. & \cite{scaleleader} \\

\end{longtable}

\newpage
\newpage
\small
\setlongtables
\begin{longtable}{p{0.32\textwidth} c p{0.45\textwidth} p{0.08\textwidth}}
\caption{\normalsize \textbf{Indicator Scores for Stability AI}} \\
\textbf{Indicator} & \textbf{Score} & \textbf{Justification} & \textbf{Source} \\
\toprule
\endfirsthead

\multicolumn{4}{c}{Indicator Scores for Stability AI -- \textit{Continued from previous page}} \\
\textbf{Indicator} & \textbf{Score} & \textbf{Justification} & \textbf{Source} \\
\toprule
\endhead

\multicolumn{4}{r}{\textit{Continued on next page}} \\
\endfoot

\bottomrule
\endlastfoot

Internal red-teaming & \tikzxmark & While the model card for Stable Diffusion details red-teaming as part of integrity evaluation, the company does not distinguish between internal and external red-teaming in its description. & \cite{stabilitycard} \\
\midrule
External red-teaming & \tikzcmark & Stability AI participated in the Generative AI Red Teaming Challenge & \cite{humaneintelligence2023grt} \\
\midrule
Red teaming coverage of risks & \tikzxmark & No relevant evidence found. & - \\
\midrule
Information sharing with companies & \tikzxmark & While Stability AI engages in commercial partnerships with other companies,  we do not award this point given that the primary objective of these partnerships is not to advance the trust and safety of AI. & \cite{stabilityamazon} \\
\midrule
Information sharing with government & \tikzxmark & Although Stability AI participated in the U.S. Senate AI Insight Forum, we do not award them this point because they did not initiate engagement and this engagement is not ongoing. & \cite{stabilitysenate} \\
\midrule
Forum or mechanism for information sharing & \tikzcmark & Stability AI is a member of the U.S. AI Safety Institute Consortium. & \cite{nistaisic} \\
\midrule
Forum or mechanism shares information on risks & \tikzcmark & We award this point for the commitments made in relation to sharing of information on CSAM. & \cite{stabilitysafety} \\
\midrule
Model weight cybersecurity practices & \tikzxmark & No relevant evidence found. & - \\
\midrule
Insider threat detection program & \tikzxmark & No relevant evidence found. & - \\
\midrule
Limiting weight-level access to relevant personnel & \tikzxmark & No relevant evidence found. & - \\
\midrule
Establish bounties, contests, or prizes & \tikzxmark & No relevant evidence found. & - \\
\midrule
Include AI systems in their existing bug bounty programs & \tikzxmark & No relevant evidence found. & - \\
\midrule
Robust provenance or watermarking for audio & \tikzxmark & No relevant evidence found. & - \\
\midrule
Robust provenance or watermarking for visual content & \tikzcmark & We award this point for the integration of Content Credentials into the Stability AI API. & \cite{stabilityapi} \\
\midrule
Develop tools or APIs to determine if a particular piece of content was created within their tools & \tikzcmark & We award this point for the integration of Content Credentials into the Stability AI API. & \cite{stabilityapi} \\
\midrule
Work with industry peers and standards-setting bodies towards developing a technical framework & \tikzcmark & We award this point for their collaboration with the Content Authenticity Initiative and their work on adopting the C2PA standard for metadata. & \cite{stabilitynist} \\
\midrule
Report capabilities & \tikzcmark & Stability AI reports model capabilities through benchmark results in the Stable Diffusion 3 technical report & \cite{stabilitytech} \\
\midrule
Report limitations & \tikzcmark & Model card for Stable Diffusion 3.5 on Hugging Face describes some limitations. & \cite{stabilitycard} \\
\midrule
Report domains of appropriate use & \tikzcmark & Model card on Hugging Face for Stable Diffusion 3.5 details intended uses. & \cite{stabilitycard} \\
\midrule
Report domains of inappropriate use & \tikzcmark & Stability AI reports domains of inappropriate use in their Acceptable Use Policy & \cite{stabilitypolicy} \\
\midrule
Report safety evaluations & \tikzxmark & No relevant evidence found. & - \\
\midrule
Report on societal risks & \tikzxmark & No relevant evidence found. & - \\
\midrule
Report on adversarial testing & \tikzxmark & No relevant evidence found. & - \\
\midrule
Empower trust and safety teams & \tikzxmark & No evidence found; external partnerships do not clearly empower their trust and safety team. & \cite{stabilitycollab} \\
\midrule
Advance AI safety research & \tikzxmark & No relevant evidence found that Stability explicitly advances AI safety, though the open release of model weights with documentation of datasets does enable others to do better AI safety research. & \cite{stabilitysafemain} \\
\midrule
Advance privacy & \tikzxmark & No relevant evidence found. & - \\
\midrule
Protect children & \tikzcmark & We award this point for their partnerships with Thorn and All Tech is Human on CSAM. & \cite{stabilitysafety} \\
\midrule
Support research and development of frontier AI systems that can help meet society’s greatest challenges & \tikzxmark & No relevant evidence found. & - \\
\midrule
Support initiatives that foster the education and training of students and workers & \tikzxmark & No relevant evidence of significant support of initiatives beyond a one-off six-week program led by HUG Studios that we deem insufficient to award this point. While Stability AI has a learning hub around setting up and using its models, these courses and guides do not adequately provide the training students and workers need to prosper from the benefits of AI. & \cite{stabilityhug} \\
\midrule
Support initiatives that help citizens understand the nature, capabilities, limitations, and impact of the technology & \tikzxmark & No relevant evidence found. & - \\

\end{longtable}

\newpage
\section{Appendix C: Company-Model Mapping}
We map from the 16 signatory companies to their flagship models (or systems if the company generally does not build models) as follows: (Adobe, Firefly Image 3), (Amazon, Nova), (Anthropic, Claude 3.5 Sonnet (new)), (Apple, Apple Intelligence Foundation Language Models), (Cohere, Command R+), (IBM, Granite 3.0), (Inflection, Infection 3.0/Pi 3.0), (Google, Gemini 1.0), (Meta, Llama 3.3), (Microsoft, Phi-4), (Nvidia, Nemotron-4 340B), (OpenAI, o1), (Palantir, AIP), (Salesforce, xgen), (Scale AI, Donovan), (Stability AI, Stable Diffusion 3.5).
\newpage
\section{Appendix D: Scoring Criteria}

\small
\setlongtables
\begin{longtable}{p{0.2\textwidth} p{0.4\textwidth} p{0.4\textwidth}}
\caption{\normalsize \textbf{Scoring Criteria for Indicators}} \\
\textbf{Indicator} & \textbf{Commitment Text} & \textbf{Criteria} \\
\toprule
\endfirsthead

\multicolumn{3}{c}{Scoring Criteria for Indicators -- \textit{Continued from previous page}} \\
\textbf{Indicator} & \textbf{Commitment Text} & \textbf{Criteria} \\
\toprule
\endhead

\multicolumn{3}{r}{\textit{Continued on next page}} \\
\endfoot

\bottomrule
\endlastfoot

Internal red-teaming & ``Commit to internal ... red-teaming of models or systems in areas including misuse, societal risks, and national security concerns, such as bio, cyber, and other safety areas.`` & The company differentiates internal and external red-teaming initiatives.
Its internal red-teaming efforts cover, at minimum, areas of misuse, societal risks, and national security concerns. These are areas listed in the commitment. \\
\midrule
External red-teaming & ``Companies commit to advancing this area of research, and to developing a multi-faceted, specialized, and detailed red-teaming regime, including drawing on independent domain experts, for all major public releases of new models within scope.`` & We consider a detailed red-teaming regime to be a structured, organized exercise with the sole focus of external red-teaming. We consider a disclosure about a detailed red-teaming regime with independent domain experts to be a description of the role and activities of the external red team.
The company differentiates internal and external red-teaming initiatives. 
A bug bounty does not constitute an external red teaming regime. \\
\midrule
Red teaming coverage of risks & ``In designing the regime, they will ensure that they give significant attention to the following:
\begin{itemize}
    \item Bio, chemical, and radiological risks, such as the ways in which systems can lower barriers to entry for weapons development, design, acquisition, or use
    \item Cyber capabilities, such as the ways in which systems can aid vulnerability discovery, exploitation, or operational use, bearing in mind that such capabilities could also have useful
    \item Cyber capabilities, such as the ways in which systems can aid vulnerability discovery, exploitation, or operational use, bearing in mind that such capabilities could also have useful defensive applications and might be appropriate to include in a system. 
    \item The effects of system interaction and tool use, including the capacity to control physical systems
    \item The capacity for models to make copies of themselves or “self-replicate”.
    \item Societal risks, such as bias and discrimination`` 
\end{itemize} & The company provides full coverage of the risk areas outlined in the voluntary commitments in their red-teaming efforts.  
\\
\midrule
Information sharing with companies & ``Work toward information sharing among companies ... regarding trust and safety risks, dangerous or emergent capabilities, and attempts to circumvent safeguards`` & 
The information shared must be outside of publicly available knowledge.
The information shared must relate to trust and safety risks, AI system capabilities, or attempts to circumvent the safeguards of their models.
The primary objective of information-sharing is to advance AI trust and safety. Information-sharing through commercial partnerships do not satisfy this commitment.
\\
\midrule
Information sharing with government & ``Work toward information sharing among companies and governments regarding trust and safety risks, dangerous or emergent capabilities, and attempts to circumvent safeguards`` & The information shared must be outside of public knowledge.
Information-sharing is facilitated through a sustained and proactive mechanism. Senate testimonies do not fullfill this commitment because the companies did not initiate the enegagement and the engagement is not ongoing. The primary objective of information-sharing is to advance AI trust and safety. The information shared must relate to trust and safety risks, AI system capabilities, or attempts to circumvent safeguards of their models. \\
\midrule
Forum or mechanism for information sharing & ``They commit to establish or join a forum or mechanism through which they can develop, advance, and adopt shared standards and best practices for frontier AI safety, such as the NIST AI Risk Management Framework or future standards related to red-teaming, safety, and societal risks.`` & The company establishes or joins a forum that is dedicated to developing standards and best practices for frontier AI safety. \\
\midrule
Forum or mechanism shares information on risks & ``The forum or mechanism can facilitate the sharing of information on advances in frontier capabilities and emerging risks and threats, such as attempts to circumvent safeguards, and can facilitate the development of technical working groups on priority areas of concern.`` & The forum restricts who can join and what they do with the shared information. Information on risks should be sensitive information beyond public knowledge that must be controlled and restricted to forum members to prevent misuse or security vulnerabilities. \\
\midrule
Model weight cybersecurity practices & ``In addition, it requires storing and working with the weights in an appropriately secure environment to reduce the risk of unsanctioned release.`` & The company implements security controls for their models - specifically, the controls provide protection around the model weights. General best practices for cloud security or cybersecurity do not fulfill this commitment. \\
\midrule
Insider threat detection program & ``This includes ... establishing a robust insider threat detection program consistent with protections provided for their most valuable intellectual property and trade secrets.`` & The company implements insider threat detection programs in relation to their models or AI services. \\
\midrule
Limiting weight-level access to relevant personnel & ``This includes limiting access to model weights to those whose job function requires it ...`` & The company restricts access to model weights to only authorized personnel who require it for their role. Access control is specific to model weight. Restricting access to data and code do not fulfill this commitment as types of assets distinct from model weights. \\
\midrule
Establish bounties, contests, or prizes & ``They commit to establishing for systems within scope bounty systems, contests, or prizes to incent the responsible disclosure of weaknesses, such as unsafe behaviors, or to include AI systems in their existing bug bounty programs`` & The company establishes a bug bounty, contest, or prize. The bounty incentivizes the responsible disclosure of model vulnerabilities and safety issues. The company allows for external participation in the bug bounty. Company-wide bounty programs do not count. The bug bounty, contest, or prize covers their flagship model or AI system. The company provides sufficient information for the public to gauge the activities of the bug bounty, contest, or prize, and mechanisms for participation. \\
\midrule
Include AI systems in their existing bug bounty programs & ``They commit ... to include AI systems in their existing bug bounty programs`` &  The company expands their existing bug bounty to include AI systems. This could include covering additional components of their flagship model or including additional categories of vulnerabilities. \\
\midrule
Robust provenance or watermarking for audio & ``To further this goal, they agree to develop robust mechanisms, including provenance and/or watermarking systems for audio or visual content created by any of their publicly available systems within scope introduced after the watermarking system is developed.`` & The company develops and implements provenance and/or watermarking systems for audio content created by any of their publicly available AI models.\\
\midrule
Robust provenance or watermarking for visual content & ``To further this goal, they agree to develop robust mechanisms, including provenance and/or watermarking systems for audio or visual content created by any of their publicly available systems within scope introduced after the watermarking system is developed.`` & The company develops and implements provenance and/or watermarking systems for visual content created by any of their publicly available AI models. \\
\midrule
Develop tools or APIs to determine if a particular piece of content was created within their tools & ``They will also develop tools or APIs to determine if a particular piece of content was created with their system.`` & The company develops tools to verify if a piece of content was created with their tools. \\
\midrule
Work with industry peers and standards-setting bodies towards developing a technical framework & ``More generally, companies making this commitment pledge to work with industry peers and standards-setting bodies as appropriate towards developing a technical framework to help users distinguish audio or visual content generated by users from audio or visual content generated by AI.`` & The company collaborates with others to work towards developing a standard technical framework for content authentication in AI. \\
\midrule
Report capabilities & ``Publicly report model or system capabilities ...``. & The company provides public documentation of their flagship model and system capabilities. \\
\midrule
Report limitations & ``These reports should include ... significant limitations in performance that have implications for the domains of appropriate use`` & The company documents limitations that are specific to their model and systems, and not limitations about AI systems generally.\\
\midrule
Report domains of appropriate use & ``Publicly report ... domains of appropriate and inappropriate use`` & The company publicly reports the intended uses for their models and systems. \\
\midrule
Report domains of inappropriate use & ``Publicly report ... domains of appropriate and inappropriate use`` & The company publicly discloses the unintended and prohibited uses for their models and systems. \\
\midrule
Report safety evaluations & ``These reports should include the safety evaluations conducted (including in areas such as dangerous capabilities, to the extent that these are responsible to publicly disclose)`` & The company publicly reports the safety evaluations conducted on their models and systems. These evaluations target severe safety risks such as CBRN and child safety and not just model toxicity and bias. \\
\midrule
Report on societal risks & ``These reports should include ... discussion of the model’s effects on societal risks such as fairness and bias`` & The company describes the effects of model on societal risks, such as fairness and bias, in public reports. \\
\midrule
Report on adversarial testing used to determine appropriateness of deployment & ``These reports should include ... the results of adversarial testing conducted to evaluate the model’s fitness for deployment.`` & The company publicly reports the results of adversarial testing conducted to evaluate the model's fitness for deployment. \\
\midrule
Empower trust and safety teams & ``Companies commit generally to empowering trust and safety teams`` & The company empowers an explicitly mentioned trust and safety team. Empowerment entails providing the team with sufficient resources and authority to monitor and address potential risks such as bias and misinformation. \\
\midrule
Advance AI safety research & ``Companies commit generally to ... advancing AI safety research`` & The company produces research or develops research tools to implement safeguards and conduct safety evaluations. \\
\midrule
Advance privacy & ``Companies commit generally to ... advancing privacy`` & Efforts to reduce privacy risks must be associated with prompt-conditioned LM generation and/or the storage of user data. \\
\midrule
Protect children & ``Companies commit generally to ... protecting children, and working to proactively manage the risks of AI so that its benefits can be realized.`` & "The company partners with organizations such as Thorn and NCMEC to combat CSAM. It designs AI services and releases tools to reduce the risk of AI misuse for child exploitation. 
Funding external research on CSAM does not constitute protecting children." \\
\midrule
Support research and development of frontier AI systems that can help meet society's greatest challenges & ``Companies making this commitment agree to support research and development of frontier AI systems that can help meet society’s greatest challenges, such as climate change mitigation and adaptation, early cancer detection and prevention, and combating cyber threats.`` & The contributions the company makes must extend beyond funding to advance research through the deployment of their flagship models, close collaboration, or resource sharing. Initiatives must be driven by public benefit rather than commercial gain.
Efforts should target fundamental and widely recognized societal issues. \\
\midrule
Support initiatives that foster the education and training of students and workers & ``Companies also commit to supporting initiatives that foster the education and training of students and workers to prosper from the benefits of AI`` & The company supports initiatives to train students and workers in developing AI literacy and the skills to harness the benefits of AI. This support can be in the form of funding for educational programs dedicated to AI literacy. These educational initiatives should be accessible to all students and workers. These initiatives should be focused on AI literacy and not investments with potential downstream educational impact. These initiatives should not be profit-driven. These investments should be long-term rather than a one-time occurrence. \\
\midrule
Support initiatives that help citizens understand the technology & ``Companies also commit ... to helping citizens understand the nature, capabilities, limitations, and impact of the technology`` & The goal of these initiatives must be to improve citizens' understanding of the nature, capabilities, limitations, and impact of the technology. 
Initiatives that are focused solely on the public engagement or input process without an educational component are out of scope.
These initiatives should provide full coverage of the nature,  capabilities, limitations, and impact of the technology.
These initiatives should be sustained, rather than a one-time occurrence. \\

\end{longtable}

\bibliography{main}

\begin{thebibliography}{215}
\providecommand{\natexlab}[1]{#1}

\bibitem[{Adobe(2023)}]{adobechild}
Adobe. 2023.
\newblock Adobe’s Commitment to Child Safety.

\bibitem[{Adobe(2024{\natexlab{a}})}]{adobeguide}
Adobe. 2024{\natexlab{a}}.
\newblock Adobe Generative AI User Guidelines.

\bibitem[{Adobe(2024{\natexlab{b}})}]{adobefirefly3}
Adobe. 2024{\natexlab{b}}.
\newblock Adobe Introduces Firefly Image 3 Foundation Model to Take Creative Exploration and Ideation to New Heights.

\bibitem[{Adobe(2024{\natexlab{c}})}]{adobeskill}
Adobe. 2024{\natexlab{c}}.
\newblock Adobe Introduces New Global Initiative Aimed at Helping 30 Million Next-Generation Learners Develop AI Literacy, Content Creation and Digital Marketing Skills by 2030.

\bibitem[{Adobe(2024{\natexlab{d}})}]{adobelimit}
Adobe. 2024{\natexlab{d}}.
\newblock Known Limitations in Firefly.

\bibitem[{Adobe(2024{\natexlab{e}})}]{adobeaca}
Adobe. 2024{\natexlab{e}}.
\newblock Media Alert: Adobe Introduces Adobe Content Authenticity Web App to Champion Creator Protection and Attribution.

\bibitem[{Adobe(n.d.)}]{adobecontentauth}
Adobe. n.d.
\newblock Content Authenticity.

\bibitem[{{{A}{I} Lab Watch}(n.d.)}]{ailabwatchCommitments}
{{A}{I} Lab Watch}. n.d.
\newblock {C}ommitments.
\newblock \url{https://ailabwatch.org/resources/commitments/}.

\bibitem[{{Amazon}(2025)}]{amazonbugbounty}
{Amazon}. 2025.
\newblock Amazon Vulnerability Research Program.
\newblock \url{https://hackerone.com/amazonvrp?type=team}.
\newblock Accessed: 2025-05-24.

\bibitem[{{Amazon Science}()}]{amazonprivacy}
{Amazon Science}. ????
\newblock Security, Privacy and Abuse Prevention Research.

\bibitem[{{Amazon Science}(2024)}]{amazonscience}
{Amazon Science}. 2024.
\newblock Amazon Nova and Our Commitment to Responsible AI.

\bibitem[{{Amazon Web Services}()}]{amazoncard}
{Amazon Web Services}. ????
\newblock Amazon Nova Micro, Lite, and Pro - AWS AI Service Cards.

\bibitem[{{Amazon Web Services}(2024)}]{amazontitan}
{Amazon Web Services}. 2024.
\newblock Amazon Titan Text Premier - AWS AI Service Cards.

\bibitem[{Anca~Dragan and Dafoe(2024)}]{frontiersafety}
Anca~Dragan, H.~K.; and Dafoe, A. 2024.
\newblock Introducing the Frontier Safety Framework.

\bibitem[{Anthropic(2023{\natexlab{a}})}]{anthropiccollective}
Anthropic. 2023{\natexlab{a}}.
\newblock Collective Constitutional AI: Aligning a Language Model with Public Input.

\bibitem[{Anthropic(2023{\natexlab{b}})}]{anthropicviews}
Anthropic. 2023{\natexlab{b}}.
\newblock Core Views on AI Safety: When, Why, What, and How.

\bibitem[{Anthropic(2023{\natexlab{c}})}]{anthropicsecurity}
Anthropic. 2023{\natexlab{c}}.
\newblock Frontier Model Security.

\bibitem[{Anthropic(2023{\natexlab{d}})}]{anthropicfrontier}
Anthropic. 2023{\natexlab{d}}.
\newblock Frontier Threats Red Teaming for AI Safety.

\bibitem[{Anthropic(2024{\natexlab{a}})}]{anthropicchild}
Anthropic. 2024{\natexlab{a}}.
\newblock Aligning on child safety principles.

\bibitem[{Anthropic(2024{\natexlab{b}})}]{anthropicredteam}
Anthropic. 2024{\natexlab{b}}.
\newblock Challenges in Red Teaming AI Systems.

\bibitem[{Anthropic(2024{\natexlab{c}})}]{anthropiccard}
Anthropic. 2024{\natexlab{c}}.
\newblock The Claude 3 Model Family: Opus, Sonnet, Haiku.

\bibitem[{Anthropic(2024{\natexlab{d}})}]{anthropicaddendum}
Anthropic. 2024{\natexlab{d}}.
\newblock Claude 3.5 Sonnet Model Card Addendum.

\bibitem[{Anthropic(2024{\natexlab{e}})}]{anthropicbounty}
Anthropic. 2024{\natexlab{e}}.
\newblock Expanding our model safety bug bounty program.

\bibitem[{Anthropic(2024{\natexlab{f}})}]{anthropicwh}
Anthropic. 2024{\natexlab{f}}.
\newblock Tracking Voluntary Commitments.

\bibitem[{{Anthropic}(2025)}]{anthropiccourses}
{Anthropic}. 2025.
\newblock Anthropic Courses.
\newblock Accessed: 2025-05-24.

\bibitem[{Apple(2024{\natexlab{a}})}]{appleacademy}
Apple. 2024{\natexlab{a}}.
\newblock Apple Developer Academy Introduces AI Training for All Students and Alumni.

\bibitem[{Apple(2024{\natexlab{b}})}]{appleprivacy}
Apple. 2024{\natexlab{b}}.
\newblock Explore advancements in Machine Learning.

\bibitem[{Barrett, Newman, and Nonnecke(2023)}]{barrett2023AI-RMS}
Barrett, A.; Newman, J.; and Nonnecke, B. 2023.
\newblock AI Risk-Management Standards Profile for General-Purpose AI Systems (GPAIS) and Foundation Models.
\newblock \url{https://cltc.berkeley.edu/wp-content/uploads/2023/11/Berkeley-GPAIS-Foundation-Model-Risk-Management-Standards-Profile-v1.0.pdf#page=83}.

\bibitem[{Behnchen(2023)}]{msskills}
Behnchen, K. 2023.
\newblock Microsoft Launches New AI Skills Training and Resources as part of Skill for Jobs Initiative.

\bibitem[{Berry(2024{\natexlab{a}})}]{nvidiaaisic}
Berry, R. 2024{\natexlab{a}}.
\newblock National Institute of Standards and Technology Launches Artificial Intelligence Safety Institute Consortium.

\bibitem[{Berry(2024{\natexlab{b}})}]{nvidiaglobal}
Berry, R. 2024{\natexlab{b}}.
\newblock Nvidia Partners for Globally Inclusive AI in U.S. Government Initiative.

\bibitem[{Bommasani et~al.(2024)Bommasani, Klyman, Kapoor, Longpre, Xiong, Maslej, and Liang}]{bommasani2024foundationmodeltransparencyindex}
Bommasani, R.; Klyman, K.; Kapoor, S.; Longpre, S.; Xiong, B.; Maslej, N.; and Liang, P. 2024.
\newblock The Foundation Model Transparency Index v1.1: May 2024.
\newblock arXiv:2407.12929.

\bibitem[{Bommasani et~al.(2023{\natexlab{a}})Bommasani, Klyman, Longpre, Kapoor, Maslej, Xiong, Zhang, and Liang}]{bommasani2023fmti}
Bommasani, R.; Klyman, K.; Longpre, S.; Kapoor, S.; Maslej, N.; Xiong, B.; Zhang, D.; and Liang, P. 2023{\natexlab{a}}.
\newblock The Foundation Model Transparency Index.
\newblock \emph{ArXiv}, abs/2310.12941.

\bibitem[{Bommasani et~al.(2023{\natexlab{b}})Bommasani, Klyman, Zhang, and Liang}]{bommasani2023eu-ai-act}
Bommasani, R.; Klyman, K.; Zhang, D.; and Liang, P. 2023{\natexlab{b}}.
\newblock Do Foundation Model Providers Comply with the EU AI Act?

\bibitem[{Bommasani et~al.(2023{\natexlab{c}})Bommasani, Zhang, Lee, and Liang}]{bommasani2023transparency}
Bommasani, R.; Zhang, D.; Lee, T.; and Liang, P. 2023{\natexlab{c}}.
\newblock Improving Transparency in AI Language Models: A Holistic Evaluation.
\newblock \emph{Foundation Model Issue Brief Series}.

\bibitem[{Broxmeyer(2024)}]{meta2024communityforums}
Broxmeyer, J. 2024.
\newblock Leading the Way in Governance Innovation With Community Forums on AI.
\newblock Accessed: 2025-05-24.

\bibitem[{{C2PA}(2024)}]{amazonc2pa}
{C2PA}. 2024.
\newblock Amazon Joins the C2PA Steering Committee.

\bibitem[{Caulfield(2024)}]{nvidiasummit}
Caulfield, B. 2024.
\newblock NVIDIA AI Summit Highlights Game-Changing Energy Efficiency and AI-Driven Innovation.

\bibitem[{Charlet(2024)}]{googleprivacy}
Charlet, K. 2024.
\newblock Designing for Privacy in an {AI} World.

\bibitem[{Chegg(2023)}]{scalechegg}
Chegg. 2023.
\newblock Chegg Partners with Scale AI to Enhance Learning Experience for Students.

\bibitem[{Chockalingam and Varshney(2023)}]{nvidiasafe}
Chockalingam, A.; and Varshney, T. 2023.
\newblock NVIDIA Enables Trustworthy, Safe, and Secure Large Language Model Conversational Systems.

\bibitem[{Clegg(2024)}]{metadavos}
Clegg, N. 2024.
\newblock On AI, Progress and Vigilance Can Go Hand in Hand.

\bibitem[{Cohere(2023)}]{coheresenate}
Cohere. 2023.
\newblock U.S. Senate AI Insight Forum: Innovation - Cohere’s Written Submissions.

\bibitem[{Cohere(2024{\natexlab{a}})}]{cohereuse}
Cohere. 2024{\natexlab{a}}.
\newblock Command R and Command R+ Model Card.

\bibitem[{Cohere(2024{\natexlab{b}})}]{coheredocs}
Cohere. 2024{\natexlab{b}}.
\newblock The Command R Model (Details and Application).

\bibitem[{Cohere(2024{\natexlab{c}})}]{cohereaya}
Cohere. 2024{\natexlab{c}}.
\newblock Introducing Aya.

\bibitem[{{Cohere}(2025)}]{coherellmu}
{Cohere}. 2025.
\newblock LLM University.
\newblock Accessed: 2025-05-24.

\bibitem[{Cohere(n.d.{\natexlab{a}})}]{coheresecurity}
Cohere. n.d.{\natexlab{a}}.
\newblock Industry-leading AI security and data protection.

\bibitem[{Cohere(n.d.{\natexlab{b}})}]{coheredisclosure}
Cohere. n.d.{\natexlab{b}}.
\newblock Responsible Disclosure Policy.

\bibitem[{{Common Sense Media}(2024)}]{openaicommonsense}
{Common Sense Media}. 2024.
\newblock Common Sense Media and OpenAI Launch Free AI Training Course for K-12 Educators.

\bibitem[{Crampton(2023)}]{crampton2023reflecting}
Crampton, N. 2023.
\newblock Reflecting on our responsible AI program: Three critical elements for progress.
\newblock Accessed: 2025-05-24.

\bibitem[{CSIS(2023)}]{scalecsis}
CSIS. 2023.
\newblock CSIS Futures Lab Announces Partnership with Scale AI.

\bibitem[{Cuadros et~al.(2024)Cuadros, Delobelle, Susa, Joulin, Apostoloff, Zappella, and Lopez}]{applewhisper}
Cuadros, X.~S.; Delobelle, P.; Susa, R.~M.; Joulin, A.; Apostoloff, N.; Zappella, L.; and Lopez, P.~R. 2024.
\newblock Whispering Experts: Toxicity Mitigation in Pre-trained Language Models by Dampening Expert Neurons.

\bibitem[{Dominguez(2024)}]{stabilityamazon}
Dominguez, D. 2024.
\newblock Stability AI Announces Integration of Top Text-to-Image Models with Amazon Bedrock.

\bibitem[{Dylan~Slack(2023)}]{scaleevaluation}
Dylan~Slack, D. S. K. P. S.~H., Jean~Wang. 2023.
\newblock A HOLISTIC APPROACH FOR TEST AND EVALUATION OF LARGE LANGUAGE MODELS.
\newblock Technical report, {Scale AI}.

\bibitem[{Engineering et~al.(2024)Engineering, Architecture~(SEAR), Learning, and (AIML)}]{applepcc}
Engineering, A.~S.; Architecture~(SEAR), C. O. S. C. O. S. E.~A., User~Privacy; Learning, M.; and (AIML), A. 2024.
\newblock Private Cloud Compute: A new frontier for AI privacy in the cloud.

\bibitem[{Engineering and (SEAR)(2024)}]{applebounty}
Engineering, A.~S.; and (SEAR), A. 2024.
\newblock Security research on Private Cloud Compute.

\bibitem[{Erramilli(2024)}]{salesforce_insider_threats_2024}
Erramilli, V. 2024.
\newblock How Salesforce Helps Protect You from Insider Threats.

\bibitem[{Esser et~al.(2024)Esser, Kulal, Blattmann, Entezari, Müller, Saini, Levi, Lorenz, Sauer, Boesel, Podell, Dockhorn, English, Lacey, Goodwin, Marek, and Rombach}]{stabilitytech}
Esser, P.; Kulal, S.; Blattmann, A.; Entezari, R.; Müller, J.; Saini, H.; Levi, Y.; Lorenz, D.; Sauer, A.; Boesel, F.; Podell, D.; Dockhorn, T.; English, Z.; Lacey, K.; Goodwin, A.; Marek, Y.; and Rombach, R. 2024.
\newblock Scaling Rectified Flow Transformers for High-Resolution Image Synthesis.
\newblock arXiv:2403.03206.

\bibitem[{Fabian and Crisp(2023)}]{fabian2023redteam}
Fabian, D.; and Crisp, J. 2023.
\newblock Why Red Teams Play a Central Role in Helping Organizations Secure AI Systems.
\newblock Technical report, Google.
\newblock Accessed: 2025-05-24.

\bibitem[{Finkle(2023)}]{nvidiawashington}
Finkle, N. 2023.
\newblock NVIDIA Lends Support to Washington’s Efforts to Ensure AI Safety.

\bibitem[{{Frontier Model Forum}(2024{\natexlab{a}})}]{fmfaboutus}
{Frontier Model Forum}. 2024{\natexlab{a}}.
\newblock About Us.
\newblock \url{https://www.frontiermodelforum.org/about-us}.

\bibitem[{{Frontier Model Forum}(2024{\natexlab{b}})}]{fmfCassessments}
{Frontier Model Forum}. 2024{\natexlab{b}}.
\newblock Company Assessments.
\newblock \url{https://www.frontiermodelforum.org/technical-reports/frontier-capability-assessments}.
\newblock Accessed: 2025-09-23.

\bibitem[{{Frontier Model Forum}(2024{\natexlab{c}})}]{fmfimplementingframeworks}
{Frontier Model Forum}. 2024{\natexlab{c}}.
\newblock Introducing the FMF’s Technical Report Series on Frontier AI Frameworks.
\newblock \url{https://www.frontiermodelforum.org/updates/introducing-the-fmfs-technical-report-series-on-frontier-ai-safety-frameworks}.
\newblock Accessed: 2025-09-23.

\bibitem[{{F}rontier~{M}odel {F}orum(n.d.)}]{frontiermodelforum}
{F}rontier~{M}odel {F}orum. n.d.
\newblock {F}rontier {M}odel {F}orum: {A}dvancing frontier {A}{I} safety and security.

\bibitem[{Gallegos et~al.(2024)Gallegos, Rossi, Barrow, Tanjim, Kim, Dernoncourt, Yu, Zhang, and Ahmed}]{adobefair}
Gallegos, I.~O.; Rossi, R.~A.; Barrow, J.; Tanjim, M.~M.; Kim, S.; Dernoncourt, F.; Yu, T.; Zhang, R.; and Ahmed, N.~K. 2024.
\newblock Bias and Fairness in Large Language Models: A Survey.
\newblock arXiv:2309.00770.

\bibitem[{{Gemini Team}(2024)}]{geminiteam2024gemini}
{Gemini Team}. 2024.
\newblock Gemini: A Family of Highly Capable Multimodal Models.
\newblock arXiv:2312.11805.

\bibitem[{Google(2023)}]{googleaiprogress}
Google. 2023.
\newblock {AI} Principles 2023 Progress Update.

\bibitem[{Google(2024)}]{gemmapolicy}
Google. 2024.
\newblock Gemma Prohibited Use Policy.

\bibitem[{{G}oogle(2024)}]{googlesecurity}
{G}oogle. 2024.
\newblock How Google protects its production services.

\bibitem[{{Google}(2025)}]{gemmacard}
{Google}. 2025.
\newblock Gemma Model Card.
\newblock Accessed: 2025-05-24.

\bibitem[{Google(n.d.)}]{googlewhitehouse}
Google. n.d.
\newblock Fulfilling the Voluntary Industry Commitments on {AI}.

\bibitem[{{Google DeepMind}(n.d.)}]{synthid}
{Google DeepMind}. n.d.
\newblock Synth{ID}.

\bibitem[{Granite(2024)}]{ibmguide}
Granite, I. 2024.
\newblock Responsible Use Guide.

\bibitem[{{Group of Seven}(2023)}]{g72023vc}
{Group of Seven}. 2023.
\newblock Hiroshima Process International Code of Conduct for Organizations Developing Advanced AI Syste.

\bibitem[{hEigeartaigh et~al.(2023)hEigeartaigh, Lannquist, Marcoci, Sevilla, Ruiz, Chaudhary, Schreier, Stein-Perlman, and Ladish}]{ohEigeartaigh2023ai-safety}
hEigeartaigh, S.~O.; Lannquist, Y.; Marcoci, A.; Sevilla, J.; Ruiz, M. A.~U.; Chaudhary, Y.; Schreier, T.; Stein-Perlman, Z.; and Ladish, J. 2023.
\newblock Do companies’ AI Safety Policies meet government best practice?
\newblock \url{https://www.lcfi.ac.uk/news-events/news/ai-safety-policies}.

\bibitem[{Heikkilä(2024)}]{heikkila2024MIT}
Heikkilä, M. 2024.
\newblock {A}{I} companies promised to self-regulate one year ago. {W}hat’s changed?

\bibitem[{Hutson(2024)}]{openairesilience}
Hutson, T. 2024.
\newblock Microsoft and OpenAI launch Societal Resilience Fund.

\bibitem[{IBM(2023)}]{ibmalliance}
IBM. 2023.
\newblock AI Alliance Launches as an International Community of Leading Technology Developers, Researchers, and Adopters Collaborating Together to Advance Open, Safe, Responsible AI.

\bibitem[{{IBM}(2023{\natexlab{a}})}]{ibmtraining}
{IBM}. 2023{\natexlab{a}}.
\newblock IBM Commits to Train 2 Million in Artificial Intelligence in Three Years, with a Focus on Underrepresented Communities.

\bibitem[{{IBM}(2023{\natexlab{b}})}]{ibmclimate}
{IBM}. 2023{\natexlab{b}}.
\newblock IBM Furthers Commitment to Climate Action Through New Sustainability Projects and Free Training in Green and Technology Skills for Vulnerable Communities.

\bibitem[{IBM(2024)}]{ibmprivacy}
IBM. 2024.
\newblock AI Privacy Toolkit.

\bibitem[{{IBM}(2025{\natexlab{a}})}]{ibmterms}
{IBM}. 2025{\natexlab{a}}.
\newblock IBM watsonx.ai Runtime as a Service.
\newblock Accessed: 2025-05-24.

\bibitem[{{IBM}(2025{\natexlab{b}})}]{ibmwatsonx}
{IBM}. 2025{\natexlab{b}}.
\newblock IBM watsonx.governance.
\newblock Accessed: 2025-05-24.

\bibitem[{IBM(n.d.)}]{ibmverify}
IBM. n.d.
\newblock IBM Verify Privileged Identity.

\bibitem[{{IBM Research}(2024)}]{ibmgranitecard}
{IBM Research}. 2024.
\newblock Granite-3.0-8B-Instruct.
\newblock Accessed: 2025-05-24.

\bibitem[{{Inflection AI}(2023{\natexlab{a}})}]{inflectionsafety}
{Inflection AI}. 2023{\natexlab{a}}.
\newblock Our policy on frontier safety.

\bibitem[{{Inflection AI}(2023{\natexlab{b}})}]{inflectionwh}
{Inflection AI}. 2023{\natexlab{b}}.
\newblock The precautionary principle: partnering with the White House on AI safety.

\bibitem[{{Inflection AI}(2023{\natexlab{c}})}]{inflectionpolicy}
{Inflection AI}. 2023{\natexlab{c}}.
\newblock Privacy Policy.

\bibitem[{{Inflection AI}(2025)}]{inflectiondocs}
{Inflection AI}. 2025.
\newblock Inflection AI Developer Documentation.
\newblock Accessed: 2025-05-24.

\bibitem[{Intelligence(2024)}]{amazontech}
Intelligence, A. A.~G. 2024.
\newblock The Amazon Nova Family of Models: Technical Report and Model Card.

\bibitem[{Intelligence(2023)}]{humaneintelligence2023grt}
Intelligence, H. 2023.
\newblock Generative AI Red Teaming Challenge.
\newblock Accessed: 2025-01-22.

\bibitem[{{ISED Canada}(2023)}]{canada2023vc}
{ISED Canada}. 2023.
\newblock Voluntary Code of Conduct on the Responsible Development and Management of Advanced Generative AI Systems.

\bibitem[{J.~Alberto Aragòn-Correa and Vogel(2020)}]{aragoncorrea2020effects}
J.~Alberto Aragòn-Correa, A. A.~M.; and Vogel, D. 2020.
\newblock The Effects of Mandatory and Voluntary Regulatory Pressures on Firms’ Environmental Strategies: A Review and Recommendations for Future Research.
\newblock \emph{Academy of Management Annals}.

\bibitem[{Jasper(2024)}]{googlechild}
Jasper, S. 2024.
\newblock An Update on Our Child Safety Efforts and Commitments.

\bibitem[{Johnson(2024)}]{googleorg}
Johnson, M. 2024.
\newblock Google.org announces new {A}{I} funding for students and educators.

\bibitem[{Jones()}]{adamjonesCompaniesDoing}
Jones, A. ????
\newblock {H}ow are {A}{I} companies doing with their voluntary commitments on vulnerability reporting?
\newblock \url{https://adamjones.me/blog/ai-vulnerability-reporting/}.

\bibitem[{Jr.(2023)}]{nvidiasecurity}
Jr., D.~R. 2023.
\newblock Six Steps Toward AI Security.

\bibitem[{Klyman(2024)}]{klyman2024acceptableusepoliciesfoundation}
Klyman, K. 2024.
\newblock Acceptable Use Policies for Foundation Models.
\newblock arXiv:2409.09041.

\bibitem[{Kogen(2024)}]{kogen2024rdr}
Kogen, L. 2024.
\newblock From Statistics to Stories: Indices and Indicators as Communication Tools for Social Change.
\newblock \emph{The International Journal of Press/Politics}, 29(4): 1090--1108.

\bibitem[{Krishan(2023)}]{msgovai}
Krishan, N. 2023.
\newblock Microsoft Rolls Out Generative AI Roadmap for Government Services.

\bibitem[{{Lakera Team}(2024)}]{coherelakera}
{Lakera Team}. 2024.
\newblock Lakera and Cohere Set the Bar for New Enterprise LLM Security Standards.

\bibitem[{Laurie~Richardson(2023)}]{googlebounty}
Laurie~Richardson, R.~H. 2023.
\newblock Acting on our commitment to safe and secure AI.

\bibitem[{Liguori and MacCárthaigh(2024)}]{amazonsecure}
Liguori, A.; and MacCárthaigh, C. 2024.
\newblock A Secure Approach to Generative AI with AWS.

\bibitem[{Longpre et~al.(2024)Longpre, Kapoor, Klyman, Ramaswami, Bommasani, Blili-Hamelin, Huang, Skowron, Yong, Kotha, Zeng, Shi, Yang, Southen, Robey, Chao, Yang, Jia, Kang, Pentland, Narayanan, Liang, and Henderson}]{longpre2024safeharbor}
Longpre, S.; Kapoor, S.; Klyman, K.; Ramaswami, A.; Bommasani, R.; Blili-Hamelin, B.; Huang, Y.; Skowron, A.; Yong, Z.-X.; Kotha, S.; Zeng, Y.; Shi, W.; Yang, X.; Southen, R.; Robey, A.; Chao, P.; Yang, D.; Jia, R.; Kang, D.; Pentland, S.; Narayanan, A.; Liang, P.; and Henderson, P. 2024.
\newblock A Safe Harbor for AI Evaluation and Red Teaming.
\newblock \emph{ArXiv}, abs/2403.04893.

\bibitem[{{Marah Abdin et al.}(2024)}]{msphi4}
{Marah Abdin et al.} 2024.
\newblock Phi-4 Technical Report.
\newblock Technical report.

\bibitem[{Martineau(2024)}]{ibmredteam}
Martineau, K. 2024.
\newblock What is Red Teaming for Generative AI.

\bibitem[{Meta()}]{metabounty}
Meta. ????
\newblock Meta Bug Bounty Program.

\bibitem[{Meta(2022{\natexlab{a}})}]{metaclimate}
Meta. 2022{\natexlab{a}}.
\newblock How AI is Helping Address the Climate Crisis.

\bibitem[{Meta(2022{\natexlab{b}})}]{metaalliance}
Meta. 2022{\natexlab{b}}.
\newblock Meta’s AI Learning Alliance aims to expand pipeline of underrepresented students in AI.

\bibitem[{Meta(2023{\natexlab{a}})}]{metapurple}
Meta. 2023{\natexlab{a}}.
\newblock Announcing Purple Llama: Towards open trust and safety in the new world of generative AI.

\bibitem[{Meta(2023{\natexlab{b}})}]{metasafety}
Meta. 2023{\natexlab{b}}.
\newblock Overview of Meta AI safety policies prepared for the UK AI Safety Summit.

\bibitem[{Meta(2024)}]{metallama}
Meta. 2024.
\newblock Expanding our open source large language models responsibly.

\bibitem[{{Meta}(2024{\natexlab{a}})}]{meta2024genaiguide}
{Meta}. 2024{\natexlab{a}}.
\newblock Generative AI Privacy Guide.
\newblock Accessed: 2025-05-24.

\bibitem[{{Meta}(2024{\natexlab{b}})}]{meta2024labeling}
{Meta}. 2024{\natexlab{b}}.
\newblock Labeling AI-Generated Images on Facebook, Instagram and Threads.
\newblock Accessed: 2025-05-24.

\bibitem[{Meta(2024)}]{metaaudio}
Meta. 2024.
\newblock Proactive Detection of Voice Cloning with Localized Watermarking.

\bibitem[{{Meta}(n.d.)}]{meta_privacy_progress}
{Meta}. n.d.
\newblock Privacy Progress.
\newblock Accessed: 2025-05-24.

\bibitem[{{Meta AI}(2023)}]{meta2023stablesignature}
{Meta AI}. 2023.
\newblock Stable Signature: A New Method for Watermarking Images Created by Open-Source Generative AI Models.
\newblock Accessed: 2025-05-24.

\bibitem[{{Meta AI}(2024)}]{meta2024llamacard}
{Meta AI}. 2024.
\newblock Llama 3.3 Model Card.
\newblock Accessed: 2025-05-24.

\bibitem[{Microsoft(2023)}]{mssafety}
Microsoft. 2023.
\newblock An update prepared for the UK AI Safety Summit.

\bibitem[{Microsoft(2024{\natexlab{a}})}]{msfederal}
Microsoft. 2024{\natexlab{a}}.
\newblock Accelerate your federal AI journey with Microsoft.

\bibitem[{Microsoft(2024{\natexlab{b}})}]{msaudio}
Microsoft. 2024{\natexlab{b}}.
\newblock Introducing the Watermark Algorithm for Synthetic Voice.

\bibitem[{Microsoft(2024{\natexlab{c}})}]{mschild}
Microsoft. 2024{\natexlab{c}}.
\newblock Microsoft Joins Thorn and All Tech Is Human to enact strong child safety commitments for generative AI.

\bibitem[{Microsoft(2024{\natexlab{d}})}]{msprivacy}
Microsoft. 2024{\natexlab{d}}.
\newblock Microsoft Trustworthy AI: Unlocking Human Potential Starts with Trust.

\bibitem[{Microsoft(2024{\natexlab{e}})}]{msphicard}
Microsoft. 2024{\natexlab{e}}.
\newblock Phi-4 Model Card.

\bibitem[{Microsoft(2024{\natexlab{f}})}]{msreport}
Microsoft. 2024{\natexlab{f}}.
\newblock Responsible AI Transparency Report.
\newblock Technical report, Microsoft.

\bibitem[{{Microsoft}(2025)}]{microsoft2025copilotbounty}
{Microsoft}. 2025.
\newblock Microsoft Copilot Bounty Program.
\newblock Accessed: 2025-05-24.

\bibitem[{Microsoft(2025)}]{mstoolkit}
Microsoft. 2025.
\newblock Microsoft Education AI Toolkit.
\newblock Technical report.

\bibitem[{{Microsoft Research}({\natexlab{a}})}]{mshealth}
{Microsoft Research}. ????{\natexlab{a}}.
\newblock AI for Health.

\bibitem[{{Microsoft Research}({\natexlab{b}})}]{msnoah}
{Microsoft Research}. ????{\natexlab{b}}.
\newblock The Prompt: with Trevor Noah.

\bibitem[{MLCommons(2023)}]{inflectionml}
MLCommons. 2023.
\newblock MLCommons Announces the Formation of AI Safety Working Group.

\bibitem[{Munoz et~al.(2024)Munoz, Minnich, Lutz, Lundeen, Dheekonda, Chikanov, Jagdagdorj, Pouliot, Chawla, Maxwell, Bullwinkel, Pratt, de~Gruyter, Siska, Bryan, Westerhoff, Kawaguchi, Seifert, Kumar, and Zunger}]{mspyrit}
Munoz, G. D.~L.; Minnich, A.~J.; Lutz, R.; Lundeen, R.; Dheekonda, R. S.~R.; Chikanov, N.; Jagdagdorj, B.-E.; Pouliot, M.; Chawla, S.; Maxwell, W.; Bullwinkel, B.; Pratt, K.; de~Gruyter, J.; Siska, C.; Bryan, P.; Westerhoff, T.; Kawaguchi, C.; Seifert, C.; Kumar, R. S.~S.; and Zunger, Y. 2024.
\newblock PyRIT: A Framework for Security Risk Identification and Red Teaming in Generative AI Systems.
\newblock arXiv:2410.02828.

\bibitem[{Murthy(2023)}]{scalellm}
Murthy, A. 2023.
\newblock Scale Unlocks Open-Source LLMs with New Platform and Partnership with Meta.

\bibitem[{{National Institute of Standards and Technology}(2024)}]{nist2024managing}
{National Institute of Standards and Technology}. 2024.
\newblock Managing Misuse Risk for Dual-Use Foundation Models.
\newblock Technical Report NIST AI 600-1, NIST.

\bibitem[{Nevo et~al.(2024)Nevo, Lahav, Karpur, Bar-On, Bradley, and Alstott}]{nevo2024securing}
Nevo, S.; Lahav, D.; Karpur, A.; Bar-On, Y.; Bradley, H.-A.; and Alstott, J. 2024.
\newblock \emph{Securing AI model weights: Preventing theft and misuse of frontier models}.
\newblock 1. Rand Corporation.

\bibitem[{NIST(2024)}]{nistaisic}
NIST. 2024.
\newblock {NIST} {AI} Safety Institute Consortium Members.

\bibitem[{{NIST}(2024)}]{nistsharing}
{NIST}. 2024.
\newblock U.S. AI Safety Institute Signs Agreements Regarding AI Safety Research, Testing and Evaluation With Anthropic and OpenAI.

\bibitem[{Nvidia()}]{nvidiamorpheus}
Nvidia. ????
\newblock NVIDIA Morpheus.

\bibitem[{Nvidia(2024{\natexlab{a}})}]{nvidiabase}
Nvidia. 2024{\natexlab{a}}.
\newblock Nemotron-4-340B-Base Model Card.

\bibitem[{Nvidia(2024{\natexlab{b}})}]{nvidiainstruct}
Nvidia. 2024{\natexlab{b}}.
\newblock Nemotron-4-340B-Instruct Model Card.

\bibitem[{Nvidia(2024{\natexlab{c}})}]{nvidiatech}
Nvidia. 2024{\natexlab{c}}.
\newblock Nemotron-4-340B Technical Report.
\newblock \emph{arXiv preprint}.

\bibitem[{Nvidia(2024{\natexlab{d}})}]{nvidialicense}
Nvidia. 2024{\natexlab{d}}.
\newblock NVIDIA Open Model License Agreement.

\bibitem[{on~AI~Staff(2024)}]{inflectionpai}
on~AI~Staff, P. 2024.
\newblock PAI Welcomes Four New Partners: Ada Lovelace, EY, Inflection AI, and North American Broadcasters Association.

\bibitem[{OpenAI(2023{\natexlab{a}})}]{openaibounty}
OpenAI. 2023{\natexlab{a}}.
\newblock Announcing OpenAI’s Bug Bounty Program.

\bibitem[{OpenAI(2023{\natexlab{b}})}]{openaired}
OpenAI. 2023{\natexlab{b}}.
\newblock OpenAI Red Teaming Network.

\bibitem[{OpenAI(2023{\natexlab{c}})}]{openaiapproach}
OpenAI. 2023{\natexlab{c}}.
\newblock Our Approach to AI Safety.

\bibitem[{OpenAI(2023{\natexlab{d}})}]{openaifrontier}
OpenAI. 2023{\natexlab{d}}.
\newblock Our Approach to Frontier Risk.

\bibitem[{OpenAI(2024{\natexlab{a}})}]{openaiprivacy}
OpenAI. 2024{\natexlab{a}}.
\newblock Consumer privacy at OpenAI.

\bibitem[{OpenAI(2024{\natexlab{b}})}]{openaicommittee}
OpenAI. 2024{\natexlab{b}}.
\newblock OpenAI Board Forms Safety and Security Committee.

\bibitem[{OpenAI(2024{\natexlab{c}})}]{openaio1}
OpenAI. 2024{\natexlab{c}}.
\newblock OpenAI o1 System Card.
\newblock Technical report.

\bibitem[{OpenAI(2024{\natexlab{d}})}]{openaisafety}
OpenAI. 2024{\natexlab{d}}.
\newblock OpenAI Safety Update.

\bibitem[{OpenAI(2024{\natexlab{e}})}]{openaiusage}
OpenAI. 2024{\natexlab{e}}.
\newblock OpenAI Usage policies.

\bibitem[{OpenAI(2024{\natexlab{f}})}]{openaisecure}
OpenAI. 2024{\natexlab{f}}.
\newblock Securing Research Infrastructure for Advanced AI.

\bibitem[{OpenAI(2024{\natexlab{g}})}]{openaiwater}
OpenAI. 2024{\natexlab{g}}.
\newblock Understanding the Source of What We See and Hear Online.

\bibitem[{OpenAI(n.d.)}]{openaihealth}
OpenAI. n.d.
\newblock Color Health.

\bibitem[{Orlando~Lugo(2024)}]{salesforcehack}
Orlando~Lugo, S.~T. 2024.
\newblock Ethical Hacking Practices Prove Successful in Building Trusted AI Products.

\bibitem[{Palantir(2023)}]{palantirethics}
Palantir. 2023.
\newblock Palantir Technologies’ Approach to AI Ethics.

\bibitem[{Palantir(2024{\natexlab{a}})}]{palantirprivacy}
Palantir. 2024{\natexlab{a}}.
\newblock Human Rights and Technology.

\bibitem[{Palantir(2024{\natexlab{b}})}]{palantirncmec}
Palantir. 2024{\natexlab{b}}.
\newblock Palantir \& NCMEC.

\bibitem[{Palantir(2024{\natexlab{c}})}]{palantirtes}
Palantir. 2024{\natexlab{c}}.
\newblock Palantir and Green Energy Pioneer TES Forge Long-Term Partnership to Drive Global Decarbonization.

\bibitem[{Palantir(2024{\natexlab{d}})}]{palantiromb}
Palantir. 2024{\natexlab{d}}.
\newblock Palantir's Response to OMB on AI Governance, Innovation, and Risk Management.

\bibitem[{Palantir(n.d.{\natexlab{a}})}]{palantiraip}
Palantir. n.d.{\natexlab{a}}.
\newblock AIP.

\bibitem[{Palantir(n.d.{\natexlab{b}})}]{palantirai}
Palantir. n.d.{\natexlab{b}}.
\newblock Artificial Intelligence Platform Now.

\bibitem[{Palantir(n.d.{\natexlab{c}})}]{palantirscholar}
Palantir. n.d.{\natexlab{c}}.
\newblock The Palantir Future Global Scholarship Program.

\bibitem[{Palantir(n.d.{\natexlab{d}})}]{palantirterms}
Palantir. n.d.{\natexlab{d}}.
\newblock Palantir terms and conditions.

\bibitem[{Pearce and Lucas(2023)}]{nvidiared}
Pearce, W.; and Lucas, J. 2023.
\newblock Nvidia AI Red Team: An Introduction.

\bibitem[{Philomin(2024)}]{amazonprogress}
Philomin, V. 2024.
\newblock A Progress Update on Our Commitment to Safe, Responsible Generative AI.

\bibitem[{Poccia(2024)}]{amazonnova}
Poccia, D. 2024.
\newblock Introducing Amazon Nova foundation models: Frontier intelligence and industry leading price performance.

\bibitem[{Rafieyan(2024)}]{openaiinsider}
Rafieyan, D. 2024.
\newblock OpenAI is Hiring Someone to Investigate its Own Employees.

\bibitem[{Ram Shankar Siva~Kumar(2024)}]{msredteam}
Ram Shankar Siva~Kumar, A. R.~T., Data~Cowboy. 2024.
\newblock Announcing Microsoft’s open automation framework to red team generative AI Systems.

\bibitem[{{Ranking Digital Rights}(2020)}]{rdr2020index}
{Ranking Digital Rights}. 2020.
\newblock 2020 Ranking Digital Rights Corporate Accountability Index.

\bibitem[{Rao(2023)}]{adobecommit}
Rao, D. 2023.
\newblock Building safe, secure, and trustworthy AI: Adobe’s commitments to our customers and community.

\bibitem[{Rao(2024)}]{adoberoad}
Rao, D. 2024.
\newblock The road ahead for responsible innovation.

\bibitem[{Research(2024)}]{ibmtech}
Research, I. 2024.
\newblock Granite 3.0 Technical Report.

\bibitem[{Roose(2023)}]{roose2023NYT}
Roose, K. 2023.
\newblock How Do the White House’s A.I. Commitments Stack Up?

\bibitem[{Salesforce(2023)}]{salesforcexgen}
Salesforce. 2023.
\newblock XGen-7B Technical Report.

\bibitem[{Salesforce(2024{\natexlab{a}})}]{salesforce_ibm_2024}
Salesforce. 2024{\natexlab{a}}.
\newblock IBM and Salesforce Expand Partnership to Advance Open, Trusted AI and Data Ecosystems.

\bibitem[{Salesforce(2024{\natexlab{b}})}]{salesforceeinstein}
Salesforce. 2024{\natexlab{b}}.
\newblock Salesforce Einstein Model Cards.
\newblock Technical report, Salesforce.

\bibitem[{Salesforce(2024{\natexlab{c}})}]{salesforcegrants}
Salesforce. 2024{\natexlab{c}}.
\newblock Salesforce Gives \$23M to Education to Help the AI Generation Unlock Critical Skills.

\bibitem[{Salesforce(2024{\natexlab{d}})}]{salesforceai}
Salesforce. 2024{\natexlab{d}}.
\newblock Tracking Our Progress on the White House Voluntary AI Commitments.

\bibitem[{{Scale AI}({\natexlab{a}})}]{scaledonovan}
{Scale AI}. ????{\natexlab{a}}.
\newblock Fine-Tuned LLMs for Defense.

\bibitem[{{Scale AI}({\natexlab{b}})}]{scalemodel}
{Scale AI}. ????{\natexlab{b}}.
\newblock Scale Evaluation.

\bibitem[{{Scale AI}({\natexlab{c}})}]{scalevision}
{Scale AI}. ????{\natexlab{c}}.
\newblock Test and Evaluation Vision.

\bibitem[{{Scale AI}(2022{\natexlab{a}})}]{scaleaup}
{Scale AI}. 2022{\natexlab{a}}.
\newblock Scale Acceptable Use Policy.

\bibitem[{{Scale AI}(2022{\natexlab{b}})}]{scaleukraine}
{Scale AI}. 2022{\natexlab{b}}.
\newblock Scale AI providing free datasets to help national security partners gain current insights into Russia-Ukraine conflict.

\bibitem[{{Scale AI}(2023{\natexlab{a}})}]{scaleseal}
{Scale AI}. 2023{\natexlab{a}}.
\newblock SEAL: Scale’s Safety, Evaluations and Alignment Lab.

\bibitem[{{Scale AI}(2023{\natexlab{b}})}]{scalewhite}
{Scale AI}. 2023{\natexlab{b}}.
\newblock Test and Evaluation White Paper.

\bibitem[{{Scale AI}(2024{\natexlab{a}})}]{scaledefense}
{Scale AI}. 2024{\natexlab{a}}.
\newblock Defense Llama: The LLM Purpose-Built for American National Security.

\bibitem[{{Scale AI}(2024{\natexlab{b}})}]{scaleresponsible}
{Scale AI}. 2024{\natexlab{b}}.
\newblock Responsible AI with Scale Evaluation for the Public Sector.

\bibitem[{{Scale AI}(2024{\natexlab{c}})}]{scaleexam}
{Scale AI}. 2024{\natexlab{c}}.
\newblock Submit Your Toughest Questions for Humanity's Last Exam.

\bibitem[{{Scale AI}(n.d.)}]{scaleleader}
{Scale AI}. n.d.
\newblock SEAL Leaderboards.

\bibitem[{Shaikh(2024)}]{ibmsecurity}
Shaikh, M. 2024.
\newblock Building the Future of AI Security: The IBM and Celerity Partnership.

\bibitem[{{Stability AI}({\natexlab{a}})}]{stabilitycollab}
{Stability AI}. ????{\natexlab{a}}.
\newblock Collaboration is Key to Safe Innovation.

\bibitem[{{Stability AI}({\natexlab{b}})}]{stabilitysafemain}
{Stability AI}. ????{\natexlab{b}}.
\newblock Stable Safety.

\bibitem[{{{Stability AI}}(2023{\natexlab{a}})}]{stabilityapi}
{{Stability AI}}. 2023{\natexlab{a}}.
\newblock Stability AI Previews Enhanced Image Offerings: APIs for Business \& New Product Features.

\bibitem[{{{Stability AI}}(2023{\natexlab{b}})}]{stabilitysenate}
{{Stability AI}}. 2023{\natexlab{b}}.
\newblock Statement to the U.S. Senate AI Insight Forum on Transparency, Explainability, and Copyright.

\bibitem[{{Stability AI}(2024{\natexlab{a}})}]{stabilitypolicy}
{Stability AI}. 2024{\natexlab{a}}.
\newblock Acceptable Use Policy.

\bibitem[{{Stability AI}(2024{\natexlab{b}})}]{stabilitynist}
{Stability AI}. 2024{\natexlab{b}}.
\newblock Response to Request for Information on Artificial Intelligence from the National Institute of Standards and T echnology.

\bibitem[{{{Stability AI}}(2024)}]{stabilitysafety}
{{Stability AI}}. 2024.
\newblock Stability AI Joins Thorn and All Tech Is Human to Enact Child Safety Commitments for Generative AI.

\bibitem[{{Stability AI}(2024)}]{stabilitycard}
{Stability AI}. 2024.
\newblock Stable Diffusion 3.5 Large Model Card.

\bibitem[{Stewart(2024)}]{nvidiaedu}
Stewart, L. 2024.
\newblock Golden Opportunities: California to Train Students, Educators in AI.

\bibitem[{Studios(2024)}]{stabilityhug}
Studios, H. 2024.
\newblock INNOVATION LABORATORY II.

\bibitem[{Tabassi(2023)}]{tabassi2023airmf}
Tabassi, E. 2023.
\newblock Artificial Intelligence Risk Management Framework (AI RMF 1.0).

\bibitem[{Team(2024{\natexlab{a}})}]{coherescholars}
Team, C. F.~A. 2024{\natexlab{a}}.
\newblock Cohere For AI Scholars Program: Research Journeys Start Here.

\bibitem[{Team(2024{\natexlab{b}})}]{coheregrants}
Team, C. F.~A. 2024{\natexlab{b}}.
\newblock Granting Access: Supporting Researchers to Use LLMs.

\bibitem[{Thorn(2019)}]{salesforcethorn}
Thorn. 2019.
\newblock Thorn Launches Initiative To Eliminate Child Sexual Abuse Material From The Internet.

\bibitem[{Thorn(2024)}]{inflectionthorn}
Thorn, A. T. I.~H. 2024.
\newblock Safety by Design for Generative AI: Preventing Child Sexual Abuse.

\bibitem[{{Tom Gunter et al.}(2024)}]{applefoundation}
{Tom Gunter et al.} 2024.
\newblock Apple Intelligence Foundation Language Models.
\newblock arXiv:2407.21075.

\bibitem[{Utpala, Hooker, and Chen(2023)}]{utpala2023dprompt}
Utpala, S.; Hooker, S.; and Chen, P.-Y. 2023.
\newblock Locally Differentially Private Document Generation Using Zero-Shot Prompting.
\newblock Accessed: 2025-05-24.

\bibitem[{Ventura(2024)}]{adoberedt}
Ventura, D. 2024.
\newblock Adobe Collaborates with Ethical Hackers to Build Safer, More Secure AI Tools.

\bibitem[{{White House}(2023)}]{whvc}
{White House}. 2023.
\newblock {E}nsuring {S}afe, {S}ecure, and {T}rustworthy {A}{I}.
\newblock \url{https://bidenwhitehouse.archives.gov/wp-content/uploads/2023/07/Ensuring-Safe-Secure-and-Trustworthy-AI.pdf}.

\bibitem[{{White House}(2024)}]{wh24vc}
{White House}. 2024.
\newblock White House Announces New Private Sector Voluntary Commitments to Combat Image-Based Sexual Abuse.
\newblock \url{https://bidenwhitehouse.archives.gov/ostp/news-updates/2024/09/12/white-house-announces-new-private-sector-voluntary-commitments-to-combat-image-based-sexual-abuse/}.

\bibitem[{{White House}(2025)}]{biden2025aiinfrastructure}
{White House}. 2025.
\newblock Advancing United States Leadership in Artificial Intelligence Infrastructure.
\newblock \url{https://www.federalregister.gov/documents/2025/01/17/2025-01395/advancing-united-states-leadership-in-artificial-intelligence-infrastructure}.

\bibitem[{Yue and Berrios(2024)}]{scalewmdp}
Yue, S.; and Berrios, D. 2024.
\newblock Introducing WMDP: Measuring and Mitigating Catastrophic Risk Potential from LLMs.

\bibitem[{Zarfati(2023)}]{msazure}
Zarfati, F. 2023.
\newblock Announcing new AI Safety \& Responsible AI features in Azure OpenAI Service at Ignite 2023.

\end{thebibliography}
\end{document}